\documentclass[aps,prb,twocolumn,groupedaddress]{revtex4}
\usepackage{graphicx}
\usepackage{epstopdf}
\usepackage{amsmath}
\usepackage{notes2bib}

\begin{document}
\title{Hydrogen bonding in acrylamide and its role in the scattering behavior of acrylamide-based block copolymers}

\author{Elena Patyukova}
\affiliation{Aston Institute of Materials Research, Aston University, Birmingham, B4 7ET, UK}
\email{patyukova@gmail.com}
\author{Taylor Rottreau}
\affiliation{Aston Institute of Materials Research, Aston University, Birmingham, B4 7ET, UK}
\author{Robert Evans}
\affiliation{Aston Institute of Materials Research, Aston University, Birmingham, B4 7ET, UK}
\author{Paul D.~Topham}
\affiliation{Aston Institute of Materials Research, Aston University, Birmingham, B4 7ET, UK}
\email{p.d.topham@aston.ac.uk}
\author{Martin J.~Greenall}
\affiliation{School of Mathematics and Physics, University of Lincoln, Brayford Pool, Lincoln, LN6 7TS, UK}
\email{mgreenall@lincoln.ac.uk}

\date{\today}

\begin{abstract}
Hydrogen bonding plays a role in the microphase separation behavior of many block copolymers, such as those used in lithography\cite{Kwak2017FabricationNanolithography}, where the stronger interactions due to H-bonding can lead to a smaller period for the self-assembled structures, allowing the production of higher resolution templates. However, current statistical thermodynamic models used in descriptions of microphase separation, such as the Flory-Huggins approach, do not take into account some important properties of hydrogen bonding, such as site specificity and cooperativity. In this combined theoretical and experimental study, a step is taken toward the development of a more complete theory of hydrogen bonding in polymers, using polyacrylamide as a model system.  We begin by developing a set of association models\cite{Coleman1995} to describe hydrogen bonding in amides. Both models with one association constant and two association constants are considered. This theory is used to fit IR spectroscopy data from acrylamide solutions in chloroform, thereby determining the model parameters. These parameters are then employed to calculate the scattering function of the disordered state of a diblock copolymer with one polyacrylamide block and one non-hydrogen-bonding block in the random phase approximation. It is then shown that the expression for the inverse scattering function with hydrogen bonding is the same as that without hydrogen bonding, but with the Flory-Huggins parameter $\chi$ replaced by an effective value $\chi_\text{eff}=\chi+\delta\chi_\text{HB}\left(f\right)$, where the hydrogen-bonding contribution $\delta\chi_\text{HB}$ depends on the volume fraction $f$ of the hydrogen-bonding block. We find that models with two constants give better predictions of bond energy in the acrylamide dimer and more realistic asymptotic behavior of the association constants and $\delta\chi_\text{HB}$ in the limit of high temperatures.

\end{abstract}


\maketitle

\section{Introduction}
Hydrogen bonding interactions occur very widely in nature. Although individual H-bonds are relatively weak, their effect on the physical properties of substances can be profound, and is responsible for the anomalous properties of water and the secondary structure of proteins. However, the characteristics of hydrogen bonding, such as site specificity and cooperativity, make it difficult to build a general theoretical description of H-bonding systems\cite{Mahadevi2016}.   

One of the most natural ways to describe the thermodynamics of the formation of hydrogen bonds is to treat this phenomenon as a reversible chemical reaction. This {\em association model} approach (sometimes called the ERAS model\cite{Vasiltsova2007NewMixtures}) was initially proposed to describe hydrogen bonding association in alcohols and polyalcohols \cite{Coleman1995}. In the framework of this model, it is assumed that alcohols in the liquid state form a full range of linear chain aggregates due to hydrogen-bonding association. It was shown that chemical equilibrium in this kind of system is described well by two association constants, one corresponding to the formation of a dimer (dimer association constant) and the other corresponding to addition of further molecules to the chain (multimer association constant). 

The association model of hydrogen bond formation in alcohols was later applied by Painter and Coleman to describe the miscibility of hydrogen-bonding homopolymers\cite{Coleman1995,Kuo2008}. They showed that association constants measured for low molecular weight analogs of polymer segments can be used (after rescaling in order to take into account the difference between the molar volumes of monomers and polymer segments) to describe hydrogen bonding in polymer systems. This approach has several strengths: (i) its parameters are measurable quantities, (ii) it treats non-hydrogen-bonding interactions and hydrogen-bonding interactions separately, (iii) the number of hydrogen bonded contacts is not random,  and (iv) it works as an extension of the Flory-Huggins theory of polymer melts, which is the basic theoretical platform in polymer physics\cite{Matsen2006Self-ConsistentApplications}. However, as this work was focused specifically on alcohols, hydrogen bonding in homopolymers and diblock copolymers is still often described by means of a negative Flory-Huggins parameter \cite{Dehghan2013,Han2011PhaseBonding,Sunday2016}$^,$\bibnote{For two cases of alternative approaches see\cite{Lefevre2010Self-assemblyBonds,Zhang2015SupramolecularPotentials}}.

Taking into account the virtues of the association model approach, it is useful to extend it to other classes of self-associating hydrogen-bonding compounds such as amides and acids. Here, we develop a set of association models for amides and test them by comparison with IR absorption measurements on acrylamide solutions.

The choice of acrylamide as a system to study is motivated by the current interest in the properties of its corresponding polymer, polyacrylamide. Polyacrylamide is a commercially important polymer which, in addition to its uses in chromatographic columns, soft contact lenses and cosmetics, is now finding applications in the areas of biomaterials and smart materials research \cite{Liao2017PH-Microcapsules,Sun2012HighlyHydrogels}.  A key factor in these applications is hydrogen bonding: the acrylamide group has both hydrogen donor and acceptor sites and can serve as a universal hydrogen bonding agent.  

In addition to the strengths of the association model listed above, we also believe that it will yield insights into hydrogen-bonded acrylamide aggregates that would be difficult, or even impossible, to obtain by other techniques, such as density functional theory (DFT) \cite{Duarte2005, Wang2016Hydrogen-bondingAnalysis} and molecular dynamics simulations \cite{Bako2010HydrogenFormamide}. DFT is a powerful tool that gives many valuable insights into the physics of hydrogen bonding, such as the effect of the conformation and relative positions of the molecules on the energy of the hydrogen bonding interaction. It can also be used to investigate hydrogen-bonded clusters, and such studies have been carried out for acetamide \cite{Esrafili2008TheoreticalClusters, Mahadevi2011A115}. However, to the best of our knowledge, existing studies on acrylamide focus on the structure and spectral features of isolated molecules and hydrogen-bonded dimers and do not provide any information on networks of hydrogen bonds or on the entropy of hydrogen bond formation in an ensemble of molecules.

Molecular dynamics (MD) has also been used to study hydrogen-bonded networks and, for example, has been used to investigate liquid formamide \cite{Bako2010HydrogenFormamide}. However, there are questions about the use of MD in these systems, since it was shown in the case of alcohols that molecular dynamics simulations using common force fields do not reproduce the spectral features of hydrogen-bonded aggregates in solution, even on a qualitative level \cite{Stubbs2005ElucidatingN-hexane}. Furthermore, there is always a constraint on the size of the system in molecular dynamics simulations, which limits the possibility of simulating the true size distribution of aggregates in solution. 

The main difficulty in constructing an association model of acrylamide is that there is not enough information about the ``rules'' of association and the minimal number of association constants necessary to describe association. In the literature of the interpretation of IR data from solutions of amides, all models that we are aware of use linear chain aggregates or cyclic dimers\cite{Spencer1980AmideMedium}. In the crystal phase, acrylamide is known to form two-dimensional ribbon-like networks of hydrogen bonds\cite{Udovenko2008CRYSTALACRYLAMIDE} with two bonds per oxygen, two bonds per $\text{NH}_2$ group and ribbons built from the double-bonding of cyclic dimers. However, we believe that, in both the solution and the melt, the association model may be different. For example, for relatively large acetamide clusters\cite{Mahadevi2011A115} with aggregation numbers up to $i=15$, it was shown by means of DFT simulations that clusters with ``irregular'' (as well as linear and cyclic) structure have lower energies than clusters constructed from crystal polymorphs, and we expect similar behavior for acrylamide.  

Our strategy to deal with the this uncertainty is to develop a set of models with different association rules. We start with models with one association constant (in other words, in all these models we assume that all hydrogen bonds have the same energy regardless of their position inside the aggregate). These models are then applied to interpret IR spectroscopy data from acrylamide solutions in chloroform and determine the association constants.   
Next, models with two association constants are investigated. These give better correspondence to the bond energies in acrylamide dimers predicted by DFT\cite{Wang2016Hydrogen-bondingAnalysis, Duarte2005}.

At the end of the paper, we take a first step toward the application of the association model approach to the description of hydrogen bonding in block copolymers by calculating the structure factor of a disordered melt of block copolymer with one hydrogen-bonding self-associating block and one non-hydrogen-bonding block in the random phase approximation (RPA). Next, the models and parameters determined for acrylamide are used to estimate how the Flory-Huggins parameter that appears in the RPA formula for the structure factor of the diblock copolymer with a polyacrylamide block is shifted by the presence of hydrogen bonds in these systems. It should be noted that these results can be considered only as preliminary, because, though our calculations take into account the non-randomness of hydrogen bonding contacts, they do not take into account the non-randomness of mixing in polymer systems, which should be accounted for in the future.

\section{Models with one association constant}

According to one definition, a hydrogen bond is an attractive short-ranged force between a hydrogen atom bonded to a strongly electronegative atom and another electronegative atom with a lone pair of electrons. 
\begin{figure}
\includegraphics{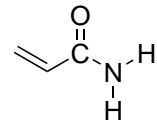}
\caption{Acrylamide molecule.}
\label{acrylamide}
\end{figure}

Acrylamide has a primary amide group (see Figure~\ref{acrylamide}), which consists of an oxygen atom (a hydrogen acceptor) and an $\text{NH}_2$ group (a hydrogen donor). The oxygen atom can potentially form two hydrogen bonds as it has two lone electron pairs. Since it has two hydrogens, the $\text{NH}_2$ group can also potentially form two bonds. However, it is also possible that the oxygen atom will predominantly form only one bond (as in alcohols) \cite{Bohmer2014StructureDebye} or that the second hydrogen on the $\text{NH}_2$ group will lose its donating properties after the first hydrogen becomes bonded \cite{Milani2010HydrogenCalculations}. Different rules of association produce aggregates of different architectures. If we allow only one bond per oxygen and one bond per $\text{NH}_2$ group, then the hydrogen-bonding association of acrylamide molecules produces linear chains. In all other cases, branched aggregates are produced. 

From both experimental studies and DFT simulations it is known that tautomerism is not present in acrylamide because imidic acid has an energy approximately $11\,\text{kcal/mol}$ higher than the ground state energy of the syn-isomer\cite{Wang2016Hydrogen-bondingAnalysis}, so we do not consider the possibility of tautomerism in our models. Isomerism is also not considered, because $95\%$ 
of the molecules are in the syn-isomer state at room temperature due to the large difference in ground state energy between the two isomers \bibnote{This estimation is made using Boltzmann distribution and corresponding ground state energies\cite{Wang2016Hydrogen-bondingAnalysis}}. In all of our models, it is also assumed that the two hydrogens are equivalent and the slight difference in electro-negativity between them is neglected.

In this section, we consider models for which it is assumed that all hydrogen bonds have the same energy regardless of their position in the aggregate and that there are no cycles of any kind. The list of models generated by these assumptions and different association rules is presented in Table~\ref{list-models-1}. 

\begin{table}
  \caption{List of models with one association constant and no cycles.}
  \label{list-models-1}
  \begin{tabular}{ |l | l | }
    \hline
    Model & association rules \\  
    \hline
    \hline
    \textbf{0} & only linear dimers \\
    \hline
    \textbf{1} & one bond per oxygen, two bonds per $\text{NH}_2$ group\\
    \hline
    \textbf{2} & two bonds per oxygen, two bonds per $\text{NH}_2$ group \\
    \hline
    \textbf{3} & one bond per oxygen, one bond per $\text{NH}_2$ group \\
    \hline
    \textbf{4} & two bonds per oxygen, one bond per $\text{NH}_2$ group \\
    \hline
  \end{tabular}
\end{table}

In order to illustrate our calculation method, model 1 is used as an exemplar. The calculations for all other models can be found in the supplementary materials.

In model 1, it is assumed that the rules of association can be formulated as ``one bond per oxygen, two bonds per $\text{NH}_2$ group''. The starting assumption is that the free energy of the solution can be written as
\begin{equation}
F=F_0+F_\text{HB},
\end{equation}
where $F_0$ is the free energy of solution without any hydrogen bonding association of the dissolved molecules and $F_\text{HB}$ is the contribution due to hydrogen bond formation. 

If there are $N$ acrylamide molecules in solution with $M$ hydrogen bonds in total between them and the energy of one hydrogen bond is $\epsilon$, one can write \cite{Veytsman1990a}
\begin{equation}
F_\text{HB}=M\epsilon-kT\ln\left[p^M\Xi\right],
\label{fhb-model1}
\end{equation}
where $p$ is the probability of formation of one bond, which can be expected to be inversely proportional to the volume of the system, so that $p=C/V$ (where $C$ is a constant), and $\Xi$ is the combinatorial number of ways to form $M$ bonds in the system. It is worth noting that here we implicitly use our assumption about the absence of cycles because we suppose that the formation of each bond contributes to the statistical weight factor $p$, which accounts for the loss of entropy due to bond formation. In the case when an additional bond is formed that leads to the creation of a cycle, the entropy loss is either absent or much smaller because the participating molecules are already held close to each other by other bonds in the aggregate. 

The expression for $\Xi$ in the case when one bond per oxygen and two bonds per $\text{NH}_2$ group is allowed has the form
\begin{equation}
\Xi=\frac{N!2^M}{\left(N-M\right)!}\frac{\left(2N-2\right)!}{\left(2N-2-M\right)!}\frac{1}{M!}.
\label{xi-1}
\end{equation}
Here, the first factor is the number of ways to choose $M$ acceptors for $M$ bonds out of $N$ molecules, taking into account that each oxygen can form only one bond, but has two bonding sites. The second factor is the number of ways to choose $M$ donors for the bonds (at this stage all atoms are treated as distinguishable) and the last factor accounts for the indistinguishability of bonds. Since $M$ and $N$ are large numbers, the $-2$ terms in the second factor will be neglected. 

Substituting $\Xi$ into Equation~\ref{fhb-model1}, using Stirling's formula and minimizing $F_\text{HB}$ with respect to $M$ yields
\begin{equation}
\frac{M}{2\left(N-M\right)\left(2N-M\right)}=\frac{K}{V},
\end{equation}
where the equilibrium association constant $K\equiv C\exp\left(-\frac{\epsilon}{kT}\right)$ has been introduced. Alternatively, in terms of concentrations
\begin{equation}
\frac{m}{2\left(n-m\right)\left(2n-m\right)}=K.
\label{m-equation}
\end{equation}
Solving this quadratic equation with respect to $m$, the dependence of the free energy on the total concentration of solution can be obtained as
\begin{equation}
\frac{F_\text{HB}}{kTV}=m+n\ln\left[\frac{\left(n-m\right)\left(2n-m\right)^2}{4n^3}\right],
\end{equation}
where
\begin{equation}
m=\frac{1+6Kn-\sqrt{1+12Kn+4K^2n^2}}{4K}.
\end{equation}
In order to couple this expression with the Flory-Huggins formula for the free energy of the system without hydrogen bonds, it can be rewritten in terms of volume fractions as
\begin{equation}
\frac{F_\text{HB}v}{kTV}=\phi_m+\phi\ln\left[\frac{\left(\phi-\phi_m\right)\left(2\phi-\phi_m\right)^2}{4\phi^3}\right],
\end{equation}
with
\begin{equation}
\phi_m=\frac{1+6K'\phi-\sqrt{1+12K'\phi+4K'^2\phi^2}}{4K'},
\end{equation}
where $v$ is a reference volume and $K'=K/v$ is a dimensionless association constant.
As the calculation method is the same for all models, only the final expressions for the free energies of the other models with one association constant are presented here, in Table~\ref{free-energies-k}. The full derivations can be found in the Supporting Information.

\begin{table}
  \caption{Free energies for models with one association constant.}
  \label{free-energies-k}
  \begin{tabular}{ |l | c | c | }
    \hline
    Model & $F_\text{HB}v/kTV$ & $\phi_m$ \\    
    \hline
    \hline
    0 & $2K'\phi_1^2+\phi\ln\frac{\phi_1}{\phi}$ & $\frac{1+16K'\phi-\sqrt{1+32K'\phi}}{32K'}$ \\    
    \hline
    1 $\&$ 4 & $\phi_m+\phi\ln\left[\frac{\left(\phi-\phi_m\right)\left(2\phi-\phi_m\right)^2}{4\phi^3}\right]$ & $\frac{1+6K'\phi-\sqrt{1+12K'\phi+4K'^2\phi^2}}{4K'}$ \\    
    \hline
    2 & $\phi_m+4\phi\ln\left(1-\frac{\phi_m}{2\phi}\right)$ & $\frac{1+4K'\phi-\sqrt{1+8K'\phi}}{2K'}$ \\    
    \hline
    3 & $\phi_m+2\phi\ln\left(1-\frac{\phi_m}{\phi}\right)$ & $\frac{1+8K'\phi-\sqrt{1+16K'\phi}}{8K'}$ \\    
    \hline
  \end{tabular}
\end{table}

Model 4 is similar to model 1 but with the roles of donors and acceptors exchanged, and is therefore described by the same equations. However, in this model, the number of free $\text{NH}_2$ groups will be different from model 1, so it is considered as a separate case here.

\section{Determination of association constants}

The association constants of alcohols were previously measured by others using IR spectroscopy with the help of the following idea \cite{Coggeshaci1951InfraredEquilibria, Coleman1992EquilibriumPolyesters}. Suppose that the hydrogen-bonding substance is dissolved in a solvent that has no specific (i.e.\ hydrogen bonding or strong polar) interactions with the solute. Then, at vanishingly small concentrations of solute, peaks corresponding to the vibrations of the hydrogen-bonding groups in isolated molecules should be seen. As the concentration is increased, new peaks should appear that correspond to hydrogen-bonded states of hydrogen-bonding groups, as hydrogen bonding changes the absorption frequency of groups participating in the bond. In consequence, the dependence of the height of the peaks corresponding to absorption by isolated molecules should have a weaker than linear dependence on the total concentration of the solution. So, if a formula can be found to describe how the concentration of the species corresponding to a given peak depends on the total concentration, then the association constants can be determined by fitting this expression to experimental results on the dependence of the peak height on the total concentration, with the association constants treated as adjustable parameters.

This procedure is now applied to the case of solutions of acrylamide in chloroform. Chloroform is chosen because it is a non-hydrogen bonding solvent that dissolves acrylamide sufficiently well to give a good range of concentrations (compared, for example, to carbon tetrachloride), and because it has a relatively high boiling temperature (compared, for example, to dichloromethane), to allow the measurements to be conducted over a sufficiently broad range of temperatures.

\begin{figure}
\includegraphics[width=0.5\textwidth]{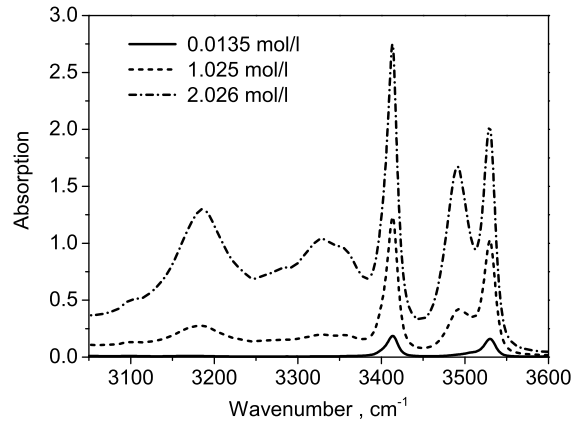}
\caption{IR spectrum of acrylamide in chloroform at $22^{\circ}\text{C}$ at different concentrations focusing on the $\text{NH}_2$  absorption region of spectrum for acrylamide.}
\label{spectrum}
\end{figure}

Figure~\ref{spectrum} shows the changes of IR absorption by acrylamide in the range $3700\,\text{cm}^{-1}$ to $3000\,\text{cm}^{-1}$ as the total concentration of the solution is increased. At low concentrations, we can see two clear peaks at $3414\,\text{cm}^{-1}$ and $3530\,\text{cm}^{-1}$, which are attributed to the in-phase and out-of-phase vibrations of the $\text{NH}_2$ group respectively \cite{Duarte2005}. It should be noted that, even at the lowest concentrations we studied, there is a shoulder on the $3530\,\text{cm}^{-1}$ peak, which we are unable to assign accurately. At larger concentrations, the peaks at $3414\,\text{cm}^{-1}$ and $3530\,\text{cm}^{-1}$ remain, but the shoulder develops and a broad conglomerate of a number of peaks at lower energies (to the left of $3414\,\text{cm}^{-1}$) appears. It is well known that hydrogen bonding of the donor N-H group leads to a red-shift of the N-H vibration frequency from that of the free group\cite{Mirkin2004StructuralSystems}, so all peaks that appear as the concentration of acrylamide solution increases are assigned to absorption by $\text{NH}_2$ groups in different hydrogen-bonded states.

However, one must be cautious with respect to the attribution of the $3414\,\text{cm}^{-1}$ and $3530\,\text{cm}^{-1}$ peaks to the ``free'' $\text{NH}_2$ groups, because the bonding state of the oxygen in the same amide group can influence the frequency of absorption. Extensive attempts to rationalize the rules according to which hydrogen bonding affects the absorption wavelengths in amides resulted in a conclusion that there are no universal rules and that each amide system should be carefully studied in order to find the contributions to the shifts in each particular case \cite{Myshakina2008DependenceBonding,Galan2014TheoreticalConformations,Lu2005DensityDimer,Esrafili200814NBonding}. DFT simulations of hydrogen bonded aggregates of acrylamide could potentially have shed some light on this matter, but we are not aware of any such work in the literature, and the closest study we could find was carried out for perfluorinated polyamides\cite{Milani2010HydrogenCalculations}. In their work, the shifts in absorption wavenumbers of the $\text{NH}_2$ group in linear and cyclic dimers and trimers were calculated. Interestingly, the shift of the out-of-phase vibrations of the free $\text{NH}_2$ groups in the linear dimer was calculated to be $6\,\text{cm}^{-1}$  and in the linear trimer to be $15\,\text{cm}^{-1}$.  When applied to the spectra of acrylamide, this would result in a contribution of the free groups of linear dimers to the $3530\,\text{cm}^{-1}$ band but would also mean that the absorption of the free group in the trimer would lie away from this band. Similar behavior was reported for acetamide clusters \cite{Esrafili2008TheoreticalClusters}. Consequently, based on the available information,  the following possibilities for peak attribution have been considered. The first possibility is that the ``free'' peaks correspond to unimers. This assumption implies that any bonding of oxygen in the amide group substantially shifts the absorption of the $\text{NH}_2$ group. The opposite possibility is that the ``free'' peaks correspond to the free $\text{NH}_2$ groups regardless of the bonding state of oxygen in the same amide group. Finally, the third possibility considered is that these peaks correspond to the absorption of the free $\text{NH}_2$ groups in only unimers and dimers, which would correspond to the case when the shift of absorption of a free $\text{NH}_2$ group in a dimer is small enough to give a contribution to the $3530\,\text{cm}^{-1}$ peak together with free molecules, but the shift of absorption in free $\text{NH}_2$ groups in larger aggregates is large enough not to give a contribution to the $3530\,\text{cm}^{-1}$ peak.

Returning to model 1, one can find now the dependence of the concentration of free molecules and free groups on the total concentration of the solution.

\begin{figure}
\includegraphics[width=0.5\textwidth]{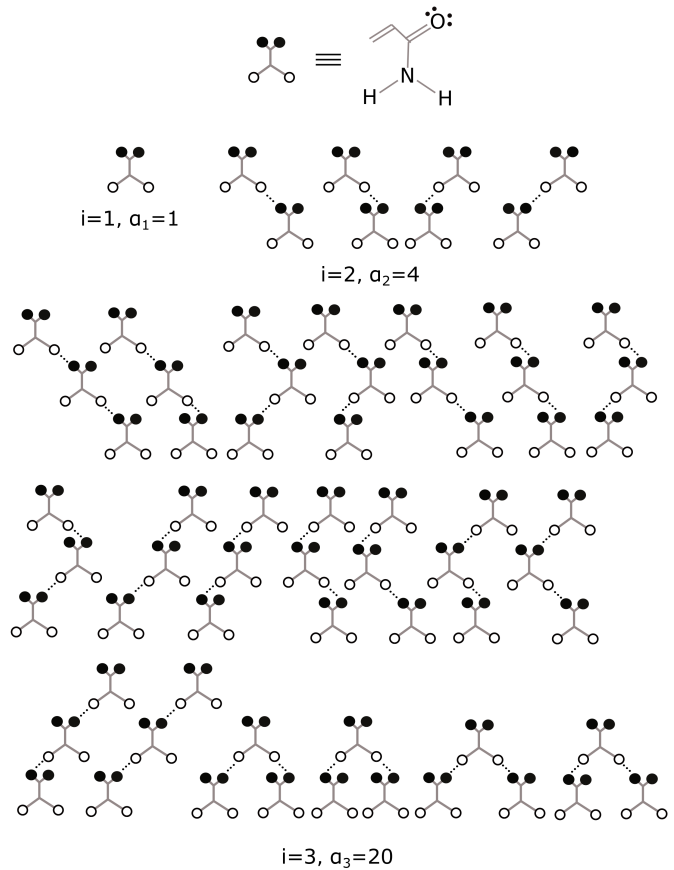}
\caption{Schematic representation of aggregates with size up to $i=3$ in model 1.}
\label{model1}
\end{figure}

Let us first find the dependence of the total concentration on the concentration of free molecules. Hydrogen bonding is described in the current work as a reversible chemical reaction that produces a range of aggregates of different structures and sizes, and it is assumed that all aggregates are tree-like and no cycles can be formed. In this case, the concentration of aggregates of size $i$ has the form $c_i=\alpha_iK^{i-1}c_1^{i}$ where $K$ is the equilibrium association constant, $c_1$ is the concentration of free molecules and $\alpha_i$ is a coefficient that depends on the size of the aggregate. With knowledge of $c_i$, the total concentration of the solution $n$ and the concentration of bonds $m$ can be calculated to be
\begin{equation}
n=\sum_{i=1}^{\infty}ic_i=c_1\sum_{i=1}^{\infty}i\alpha_i\left(Kc_1\right)^{i-1}
\label{n-def}
\end{equation}
and
\begin{equation}
m=\sum_{i=1}^{\infty}\left(i-1\right)c_i=c_1\sum_{i=1}^{\infty}\left(i-1\right)\alpha_i\left(Kc_1\right)^{i-1}.
\end{equation}
Next, the function
\begin{equation}
g\left(x\right)=\sum_{i=1}^{\infty}\alpha_ix^i
\end{equation}
is introduced, where $x=Kc_1$; then, $Km=xg'-g$, $K\left(n-m\right)=g$ and $K\left(2n-m\right)=xg'+g$. Substituting these expressions into Equation~\ref{m-equation} yields
\begin{equation}
xg'\left(x\right)-g\left(x\right)=2g\left(x\right)\left(xg'\left(x\right)+g\left(x\right)\right).
\end{equation}
The solution of this equation that remains finite as $x \to 0$ is\bibnote{In order to get this solution, we first introduce a function $f\left(z\right)=xg\left(x\right)$ and get the equation $xf'-2ff'-2f=0$. Then we go to the inverse function $x\left(f\right)$ for which $2fx'-x+2f=0$. This equation has the solution $x=-2f+A\sqrt{f}$. Finally, we solve the quadratic equation with respect to $f$.} 
\begin{equation}
g\left(x\right)=\frac{\exp{A}-4x-\sqrt{\exp{2A}-8x\exp{A}}}{8x},
\label{g}
\end{equation}
where $A$ is a constant determined by the boundary conditions (i.e.\ the value of $\alpha_1$, which is put everywhere equal to 1).
We can find $\alpha_i$ by expanding Equation~\ref{g} as a Taylor series. The general formula for $\alpha_i$ is
\begin{equation}
\alpha_{i}=2^{i-1}\cdot\frac{\left(2i\right)!}{i!\left(i+1\right)!}.
\end{equation}
The sequence $\beta_i=\frac{\left(2i\right)!}{i!\left(i+1\right)!}$ is known as the Catalan numbers (published electronically at https://oeis.org, May 2018). It is known that the Catalan numbers represent the number of different rooted binary trees with $i+1$ leaves. In the current case, there is an additional factor of $2^{i-1}$, since each molecule apart from the root can be added in two ways to form a bond with one of the free hydrogens as there are two bonding sites on the oxygen. It can then be said that the physical meaning of $\alpha_i$ is the number of ways to compose an aggregate of size $i$ out of $i$ molecules. All aggregates allowed in our model with size up to $i=3$ are shown in Figure~\ref{model1}. 

Substituting the expression for $\alpha_i$ into Equation~\ref{n-def}, a relation between the concentration of free molecules and the total concentration of the solution is obtained:
\begin{equation}
n=\frac{1-4c_1K-\sqrt{1-8c_1K}}{8c_1K^2\sqrt{1-8c_1K}}.
\end{equation}
This expression was used to fit our IR spectroscopy data in the case where the  $3530\,\text{cm}^{-1}$ peak is attributed to free molecules, because in this case we simply write $c_1=Ax$ where $A$ is some constant and $x$ is a peak height.

It is straightforward to obtain from this expression a fitting equation for the case where the peak is attributed to both free molecules and free groups in dimers. In this case $Ax=c_1+4Kc_1^2$. 

\begin{figure}
\centering
\includegraphics[width=0.5\textwidth]{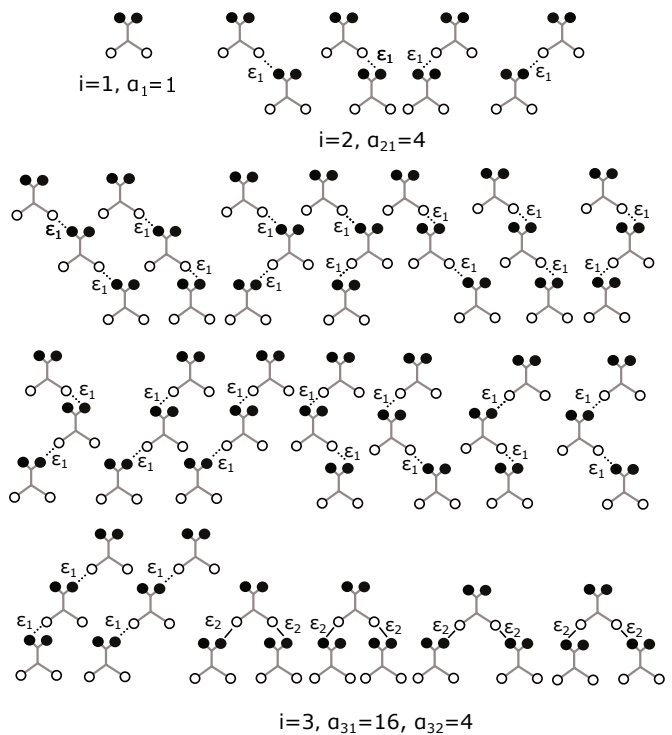}
\caption{Schematic representation of aggregates with size up to $i=3$ in the extension of model 1 when we assume that the energy of the bond is $\epsilon_1$ in the case when a second proton in a donor molecule is free and $\epsilon_2$ otherwise.}
\label{model5-scheme}
\end{figure}

Finally, the case where the peak is attributed to free $\text{NH}_2$ groups needs to be considered. In order to find the fitting equation, it is necessary to find the concentration of free groups. However, the number of free groups depends on the structure of the aggregate. The most convenient way to do these calculations is to consider the model with two association constants determined by the bonding state of the $\text{NH}_2$ group in the donor molecule (see Figure~\ref{model5-scheme}) and, after the calculations are complete, put the association constants equal to each other. So, we will assume that the energy of the bond is $\epsilon_1$ in the case when a second proton in a donor molecule is free and $\epsilon_2$ otherwise. For the sake of clarity, further calculations are omitted here (these can be found in the supporting information in Section 1.6), and the final result for the dependence of the total concentration $n$ on the number of free groups $n_f$ is
\begin{equation}
n=\frac{1-4Kn_f+16K^2n_f^2-\left(1-4Kn_f\right)\sqrt{1+16K^2n_f^2}}{8K^2n_f}.
\end{equation}

For other models all calculations can be found in the Supporting Information. 

With all this information in hand, the association constants corresponding to different models can be determined by fitting the dependence of the height of the $3530\,\text{cm}^{-1}$ peak on the total concentration of the solution. In practice, the inverse dependence $n(x)$ will be fitted for numerical convenience.

\begin{figure}
\centering
\includegraphics{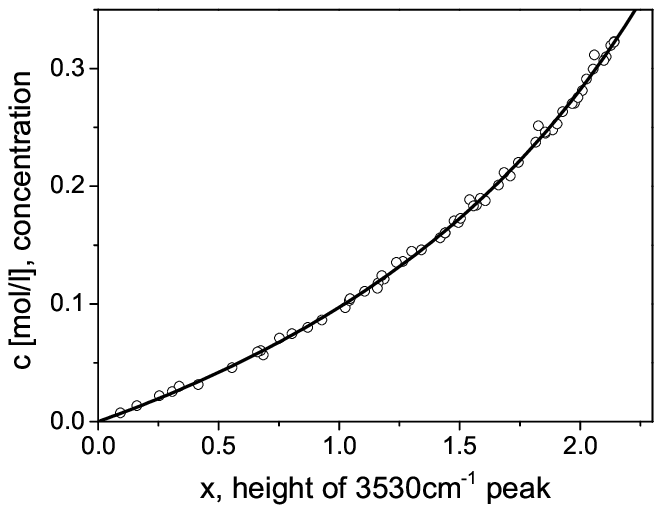}
\caption{Fitting of the dependence of the total concentration on the height of the $3530\,\text{cm}^{-1}$ IR band at $T=22^\circ\text{C}$ with model 1 with the free molecules assumption.}
\label{model-1m-22C}
\end{figure}

\begin{figure}
\centering
\includegraphics{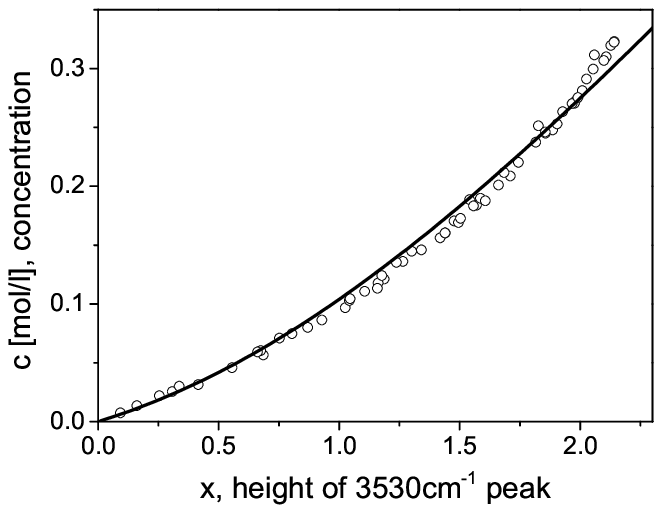}
\caption{Fitting of the dependence of the total concentration on the height of the $3530\,\text{cm}^{-1}$ IR band at $T=22^\circ\text{C}$ with model 1 with free groups assumption.}
\label{model-1g-22C}
\end{figure}

The fitting result for model 1 with the free molecules assumption is shown in Figure~\ref{model-1m-22C}, and the corresponding result with the free groups assumption is shown in Figure~\ref{model-1g-22C}. The quality of the fit is visibly better with the free molecules assumption, and this point will be discussed in more detail below.

The results of fitting IR data at $22^\circ\text{C}$ are presented in Table~\ref{K-models}. The first two columns specify the number of the model and peak attribution assumption respectively. In the third column, the values of the association constants obtained as fitting parameters are shown. In the fourth column, the values of the dimensionless constants are given, which were calculated as $K'=K/v$ where $v=0.0629\,\text{l/mol}$ is the molar volume of acrylamide based on its density\cite{Udovenko2008CRYSTALACRYLAMIDE}. The final column gives the values of the AICc parameter\bibnote{Smaller values of the AICc parameter correspond to a better quality of fit. Usually, it is assumed that a difference between the AICc parameters of two models of more than 2 is significant and more than 6 is strong. See the supporting information for further details}, which characterizes the quality of the non-linear fit\cite{Akaike1998InformationPrinciple,Sugiura1978FurtherCorrections}.
It can be seen that approximately half of the models give the same quality of fit, so based on fitting results exclusively it cannot be said which model is better. However, attributing the peak to free molecules gives a better quality of fit than attributing it to free groups.

\begin{table}
  \caption{IR data fitting results with models with one association constant.}
  \label{K-models}
  \begin{tabular}{ |l | c | c | c | l | }
    \hline
    Model & Peak & $K$ [l/mol] & $K'$ & AICc \\  
 & attr. & at $22^{\circ}$C & at $22^{\circ}$C & at $22^{\circ}$C \\
    \hline
    \textbf{0} & m \textsuperscript{\emph{a}} & 2.4 &  38.2 & -457.6 \\
    \hline
     \textbf{0} & g \textsuperscript{\emph{b}} & 1.36 &  21.6 & -298 \\
    \hline
     \textbf{1} & m  & 0.42 &  6.68 & -503.8 \\
    \hline
     \textbf{1} & g  & 2.37 &  37.68 & -388.9 \\
    \hline
    \textbf{1} & s \textsuperscript{\emph{c}}  & 0.65 &  10.30 & -503.9 \\
    \hline
     \textbf{2} & m  & 0.334 &  5.31 & -502.5 \\
    \hline
     \textbf{2} & g  & 1.06 &  16.82 & -499.7 \\
    \hline
    \textbf{2} & s  & 0.442 &  7.03 & -494.2 \\
    \hline
      \textbf{3} & m  & 0.42 &  6.68 & -499.7 \\
    \hline
     \textbf{3} & g  & 4.8 &  76.3 & -457.6 \\
    \hline
    \textbf{3} & s  & 1.02 &  16.22 & -499.3 \\
    \hline
      \textbf{4} & m  & 0.42 &  6.68 & -503.8 \\
    \hline
     \textbf{4} & g  & 1.332 &  21.18 & -495.8 \\
    \hline
    \textbf{4} & s  & 0.65 &  10.30 & -503.9 \\
    \hline
  \end{tabular}
  
  \textsuperscript{\emph{a}} free molecules assumption 
  \textsuperscript{\emph{b}} free groups assumption 
  \textsuperscript{\emph{c}} assumption that $3530\text{cm}^{-1}$ peak corresponds to free $\text{NH}_2$ groups in unimers and dimers.
\end{table}

\begin{figure}
\centering
\includegraphics{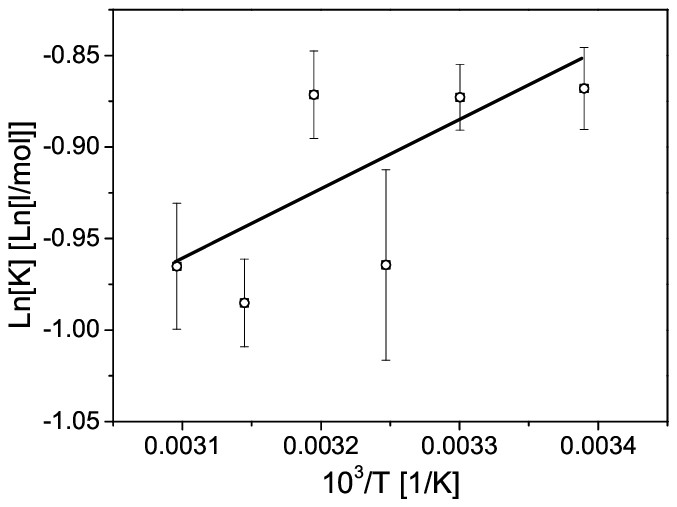}
\caption{Dependence of $\ln K$ on $1/T$ for model 1 with the free molecules assumption.}
\label{model-1m-kfit}
\end{figure}

\begin{figure}
\centering
\includegraphics{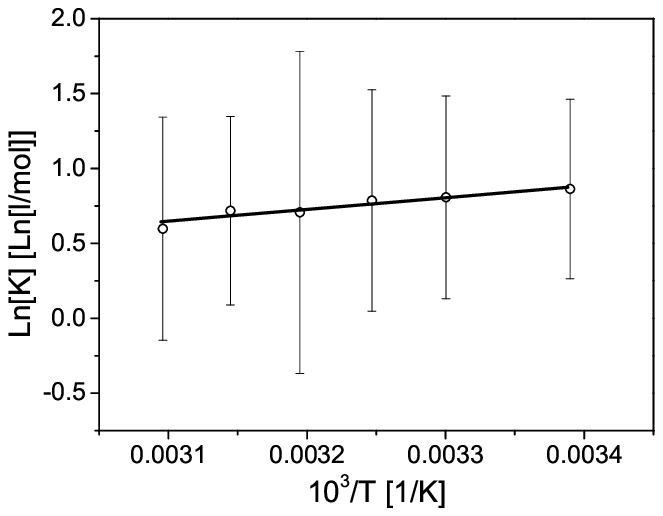}
\caption{Dependence of $\ln K$ on $1/T$ for model 1 with the free groups assumption.}
\label{model-1g-kfit}
\end{figure}

Next, the bond energies are determined from the association constants. In order to do this, we perform our measurements and fitting procedures at several different temperatures. Since $\ln K=\ln C-\frac{\epsilon}{kT}$ by definition, $\epsilon$ can be determined from a plot of $\ln K$ against $1/T$. It is noteworthy that this procedure can serve as an additional test for the model, because if the model fits the data well then $\ln K\left(1/T\right)$ is a linear function with a negative value of the energy of bond formation, $\epsilon < 0$. Examples of such plots are shown in Figures~\ref{model-1m-kfit}~and~\ref{model-1g-kfit} for model 1 with the free molecules and free groups assumptions. The corresponding quantities $C'$ and $\epsilon'$ for the dimensionless association constant $K'$  are shown in Table~\ref{K-temperature}. The first two columns in the table specify the model and peak attribution assumption. The third and fourth columns give the values of $\ln C'$ and $\epsilon$ respectively, and the values of the coefficients of determination characterizing the quality of the fit of $\ln K'$ vs.\ $1/T$ are listed in the last column.

\begin{table}
  \caption{Temperature dependence of (dimensionless) association constants and bond energies for the 'one constant' models.}
  \label{K-temperature}
  \begin{tabular}{ |l | c | c | c | l | }
    \hline
    Model & Peak attr. & $\ln C'$ & $\epsilon$, [kcal/mol] & $r^2$  \\
    \hline
    \textbf{0} & m \textsuperscript{\emph{a}} & 1.92 &  -1.03 & 0.07 \\
    \hline
     \textbf{0} & g \textsuperscript{\emph{b}} & 1.12 &  -1.15 & 0.75 \\
    \hline
     \textbf{1} & m  & 0.63 &  -0.75 & 0.56 \\
    \hline
     \textbf{1} & g  & 1.00 &  -1.55 & 0.9 \\
    \hline
    \textbf{1} & s  & 1.01 &  -0.78 & 0.66 \\
    \hline
     \textbf{2} & m  & 0.31 &  -0.8 & 0.79 \\
    \hline
     \textbf{2} & g  & 1.61 &  -0.72 & 0.38 \\
    \hline
    \textbf{2} & s  & 0.50 &  -0.86 & 0.92 \\
    \hline
      \textbf{3} & m  & 0.92 &  -0.72 & 0.38 \\
    \hline
     \textbf{3} & g  & 3.31 & -1.03 & 0.07 \\
    \hline
     \textbf{3} & s  & 1.57 & -0.72 & 0.37 \\
    \hline
      \textbf{4} & m  & 0.63 &  -0.75 & 0.56 \\
    \hline
     \textbf{4} & g  & 1.92 &  -0.675 & 0.28 \\
    \hline
    \textbf{4} & s  & 1.01 &  -0.78 &  0.66 \\
    \hline
  \end{tabular}
  
  \textsuperscript{\emph{a}} free molecules assumption 
  \textsuperscript{\emph{b}} free groups assumption 
  \textsuperscript{\emph{c}} assumption that $3530\text{cm}^{-1}$ peak corresponds to free $\text{NH}_2$ groups in unimers and dimers.
\end{table}

It can be seen that the absolute values of the predicted bond energies are close to $1\,\text{kcal/mol}$ for all models. However, according to DFT calculations, the absolute value of the bond energy in an acrylamide dimer in a vacuum can be estimated as $7-9\,\text{kcal/mol}$ \cite{Wang2016Hydrogen-bondingAnalysis,Duarte2005}, which is much larger than the values obtained for one-parameter models. 

In addition, one can see that, in the limit of infinite temperature, $T\rightarrow\infty$, we have that $K'\rightarrow C'$. Since the $C'$ values are relatively large (see Table~\ref{K-temperature}), $K'>1$ for all models even at infinite temperatures. This contradicts our intuitive expectation that the strength of hydrogen bonding substantially decreases as the temperature increases. In one of the following sections, models with two association constants are developed, which give a better quality of fit to the data, allowing to obtain a dimer bond energy closer to DFT predictions, and yield physically reasonable asymptotic behavior at high temperatures.      

\section{Properties of models with one association constant}

In the previous section, the concentrations of aggregates of all sizes were determined in models with one association constant. This allows one to study some properties of the models, such as the dependence of the total number of bonds and average size of the aggregates on the total concentration of the solution and on the value of the association constant. 

\begin{figure}
\includegraphics[width=0.5\textwidth]{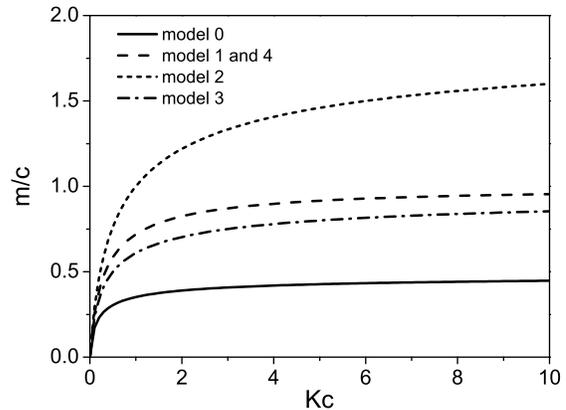}
\caption{Dependence of $m/c$ on $Kc$ at fixed total concentration for models with one association constant.}
\label{m-models}
\end{figure}

First, let us look at the dependence of the ratio of the concentration of bonds to the concentration of molecules $m/c$ on the value of the association constant at fixed total concentration $c$, which is shown in Figure~\ref{m-models}.

It can be seen that, for model 0, as $K$ increases, $m/c$ tends to a value of $0.5$ in accordance with the assumption that in this model only dimers can form. In the case of models 1, 3, and 4, the ratio $m/c$ tends to $1$ in the limit of infinite $K$. This means that, in these models, the number of bonds is always less then the number of molecules in the system. This is explained by the fact that in all of these models molecules either have one acceptor site or one donor site. In contrast, in model 2 there are two bonding sites of each type for each molecule. As a result, when  $K>1/c$, the number of bonds becomes larger than the number of molecules. In addition, since in an aggregate of size $i$ without cycles the number of bonds is always $i-1$, one can immediately conclude that the assumption about the absence of cycles is wrong when applied to model 2 with $K>1/c$. 

In fact, the restrictions on the applicability of our assumption of the absence of cycles in model 2 are even stronger, because according to our calculations for this model, the total concentration expressed as a function of concentration of unimers can be written in terms of a hypergeometric function as
\begin{equation}
\frac{c}{c_1}=\,_{3}F_{2}\left(\left[\frac{4}{3}, \frac{5}{3}, 2\right],\left[\frac{5}{2}, 3\right],\frac{27Kc_1}{2}\right).
\label{model2-conc}
\end{equation}
The right-hand side of Equation~\ref{model2-conc} is defined only for  $|\frac{27Kc_1}{2}|\leq 1$ and is an increasing function of its argument. The first of these facts means that we must have $c_1 \leq \frac{2}{27K}$, and the second means that we must also have $c_1 \geq c\,/\,_{3}F_{2}\left([\frac{4}{3}, \frac{5}{3}, 2],[\frac{5}{2}, 3],1\right)$. So, when $K>2\, _{3}F_{2}\left([\frac{4}{3}, \frac{5}{3}, 2],[\frac{5}{2}, 3],1\right)/27c=3/8c$, Equation~\ref{model2-conc} has no solutions. The ratio of $m$ and $c$ at the maximum value of $K$ is $m/c=2/3$, so the assumption about the absence of cycles fails when the number of bonds per molecule becomes larger than $2/3$.  

Another interesting property is the dependence of the average aggregate size on the values of the association constant and concentration. The average aggregation number can be calculated as
\begin{equation}
\langle i\rangle=\frac{\sum_{i=1}^{\infty}ic_i}{\sum_{i=1}^{\infty}c_i}.
\end{equation}

\begin{figure}
\includegraphics[width=0.5\textwidth]{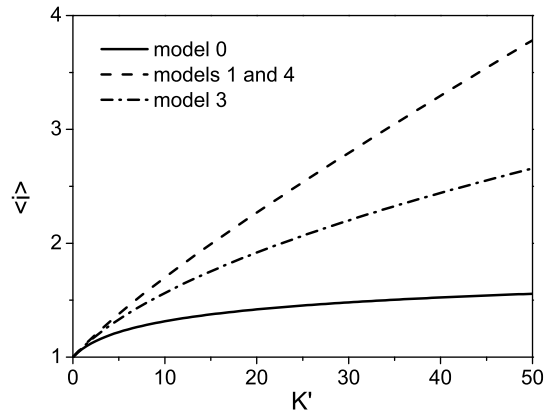}
\caption{Dependence of $\langle i \rangle$ on the value of the dimensionless association constant at volume fraction of acrylamide fixed to the largest experimental volume fraction, $\phi=0.022$.}
\label{i-av-phi1}
\end{figure}

\begin{figure}
\includegraphics[width=0.5\textwidth]{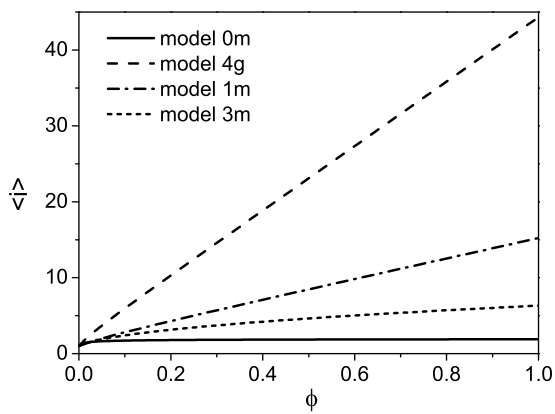}
\caption{Dependence of $\langle i \rangle$ on the volume fraction of acrylamide with association constant determined from fitting IR data at $T=22^{\circ}\text{C}$.}
\label{i-av-phi2}
\end{figure}

In Figure~\ref{i-av-phi1}, the dependence of the average aggregation number on the value of the dimensionless association constant with the volume fraction of acrylamide fixed to the largest experimental volume fraction, $\phi=0.022$, is shown for models 0, 1, 3 and 4. Figure~\ref{i-av-phi2} shows the dependence of the average aggregation number on the volume fraction of acrylamide at association constants obtained from IR measurements for models 0m, 1m, 3m, and 4g. For all models,  $\left\langle i \right\rangle$ monotonically increases as $K'$ and $\phi$ increase, and, as expected, $\left\langle i \right\rangle$ grows more quickly for models 1 and 4 than for model 3.

\begin{figure}
\includegraphics[width=0.5\textwidth]{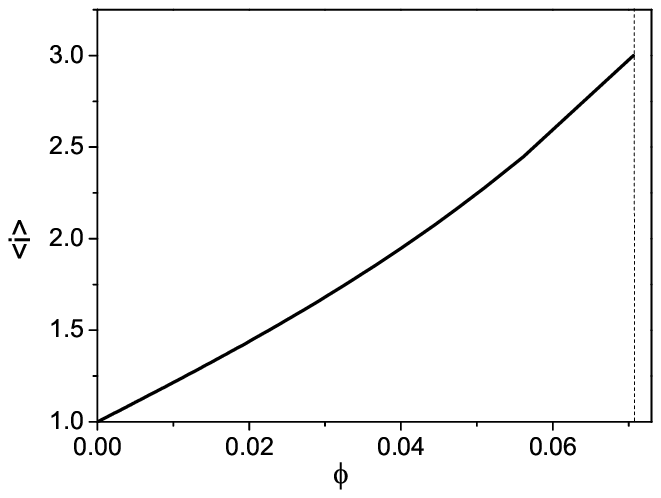}
\caption{Dependence of $\langle i \rangle$ on volume fraction of acrylamide with association constant determined from fit of IR data at $T=22^{\circ}$C for model 2m.}
\label{i-av-phi}
\end{figure}

In the case of model 2, at a fixed value of the concentration there is a maximum value of the association constant above which our assumption about the absence of cycles does not work. Correspondingly, for each value of the association constant, there is a maximum concentration of acrylamide above which model 2 again is not applicable. Within the regime where the model is valid, the average aggregation number is an increasing function of concentration and the association constant and reaches its maximum value of 3 when $Kc_1=2/27$. This behavior is illustrated in Figure~\ref{i-av-phi}, which shows the dependence of the average aggregation number in model 2 on the volume fraction of acrylamide at a fixed value of the association constant.   

\section{Models with two association constants}

In this section, we consider models in which association is characterized by two association constants. As in the case of alcohols, these constants depend on the bonding state of the other groups belonging to the molecule forming a given hydrogen bond. Again, it is assumed that cycles cannot form. Here only two-constant extensions of models 1, 3 and 4 are constructed. Model 2 has not been considered here, as this case requires significant additional study. 

\begin{figure}
\centering
\includegraphics[width=0.45\textwidth]{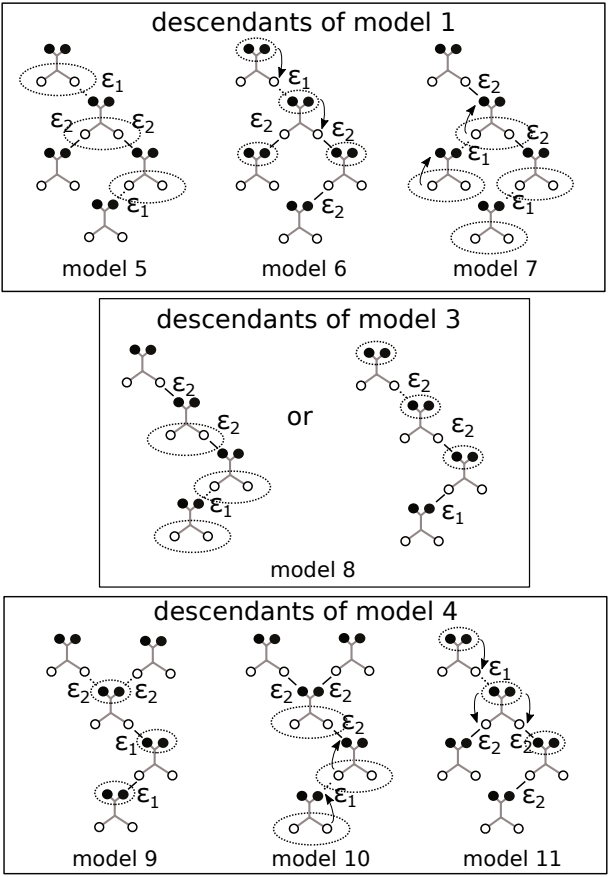}
\caption{Two-parameter models.}
\label{models-2-constants}
\end{figure}

The list of models with two association constants is presented in Table~\ref{list-models-2} and the rules for how the energy of a bond depends on its location are illustrated in Figure~\ref{models-2-constants}. 

\begin{table}[h]
  \caption{List of models with two association constants and no cycles.}
  \label{list-models-2}
  \begin{tabular}{ |l | l | }
    \hline
    Model & Association rules \\  
    \hline
    \hline
    \textbf{5} & one bond per oxygen, two bonds per $\text{NH}_2$ group,\\
    & energy of hydrogen bond is determined by the \\
    & bonding state of the neighbor hydrogen in\\
    & $\text{NH}_2$ group of the donor molecule \\
    \hline
    \textbf{6} & one bond per oxygen, two bonds per $\text{NH}_2$ group,\\
    & energy of hydrogen bond is determined by the \\
    & bonding state of acceptor in donor molecule\\
    \hline
    \textbf{7} & one bond per oxygen, two bonds per $\text{NH}_2$ group,\\
    & energy of hydrogen bond is determined by the \\
    & bonding state of the $\text{NH}_2$ group in acceptor molecule\\
    \hline
    \textbf{8} & one bond per oxygen, one bond per $\text{NH}_2$ group, \\
    & bond energy is determined by bonding state of \\
    & acceptor in donor molecule or bond energy is \\
    & determined by bonding state of donor in acceptor \\
    & molecule (both definitions give equivalent results) \\
    \hline
    \textbf{9} & two bonds per oxygen, one bond per $\text{NH}_2$ group, \\
    & bond energy depends on bonding state of acceptor \\
    & in acceptor molecule\\
    \hline
    \textbf{10} & two bonds per oxygen, one bond per $\text{NH}_2$ group, \\
    & bond energy depends on bonding state of $\text{NH}_2$ group \\
    & in acceptor molecule\\
    \hline 
    \textbf{11} & two bonds per oxygen, one bond per $\text{NH}_2$ group, \\
    & bond energy depends on bonding state of oxygen \\
    & group in donor molecule\\
    \hline 
  \end{tabular}
\end{table}

All of these cases can be treated analytically and calculations can be found in the Supporting Information. Here, just tables (see Tables~\ref{K2-models} and~\ref{K2-temperature}) of the values of the association constants for these models are shown.

\begin{table}
  \caption{IR data fitting results with models with two association constant at $22^{\circ}\text{C}$.}
  \label{K2-models}
  \begin{tabular}{ |l | c | c | c | c | c | l | }
    \hline
    Model & Peak & $K_1$ & $K_1'$ & $K_2$ & $K_2'$  & AICc  \\    
& attr. & [l/mol] & & [l/mol] & & \\
    \hline
    \textbf{5} & m  & 0.38 &  6.04 &  0.49 &  7.79 & -501.7 \\
     \hline
    \textbf{5} & g & \multicolumn{4}{c}{no good convergence} &\\
     \hline
    \textbf{5} & s & 0.69 &  11.0 & 0.61 &  9.7 & -501.7 \\
     \hline
    \textbf{6} & m  & 0.38 &  6.09  & 0.44 &  7.0 & -501.7 \\
    \hline
    \textbf{6} & g  & 0.27 &  4.29 & 1.05 &  16.69 & -493.9 \\
    \hline
    \textbf{6} & s  & 0.72 &  11.45 & 0.69 &  10.92 & -501.7 \\
     \hline
    \textbf{7} & m  & 0.38 &  6.04 & 0.45 &  7.15 & -501.7 \\
     \hline
    \textbf{7} & g  & 0.29 &  4.54 & 1.24 &  19.7 & -497.4 \\
     \hline
    \textbf{7} & s  & 0.74 &  11.8 & 0.68 &  10.8 & -501.7 \\
   \hline
    \textbf{8} & m  & 0.35 &  5.56 & 0.61 &  9.7 & -501.5 \\
    \hline
    \textbf{8} & g  & 0.33 &  5.24 & 1.36 &  21.54 & -499.5 \\
     \hline
    \textbf{8} & s  & 0.56 &  8.9 & 0.9 &  14.3 & -501.4 \\
    \hline
    \textbf{9} & m  & 0.38 &  6.07 & 0.49 &  7.73 & -501.7 \\
     \hline
    \textbf{9} & g  & 0.51 &  8.11 & 1.38 &  21.94 & -501.3 \\
    \hline
    \textbf{9} & s  & 0.70 &  11.05 & 0.61 &  9.70 & -501.7 \\
    \hline
    \textbf{10} & m  & 0.38 & 6.09 & 0.44 &  6.93 & -501.7 \\
    \hline
    \textbf{10} & g  & 0.67 &  10.59 & 1.01 &  16.06 & -501.1 \\
    \hline
     \textbf{10} & s  & 0.72 &  11.49 & 0.69 &  10.92 & -501.7 \\
    \hline
     \textbf{11} & m  & 0.38 &  6.04 & 0.45 &  7.09 & -501.7 \\
    \hline
     \textbf{11} & g & \multicolumn{4}{c}{no good convergence} &\\
    \hline
     \textbf{11} & s  &0.74 &  11.7 & 0.68 &  10.81 & -501.7 \\
    \hline
  \end{tabular}  
\end{table}

\begin{table}
  \caption{Temperature dependence of (dimensionless) association constants and bond energies for models with two association constants.}
  \label{K2-temperature}
  \begin{tabular}{ |l | c | c | c | c | c | c | l | }
    \hline
    Model & Peak & $\ln C'_1$ & $ \epsilon_1$ & $r_1^2$ & $\ln C'_2$ & $\epsilon_2$  & $r_2^2$  \\
& attr. &    & $\left[\text{kcal/mol}\right]$ & & & $\left[\text{kcal/mol}\right]$& \\
    \hline
    \textbf{5} & m  & -16.8 &  -11 &  0.76 &  6.2 & 2.4 & 0.73 \\
     \hline
    \textbf{5} & g & \multicolumn{5}{c}{no good convergence} &\\
     \hline
    \textbf{5} & s & -18.7 &  -12.5 & 0.76 &  4.7 & 1.3 & 0.56 \\
     \hline
    \textbf{6} & m  & -7.2 &  -5.2  & 0.91 & 2.2 & 0.1 & 0.03 \\
    \hline
    \textbf{6} & g  & -8.7 &  -5.9 & 0.71 &  -1.0 & -2.2 & 0.90 \\
    \hline
    \textbf{6} & s  & -9.1 &  -6.7 & 0.91 &  -0.4 & -1.6 & 0.95 \\
     \hline
    \textbf{7} & m  & -8.9 &  -6.2 & 0.9 &  4.1 & 1.2 & 0.62 \\
     \hline
    \textbf{7} & g  & -9.6 &  -6.5 & 0.8 &  -0.13 & -1.8 & 0.95 \\
     \hline
    \textbf{7} & s  & -11.6 &  -8.2 & 0.9 &  1.29 & -0.6 & 0.7 \\
   \hline
    \textbf{8} & m  & -9.5 & -6.6 & 0.89 &  2.85 & 0.3 & 0.18 \\
    \hline
    \textbf{8} & g  & -10.8 &  -7.3 & 0.83 &  -0.39 & -2 & 0.95 \\
     \hline
    \textbf{8} & s  & -11.4 &  -7.9 & 0.88 & 0.37  & -1.3 & 0.93 \\
    \hline
    \textbf{9} & m  & -16.8 &  -11 & 0.76 &  6.2 & 2.4 & 0.73 \\
     \hline
    \textbf{9} & g  & \multicolumn{5}{c}{no good convergence at higher temperatures} &\\
    \hline
    \textbf{9} & s  & -18.7 &  -12.5 & 0.76 &  4.7 & 1.3 & 0.56 \\
    \hline
    \textbf{10} & m  & -7.2 & -5.2 & 0.91 & 2.2 & 0.1 & 0.03 \\
    \hline
    \textbf{10} & g  & -8.5 &  -6.3 & 0.9 &  -1.41 & -2.4 & 0.95 \\
    \hline
     \textbf{10} & s  & -9.1 &  -6.7 & 0.91 & -0.4 & -1.6 & 0.95 \\
    \hline
     \textbf{11} & m  & -8.9 &  -6.2 & 0.9 & 4.1 & 1.2 & 0.61 \\
    \hline
     \textbf{11} & g & \multicolumn{5}{c}{no good convergence} &\\
    \hline
     \textbf{11} & s  & -11.6 &  -8.2 & 0.9 &  1.29 & -0.6 & 0.69 \\
    \hline
  \end{tabular}  
\end{table}

The first thing to note is that the quality of fitting increases as we turn to models with two association constants. Based on the combination of AICc and $r_i^2$ for the fits of $c\left(x\right)$ and $\ln K_i \left(1/T\right)$, we can conclude that models 6s (and the equivalent model 10s), 8s and 10g are good. All of these models give the value of the bond energy in a dimer as about $6-8\,\text{kcal/mol}$, which corresponds much more closely to the values calculated by DFT\cite{Wang2016Hydrogen-bondingAnalysis,Duarte2005} discussed earlier. Another important property of these models is that the values they give for both $C_1$ and $C_2$ are smaller than those found for one-constant models. So, in the limit of infinite temperature, the association constants $K_1$ and $K_2$ will have smaller values. 

Fitting also shows that model 5, which includes the assumption that bonding energy is fixed by the bonding state of the neighbor hydrogen in the $\text{NH}_2$ group, is poor. It means that either the bonding state of one hydrogen does not affect the bonding of its neighbor hydrogen (implying that model 5 reduces to model 1) or that the bonding of one of the hydrogens in the $\text{NH}_2$ group leads to the complete loss of the donor properties of the second hydrogen (so that model 5 reduces to model 3). The same observation applies to the hydrogen bonding sites on oxygen in model 9, which reduces either to model 4 or to model 3.

Similar conclusions can be made with regard to the pair of models 7 and 11, because they show worse fitting results than models 6, 8 and 10. As a result, it can be concluded that the assumption that there are only one or two bonds with energy $\epsilon_1$ at the ``top'' of each aggregate and all other bonds have energy $\epsilon_2$ is the most probable one according to the fitting results.

It is interesting to note that, in all cases, the value of $\epsilon_2$ is smaller than the value of $\epsilon_1$, so the formation of the initial dimer is a more energetically favorable process than the addition of subsequent acrylamide molecules to the aggregate. It is also interesting to note that this difference is much larger than in the similar situations in alcohols \cite{Coleman1995}.

We also can see that there is a contradiction with DFT calculations on formamide and acetamide linear clusters, which predict that the energy per bond increases with the growth of the size of the linear cluster \cite{Esrafili2008TheoreticalClusters, Kobko2003}. However, as both approaches involve their own approximations, additional study is needed to understand the reasons for this discrepancy.

\section{Structure factor of diblock copolymer with hydrogen bonding block in a disordered state}

Our motivation for developing a theory of hydrogen bonding in acrylamide is to use this model in a self-consistent field theory (SCFT) of block copolymers with hydrogen bonds. To take a first step in this direction, we calculate the structure factor of the disordered state of a diblock copolymer (of degree of polymerization $N$) with one hydrogen bonding block and one non-hydrogen bonding block in the random-phase approximation (RPA)\cite{Leibler1980TheoryCopolymers}. 

For the case of alcohols, it was shown by Painter and Coleman that the association constants measured for a monomer can be used to describe association in polymer systems. In order to do this, the association constants should be rescaled to the molar volume of a polymer segment, so that
\begin{equation}
K_i^\text{polymer}=\frac{v_\text{acrylamide}}{v_\text{segment}}K_i^\text{acrylamide}.
\end{equation}
It is worth mentioning that association constants cannot be measured directly for polymers by infrared spectroscopy, for two reasons. The first is that hydrogen bonding polymers are not soluble in ``inert'' solvents that do not have specific interactions with the polymer. The second reason is that, since the hydrogen-bonding segments are connected by covalent bonds, all hydrogen-bonding segments have hydrogen-bonding neighbors even at infinite dilution, which means that the polymer segments are not randomly mixed. This is a well-known problem and ideas have been proposed to address it in different areas of polymer theory\cite{Morse2009OnBlends, Lodge2000Self-concentrationsBlends}. Painter, Veytsman and Coleman also proposed an approach to this problem for mixtures of hydrogen-bonding homopolymers\cite{Painter1997a}. However, here we use the random mixing approximation as the simplest starting point for a discussion.

According to one of the basic assumptions of the association model approach, the hydrogen-bonding contribution can be isolated from all other contributions to the free energy, so we have $F=F_0+F_\text{HB}$ where $F_0$ is the free energy of the system (of total volume $V$) without hydrogen bonds and $F_\text{HB}$ is the contribution due to hydrogen bonding. For a melt of diblock copolymer chains where the local and mean volume fractions of the hydrogen-bonding block are $\phi_B\left(\vec{r}\right)$ and $\langle\phi_B\rangle=f$ respectively, this gives
\begin{equation}
\begin{split}
\frac{NF}{kT\rho_0V}=&-\ln\left[\frac{Q\left[w_A,w_B\right]}{V}\right]\\
+&\frac{1}{V}\int d\vec{r}\left[\chi N \phi_A\phi_B-w_A\phi_A-w_B\phi_B\right.\\
-&\eta \left(1-\phi_A-\phi_B\right)\left. +Nf_\text{HB}\left(\phi_B\right)\right],
\end{split}
\end{equation}
where $\phi_A\left(\vec{r}\right)$ is the local volume fraction of the non-hydrogen-bonding block, $\rho_0$ is the bulk segment density, $Q$ is the single-chain partition function, $w_A$ and $w_B$ are the fields corresponding to the two blocks, and $\eta$ is a Lagrange multiplier that imposes incompressibility.
With this expression in hand, the standard derivation of the scattering function in the disordered state in the RPA \cite{Leibler1980,Matsen2006Self-ConsistentApplications} can be followed to show that
\begin{equation}
\begin{split}
S^{-1}\left(k\right)=&\frac{g\left(x,1\right)}{N\text{det}\left(S\right)}-2\chi+\frac{d^2f_\text{HB}\left(f\right)}{d\phi_B^2}\\
=&S^{-1}_0\left(k\right)+\frac{d^2f_\text{HB}\left(f\right)}{d\phi_B^2},
\end{split}
\end{equation}
where\cite{Matsen2006Self-ConsistentApplications} $g(x,s)\equiv 2\left[\exp(-sx)+sx-1\right]/x^2$ and $S_0^{-1}\left(k\right)$ is the inverse scattering function of the block copolymer without hydrogen bonds.

It can therefore be seen that, in the simplest approximation, the effect of hydrogen bonds results in an increase of the effective Flory-Huggins parameter in this expression (the second derivative of the hydrogen bonding term is negative at all values of composition and association constant) and, furthermore, that the strength of this effect depends on the volume fraction of hydrogen bonding block. This effective $\chi$ parameter can then be defined as
\begin{equation}
\chi_\text{eff}=\chi-\frac{1}{2}\frac{d^2f_\text{HB}\left(f\right)}{d\phi_B^2}=\chi+\delta\chi \left(f\right).
\end{equation}

\begin{figure}
\centering
\includegraphics{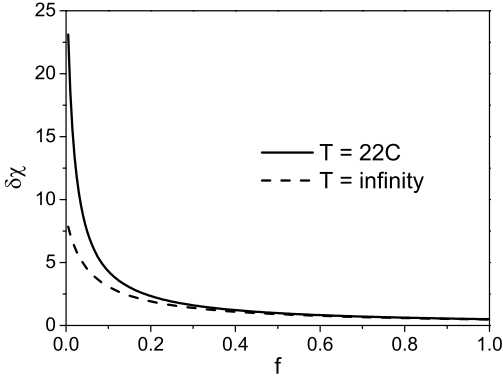}
\caption{Dependence of $\delta \chi$ on $f$ for model 1m at $T=22^\circ\text{C}$ and in the limit of infinite temperature.}
\label{chi-1m}
\end{figure}

It is interesting to note that recently it was questioned whether the Flory-Huggins parameter in the case of non-specific interactions does indeed depend on the composition of diblock copolymer, and it was shown in molecular dynamics simulations that $\chi$ can be assumed independent of volume fractions\cite{Ghasimakbari2018CorrelationsCopolymers,Mogurampelly2014Coarse-grainingSystems}. We believe that this result gives additional support to the idea of treating the contributions of non-specific and hydrogen-bonding interactions to the free energy separately.  

Now let us turn to the calculations of $\delta\chi$ for a diblock copolymer with a polyacrylamide block based on our models of hydrogen bonding association and the values of association constants deduced from fitting IR data. First, the dependence of $\delta \chi$ on $f$ for models with one association constant is considered. The graph for model 1m is shown in Figure~\ref{chi-1m}. For other models, the plots appear qualitatively the same, so they are not shown here. 

The largest absolute value of $\delta \chi$ for one-constant models is attained in the limit as $f\rightarrow 0$ and $\delta \chi$ then monotonically decreases as the fraction of hydrogen-bonding block increases. The decrease of  $\delta \chi$ with $f$ is intuitively expected, since, if more neutral segments are mixed in with the network of hydrogen-bonded segments, then more hydrogen bonds need to be broken, and more energy needs to be spent in doing so. It can also be seen that changing the temperature leads to a decrease of the maximal value of $\delta \chi$ at $f\rightarrow 0$, although this value is still very high even at infinite temperature. It is also interesting to note that at $f>0.4$ there is no change in $\delta \chi$ with temperature and that $\delta \chi$ depends only on the volume fraction $f$.

\begin{figure}
\centering
\includegraphics{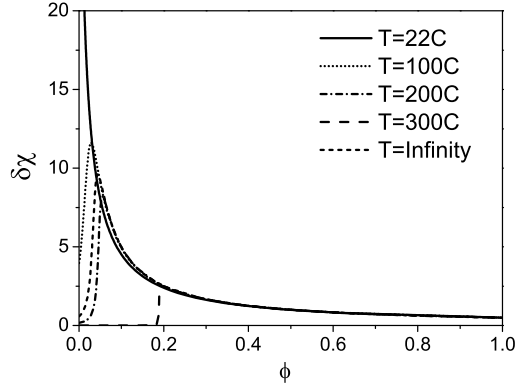}
\caption{Dependence of $\delta \chi$ on $f$ for model 6s at different temperatures.}
\label{chi-6s}
\end{figure}

\begin{figure}
\centering
\includegraphics{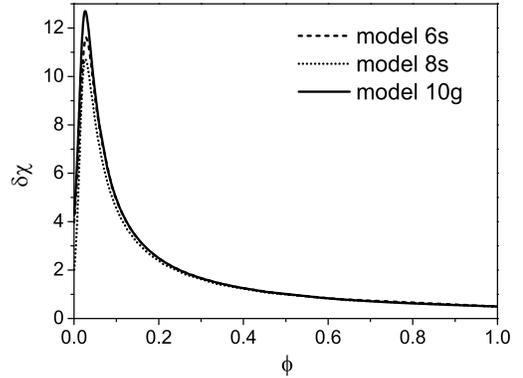}
\caption{Dependence of $\delta \chi$ on $f$ for all good models at $T=100^\circ\text{C}$.}
\label{chi-100C}
\end{figure}

In the case of models with two association constants, the behavior at small volume fractions is qualitatively different. The dependence of $\delta \chi$ on the volume fraction of the hydrogen-bonding block calculated for model 6s at different temperatures is shown in Figure~\ref{chi-6s}. As the temperature is increased, a peak at a finite value of $f$ appears, which moves to the right as the temperature grows further. However, the behavior of $\delta\chi$ at $f>0.4$ is similar to the one-parameter models: there is a gradual decrease of  $\delta \chi$ as the volume fraction of the hydrogen bonding block is increased and little difference between the curves calculated for different temperatures. It is also interesting to note that the values of $\delta\chi$ for all good two-parameter models are close to each other not only qualitatively, but also quantitatively (see Figure~\ref{chi-100C}).

It can be seen that our current predictions for $\delta \chi$ of polyacrylamide are unrealistically high. This is especially true for small values of $f$. However, these small values of $f$ are in fact never reached in polymer systems due to the non-randomness of mixing that we discussed above. In addition, it is important to re-emphasize this analysis as just an initial step on the way to the application of the association model approach to describe hydrogen bonding interactions in block copolymers.

It is also worth mentioning that the common practice of determining $\chi_\text{eff}$ by means of fitting the scattering structure factor in the disordered state in hydrogen-bonding polymers irrespective of the volume fractions probably needs to be changed\cite{Sweat2014PhaseMonomer,Kwak2017FabricationNanolithography}. Moreover, the values of $\chi_\text{eff}$ determined for hydrogen-bonding block copolymer using this route cannot be used to write the interaction free energy contribution in the form $\chi_\text{eff}\phi\left(1-\phi\right)$, since the contribution of hydrogen bonds depends on $\phi$ in a more complicated way. Instead, an appropriate expression provided by the association model approach could be used.

\section{Conclusion}

In this work, an extension of the association model approach was developed in order to describe the association of molecules with two hydrogen acceptor and two hydrogen donor sites. Models with one association constant were considered, in which it is supposed that all bonds have the same energy, and models with two association constants, in which the bond energy is determined by the local hydrogen-bonding environment. 

These models are used to fit FTIR experimental data on solutions of acrylamide in chloroform in order to determine the association constants and their temperature dependence for acrylamide, and found that several models give the same quality of fitting of experimental data.  However, models with two association constants in general give better fits than models with one association constant. Moreover, the bond energies in hydrogen bonding dimers for two-constant models are close to the predictions of DFT calculations, which is not the case in one-constant models.

It was also found that, in systems in which two bonds per acceptor site and two bonds per donor site are allowed (such as water), the assumption that cycles are absent ceases to be valid at small concentrations of the hydrogen-bonding substance and there is no non-cyclic solution of the model when the ratio between the number of bonds and number of molecules is larger than $2/3$. Interestingly, the largest average aggregation number possible in this model is equal to $3$. Based on this result, one can conclude that in such substances as acrylamide taking into account formation of cycles is essential.

Finally, the structure factor of a disordered state of diblock copolymer with one hydrogen-bonding block and one non-hydrogen-bonding block was calculated in the random phase approximation. We showed that the presence of hydrogen bonds shifts the value of the Flory-Huggins parameter that appears in the expression for the inverse scattering function, and that this shifted value depends on the volume fraction of the hydrogen bonding block, $\chi_\text{eff}=\chi+\delta\chi\left(f\right)$. The calculations showed that, in general, one-parameter and two-parameter models give similar predictions for the dependence of $\delta\chi$ on the volume fraction of the hydrogen bonding block and on temperature.
It is also interesting to note that all good two constant models give very similar quantitative predictions for $\delta\chi$, making the fact that we were unable to determine the best model unimportant from the viewpoint of practical applications.

\section{Experimental}

The FTIR experiments were conducted on a Frontier Perkin Elmer spectrometer. A Specac heatable sealed liquid cell with path length $1\,\text{mm}$ and NaCl windows was used. Acrylamide ($\ge$ 99$\%$) and chloroform (anhydrous, stabilized by amylenes, $\ge$ 99.8$\%$) were purchased from Sigma Aldrich. 

\section{Acknowledgement}

The work was supported by the Marie Sklodowska-Curie IF ``HYBOCOMIX'' (ID 704459).
The authors thank John Lane for discussions of the model selection criteria, Tom McLeish for discussions of the physics of the association model approach, and Simon Smith, Ivan Ado, Noam Zeilberger and Mark van Hoeij for discussions of the graph theoretical part of the work. The authors also thank Mark van Hoeij for the proof that the solution we found for model 2 is a unique one. 

\section{Supporting Information}

Supporting Information includes descriptions and calculations for all models and a short reference about AICc criteria for model selection.


\pagebreak
\widetext
\begin{center}
\textbf{\large Supplementary Materials: Hydrogen bonding in acrylamide and its role in the scattering behavior of acrylamide-based block copolymers}
\end{center}
\setcounter{equation}{0}
\setcounter{figure}{0}
\setcounter{table}{0}
\setcounter{page}{1}
\makeatletter
\renewcommand{\theequation}{S\arabic{equation}}
\renewcommand{\thefigure}{S\arabic{figure}}
\renewcommand{\bibnumfmt}[1]{[S#1]}
\renewcommand{\citenumfont}[1]{S#1}
\section{Models}

\subsection{Model 0}

\begin{figure}
\centering
\includegraphics[width=0.7\textwidth]{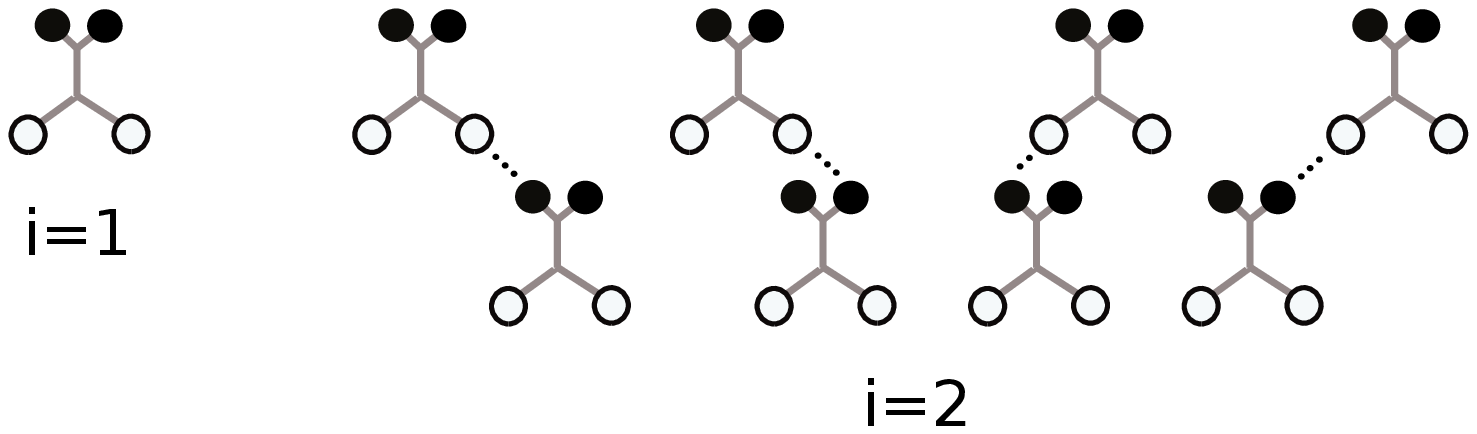}
\caption{Schematic representation of the amide group and all four ways
  of forming a hydrogen bonded linear dimer from identical
  molecules. Hydrogen bonds are represented by dots. In Model 0, we
  assume that only unimers and linear dimers are present in the solution.}
\label{model0-scheme}
\end{figure}

In model 0, we assume that the solution contains only monomers and
linear dimers (see Figure~\ref{model0-scheme}), and that chemical
equilibrium with respect to hydrogen bonding association is described
by a single association constant $K$. Then, the total concentration of
the solution and the concentration of unimers are related to each other by the equation
\begin{equation}
c=c_1+2c_2,
\label{model0-1}
\end{equation}
where the concentration of dimers is $c_2=4Kc_1^2$. The factor of $4$
appears here because there are four ways to form a dimer from two
identical molecules (due to the presence of two hydrogens in the $\text{NH}_2$ group and two lone electron pairs on the oxygen).

The free energy of hydrogen bonding of a system of volume $V$ with $N$ molecules and $M$ dimers can be written as 
\begin{equation}
F_\text{HB}=M\epsilon-kT\ln\left(p^M\Xi\right),
\end{equation}
where $\epsilon$ is the energy of a hydrogen bond, $p=C/V$ (where $C$ is a
constant) is the
probability that two molecules will meet and orient with respect to
each other to from a bond, and $\Xi$ is the combinatorial number of
ways to form $M$ dimers out of $N$ molecules, such that
\begin{equation}
\Xi=\frac{N!2^M}{\left(N-M\right)!}\frac{\left(N-M\right)!2^M}{\left(N-2M\right)!}\frac{1}{M!}=\frac{N!4^M}{\left(N-2M\right)!M!},
\end{equation}
where the first factor is the number of ways of choosing $M$ acceptor
molecules, the second factor is the number of ways of choosing $M$ donor
molecules, the factor $2^M$ takes into account that each molecule has
two hydrogens (in the case when we assume that two bonds per oxygen
are possible then an additional factor of $2^M$ appears), and the last
factor takes into account the indistinguishability of bonds.

Minimizing the free energy with respect to $M$ yields
\begin{equation}
\frac{M}{4\left(N-2M\right)^2}=\frac{K}{V},
\end{equation}
where $K=C\exp\left(-\epsilon/kT\right)$, or, in terms of concentrations,
\begin{equation}
\frac{m}{4\left(c-2m\right)^2}=K.
\end{equation}
As the concentration of unimers in the system is $c_1=c-2m=c-8Kc_1^2$, we get $m=c_2=4Kc_1^2$
and $c=c_1+8Kc_1^2$ in agreement with Equation~\ref{model0-1}.

Let us now try to determine the association constant by fitting the
dependence of the concentration on the height of the $3530\text{cm}^{-1}$ peak.

\subsubsection{Model 0m}
\begin{figure}
\centering
\includegraphics{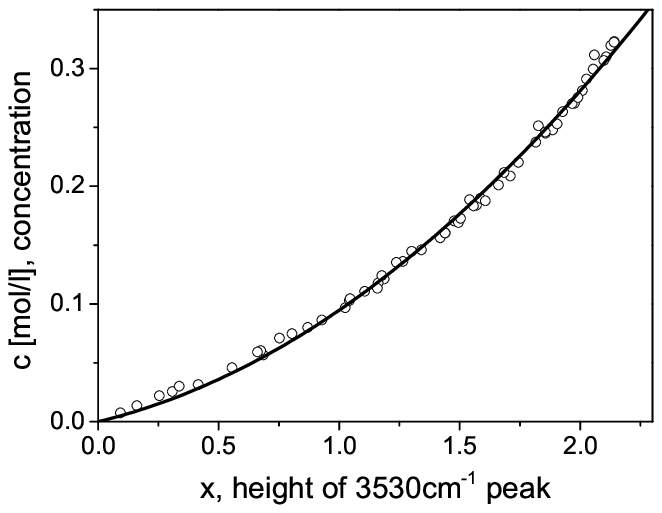}
\caption{Fit of the dependence of the total concentration on the
  height of the $3530\text{cm}^{-1}$ peak at $T=22^\circ\text{C}$ with model 0m.}
\label{model-0m-22C}
\end{figure}

First, we assume that the $3530\text{cm}^{-1}$ peak corresponds to the
out-of-phase vibrations of the $\text{NH}_2$ group in unimers. In this case,
$c_1=Ax$, where $x$ is the height of the peak and $A$ is some
constant. Substituting this in Equation~\ref{model0-1} gives the fitting equation
\begin{equation}
c=Ax+8KA^2x^2.
\end{equation}
A fit of the dependence of peak intensity on concentration at
$T=22^\circ\text{C}$ is shown in Figure~\ref{model-0m-22C}. The
quality of the
nonlinear fit can be quantified by the Akaike Information Criterion,
on which further details are given at the end of this document. The
value of this quantity for the current fit is $\text{AICc}=-457.6$.

\begin{figure}
\centering
\includegraphics{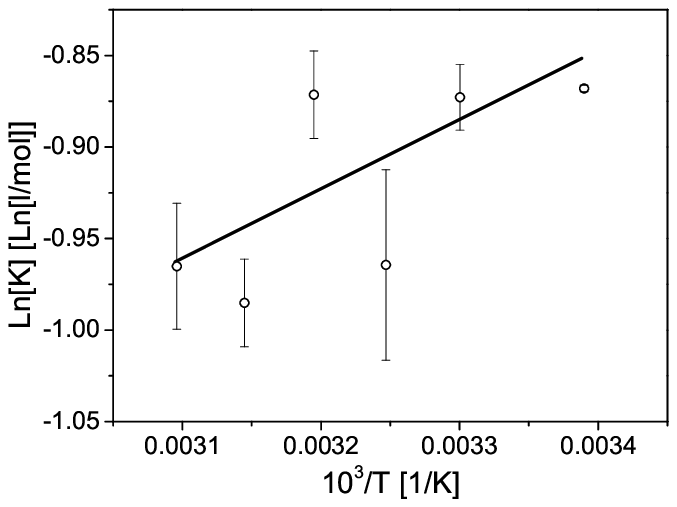}
\caption{Dependence of $\ln K$ on $1/T$ for model 0m.}
\label{model-0m-kfit}
\end{figure}

According to the definition of the association constant in our model,
$K=C\exp\left(-\epsilon/kT\right)$, $\ln K$ should depend linearly on
inverse temperature $1/T$, and this dependence, together with a linear
weighted fit, is shown in Figure~\ref{model-0m-kfit}. This yields
estimates for the model parameters of $\ln C=-0.84\pm 3 \,\ln[\text{l/mol}]$
and $\epsilon=-1.03 \pm 1.8 \,\text{kcal/mol}$. The poor quality of the fit in
this case is reflected in the low value of the coefficient
of determination, $r^2=0.07$.

\subsubsection{Model 0g}
\begin{figure}
\centering
\includegraphics{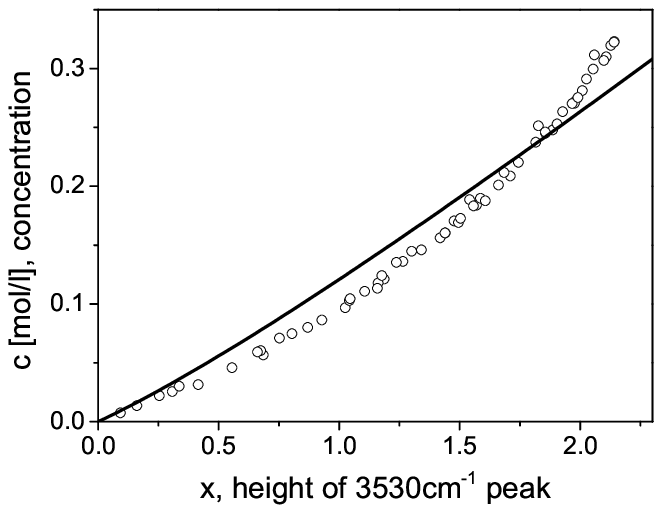}
\caption{Fit of the dependence of total concentration on height of $3530\text{cm}^{-1}$ peak at $T=22^\circ\text{C}$ with model 0g.}
\label{model-0g-22C}
\end{figure}

It is now assumed that the $3530\text{cm}^{-1}$ peak corresponds to the
out-of-phase vibrations of free $\text{NH}_2$ groups. In this case,
$Ax=c_1+4Kc_1^2$, and the fitting equation is
\begin{equation}
c=\frac{1+16AKx-\sqrt{1+16AKx}}{8K}.
\end{equation}
A fit of peak intensity versus concentration at $T=22^\circ\text{C}$ is
shown in Figure~\ref{model-0g-22C}. This fit is visibly less
successful than that for model 0m, and AICc takes the higher value of $-298$.

\begin{figure}
\centering
\includegraphics{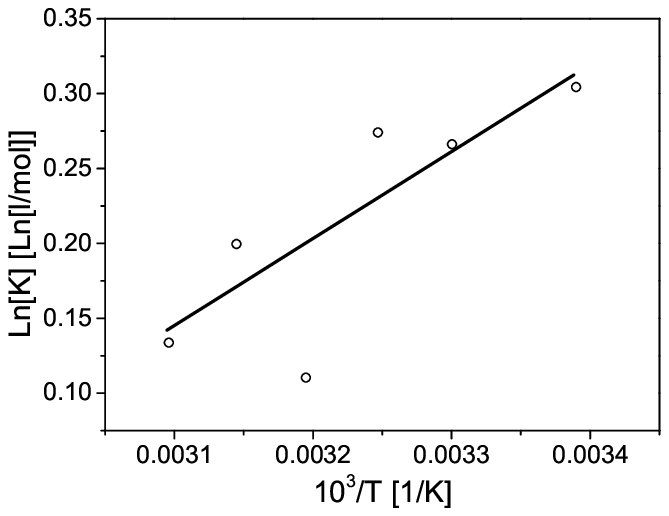}
\caption{Dependence of $\ln K$ on $1/T$ for model 0g. The error bars
  are too large to be shown.}
\label{model-0g-kfit}
\end{figure}

The dependence of $\ln K$ on $1/T$ is shown in
Figure~\ref{model-0g-kfit}. The estimates of the model parameters are
$\ln C=-1.65\pm 0.5 \,\ln[\text{l/mol}]$ and $\epsilon=-1.15 \pm 0.3\,
\text{kcal/mol}$. The quality of the fit is better than in model 0m, and this is shown by the higher value of the
coefficient of determination, $r^2=0.75$. However, we note that this
apparent improvement may be offset by the very large error bars on
$\ln K$, which probably result from the poor quality nonlinear fit in Figure~\ref{model-0g-22C}.

The free energy density due to hydrogen bonding in model 0 in terms of
the volume fraction of hydrogen bonding molecules $\phi=cv$ and the dimensionless association constant $K'=K/v$  has the form
\begin{equation}
f_\text{HB}=\frac{\left(\sqrt{1+32K'\phi}-1\right)^2}{64K'}+\phi\ln \frac{\sqrt{1+32K'\phi}-1}{16K'\phi}.
\end{equation}

\subsection{Model 1}

\begin{figure}
\centering
\includegraphics[width=0.8\textwidth]{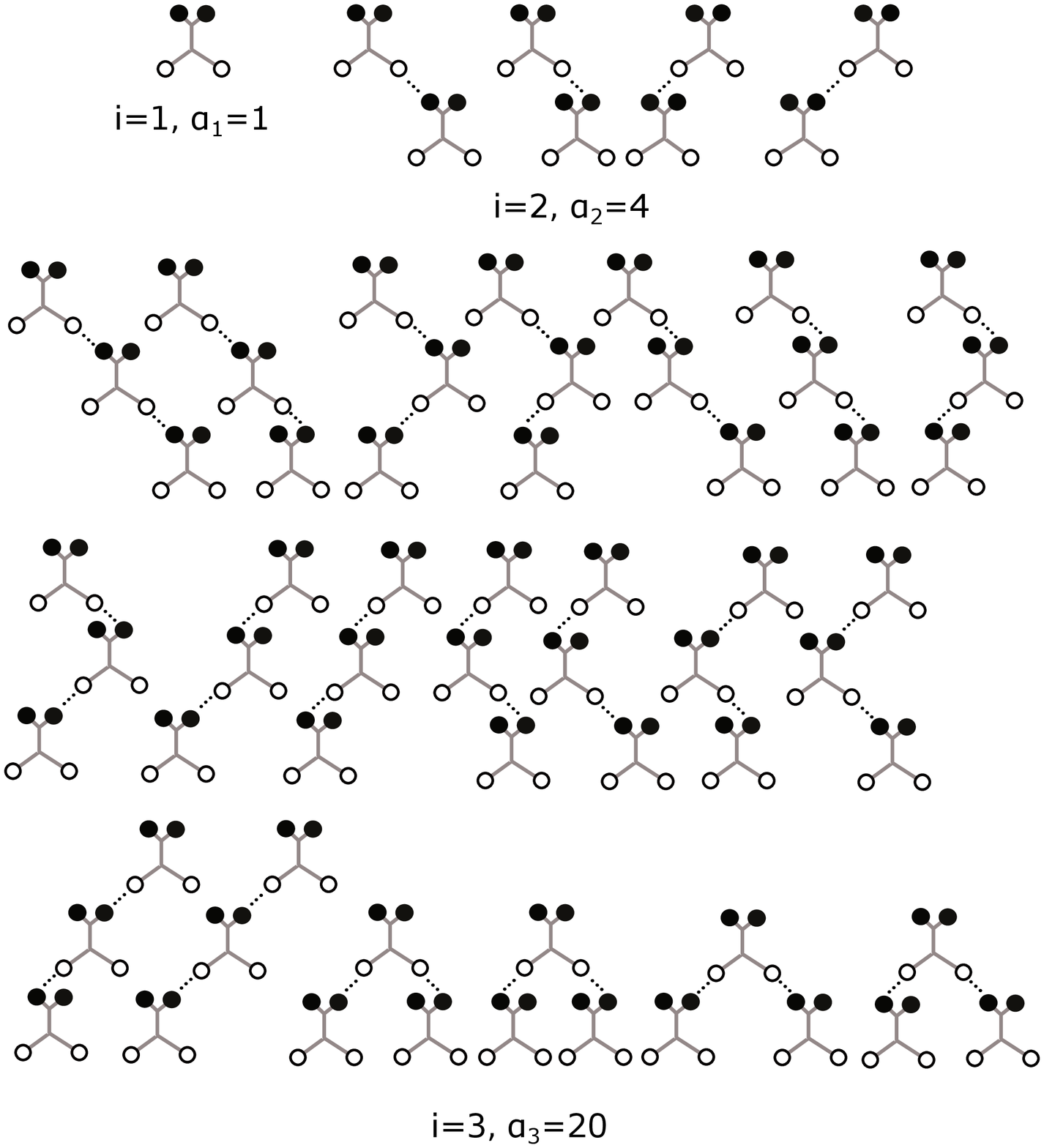}
\caption{Schematic representation of possible aggregates with size up to $i=3$ in model 1.}
\label{model1-scheme}
\end{figure}

In model 1, we assume that we have one bond per oxygen, two bonds per
$\text{NH}_2$ group, one association constant and no cycles. This
means that the aggregates are tree-shaped (see Figure~\ref{model1-scheme}).

The free energy of hydrogen bonding can be written as 
\begin{equation}
F=M\epsilon-kT\ln\left(p^M\Xi\right),
\label{free-energy1}
\end{equation}
where $M$ is the number of hydrogen bonds, $p$ is the probability that
a donor and an acceptor form a bond, and $\Xi$ is the number of ways
to form $M$ bonds, given in this case by
\begin{equation}
\Xi=\frac{N!2^M}{\left(N-M\right)!}\frac{2N!}{\left(2N-M\right)!}\frac{1}{M!},
\label{sum1}
\end{equation}
where the first factor is the number of ways to choose an acceptor,
the second is the number of ways to choose a donor, and the final
factor takes into account the fact that all bonds are identical.
Substituting Equation (\ref{sum1}) into Equation (\ref{free-energy1}) and using
Stirling's formula gives
\begin{equation}
\frac{F}{kT}=M\left(\frac{\epsilon}{kT}-\ln p\right)+N\ln\frac{\left(N-M\right)\left(2N-M\right)^2}{4N^3}+M\ln\frac{Me}{\left(N-M\right)\left(2N-M\right)}.
\label{free-energy11}
\end{equation}
After minimization with respect to $M$, we find, in terms of concentrations $m=M/V$ and $c=N/V$,
\begin{equation}
\label{1eq}
\frac{m}{2\left(c-m\right)\left(2c-m\right)}=K.
\end{equation}
Let us suppose that the concentration of aggregates of size $i$ can be expressed as  
\begin{equation}
c_i=\alpha_iK^{i-1}c_1^i,
\end{equation}
where the $\alpha_i$ are unknown coefficients. Then, for the concentration of bonds $m$ and total concentration $c$ we have  
\begin{equation}
\label{1m}
m=\sum_{i=1}^{\infty}\left(i-1\right)c_i
\end{equation}
and
\begin{equation}
\label{1c}
c=\sum_{i=1}^{\infty}ic_i.
\end{equation}
In order to find $\alpha_i$, we substitute expressions \ref{1m} and
\ref{1c} into Equation \ref{1eq} and equate coefficients in front
of like powers of $c_1$. Using this method, we can calculate the
values of the coefficients, which in this case are (starting from
$i=1$) 1, 4, 20, 672, 4224, 27456,$\ldots$. Using the On-Line Encyclopedia of
Integer Sequences (published electronically at https://oeis.org, May 2018), we can assume that the general
formula for a term of this sequence most probably has the form (in the
main paper a more elegant way to get this result is described)
\begin{equation}
\alpha_i=2^{i-1}\frac{\left(2i\right)!}{\left(i+1\right)!i!}.
\end{equation}
The sequence $\beta_i=\left(2i\right)!/\left(i+1\right)!i!$ is
known as the Catalan numbers. It is known that these numbers represent
the number of different rooted binary trees with $i + 1$ leaves. In
our case, we have an additional factor of $2^{i-1}$, since each
molecule apart from the root can be added in two ways to form a bond
with one of the free hydrogens because there are two bonding sites on the
oxygen. All aggregates allowed in this model with size up to $i = 3$
are shown in Figure~\ref{model1-scheme}. So we can say that the physical meaning of $\alpha_i$ is the number of ways to form an aggregate of size $i$ out of
$i$ molecules. 

It is also interesting to note that, by looking at
Figure~\ref{model1-scheme}, it can be seen that the aggregates can be
built recursively from each other, so a generating function
$G\left(z\right)$ can be written as 
\begin{equation}
G\left(z\right)=1+4zG\left(z\right)+4z^2G\left(z\right)^2,
\end{equation}
where $z=Kc_1$ is the multiplicative factor that is introduced when the size of the aggregate is increased by one. Solution of this equation gives 
\begin{equation}
G\left(z\right)=\frac{1-4z-\sqrt{1-8z}}{8z^2},
\end{equation}
and this generating function can be expanded with respect to $z$ to
give the values of $\alpha_i$.

With this expression for $\alpha_i$ in hand, the total concentration
of the solution can be calculated as
\begin{equation}
c=\frac{1-4 Kc_1-\sqrt{1-8 Kc_1}}{8 K^2c_1\sqrt{1-8 Kc_1} }.
\label{model1-conc}
\end{equation}

\subsubsection{Model 1m}
\begin{figure}
\centering
\includegraphics{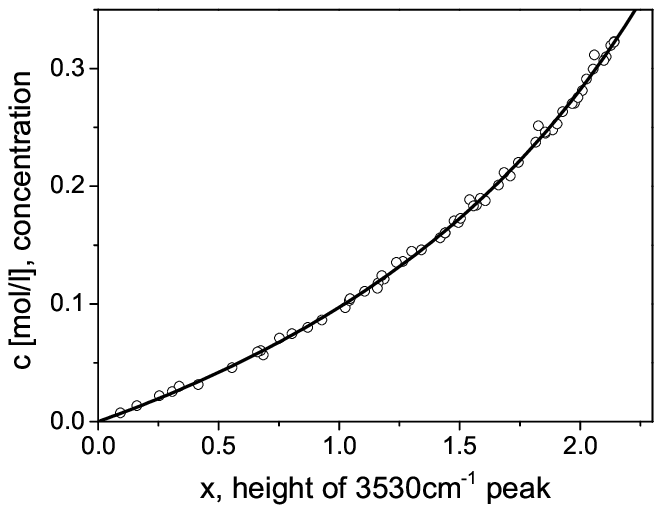}
\caption{Fit of the dependence of the total concentration on the height of $3530\text{cm}^{-1}$ peak at $T=22^\circ\text{C}$ with model 1m.}
\label{model-1m-22CS}
\end{figure}

\begin{figure}
\centering
\includegraphics{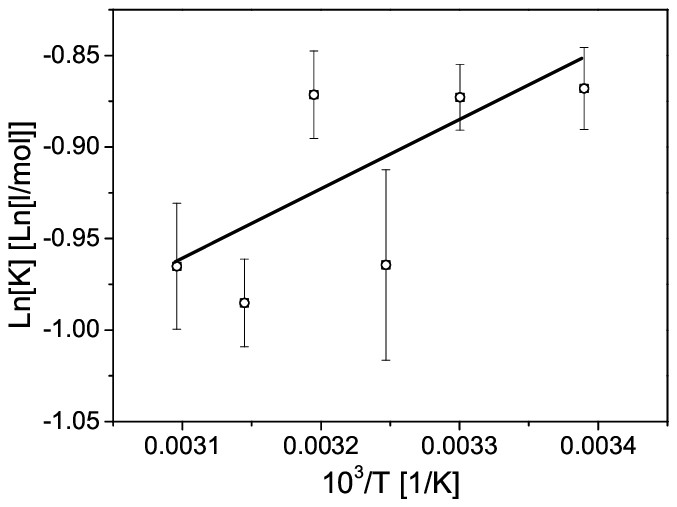}
\caption{Dependence of $\ln K$ on $1/T$ for model 1m.}
\label{model-1m-kfitS}
\end{figure}

Let us assume first that the $3530\text{cm}^{-1}$ peak corresponds to
the out-of-phase vibrations of the $\text{NH}_2$ group in unimers. In
this case, $c_1=Ax$, where $x$ is the height of the peak and $A$ is
some constant. Substituting this in Equation~\ref{model1-conc} gives the fitting equation:
\begin{equation}
c=\frac{1-4 KAx-\sqrt{1-8 KAx}}{8 K^2Ax\sqrt{1-8 KAx} }.
\end{equation}
The results of this fit are shown in Figures~\ref{model-1m-22CS}
and~\ref{model-1m-kfitS}. The value of AICc for the fit in
Figure~\ref{model-1m-22CS} is $-503.8$. The dependence of $\ln K$ on $1/T$ is shown in
Figure~\ref{model-1m-kfitS}, and this fit gives estimates of the model
parameters of $\ln C=-2.14\pm 0.6 \,\ln[\text{l/mol}]$ and
$\epsilon=-0.79 \pm 0.2 \,\text{kcal/mol}$, with a coefficient of
determination of $r^2=0.56$.  

\subsubsection{Model 1g}
\begin{figure}
\centering
\includegraphics{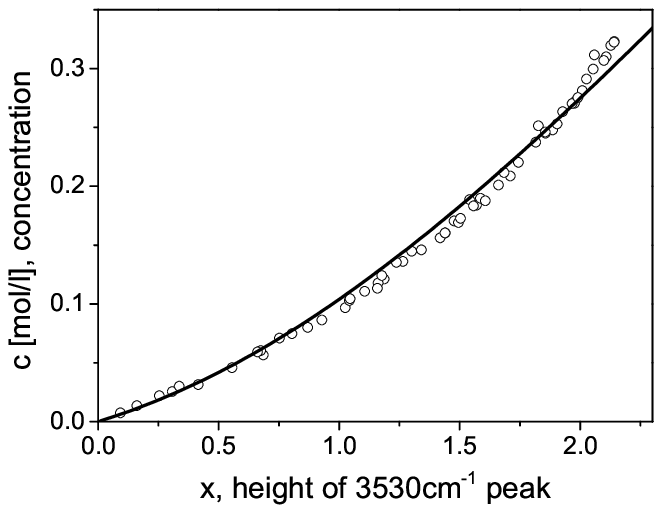}
\caption{Fit of the dependence of the total concentration on the
  height of the $3530\text{cm}^{-1}$ peak at $T=22^\circ\text{C}$ with model 1g.}
\label{model-1g-22CS}
\end{figure}

\begin{figure}
\centering
\includegraphics{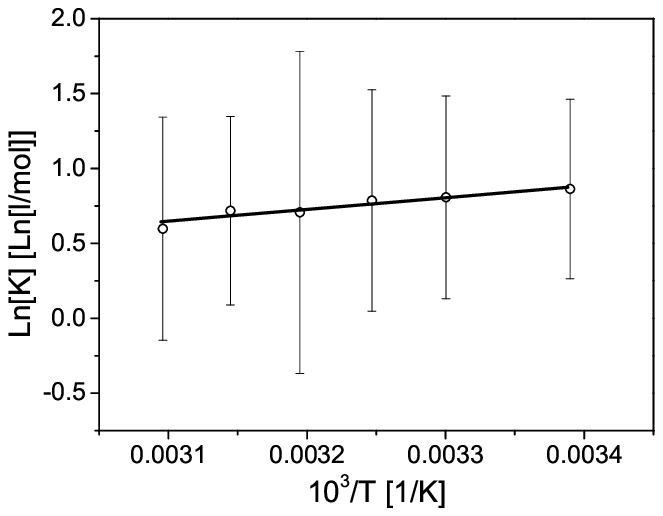}
\caption{Dependence of $\ln K$ on $1/T$ for model 1g.}
\label{model-1g-kfitS}
\end{figure}

In model 1g, it is assumed that the $3530\text{cm}^{-1}$ peak
corresponds to the out-of-phase vibrations of $\text{NH}_2$ free
groups both in free molecules and in aggregates. Therefore, in order
to find a fitting equation, we need to calculate the concentration of
free groups. However, it turns out that it is impossible to do this in
the framework of model 1 because the number of free groups depends on
the structure of the aggregate. However, model 5 (to be introduced
later) does include the relevant information about the structure of
the aggregate, and reduces to model 1 when its two association
constants are set equal to each other. Therefore, the calculation of
the fitting expression is carried out in model 5, and $K_1$ and $K_2$ are both set equal to
$K$ at the end. This gives
\begin{equation}
c=\frac{1-4AKx+16A^2K^2x^2-\left(1-4AKx\right)\sqrt{1+16A^2K^2x^2}}{8AK^2x}.
\end{equation}
The results of the fit are shown in Figures~\ref{model-1g-22CS}
and~\ref{model-1g-kfitS}. The quality of fit in
Figure~\ref{model-1g-22CS} can be characterized by the parameter
$\text{AICc}=-388.9$, and the fit shown in Figure~\ref{model-1g-kfitS}
estimates the model parameters to be $\ln C=-1.77\pm 0.4
\,\ln[\text{l/mol}]$ and $\epsilon=-1.55 \pm 0.3 \,\text{kcal/mol}$, with $r^2=0.9$.  

\subsubsection{Model 1s}
Here we assume that the $3530\text{cm}^{-1}$ peak corresponds to the
out-of-phase vibrations of $\text{NH}_2$ free groups in free molecules
and dimers. In this case, the relation between $c_1$ and $x$ takes the form $Ax=c_1+4Kc_1^2$. 

\begin{figure}
\centering
\includegraphics{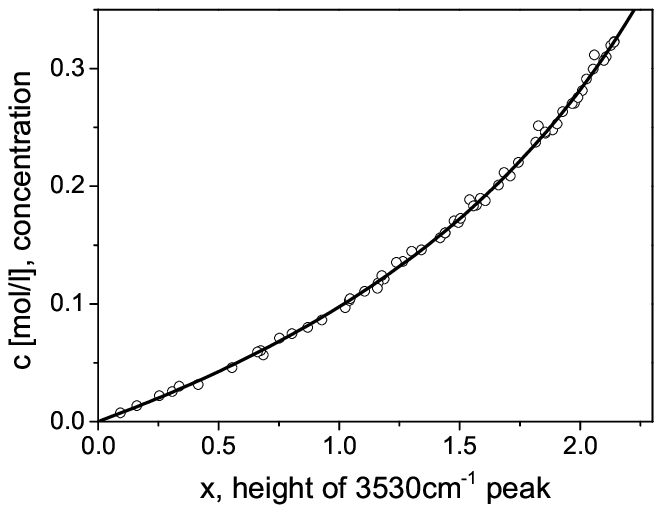}
\caption{Fit of the dependence of the total concentration on the
  height of the $3530\text{cm}^{-1}$ peak at $T=22^\circ\text{C}$ with model 1s.}
\label{model-1s-22C}
\end{figure}

\begin{figure}
\centering
\includegraphics{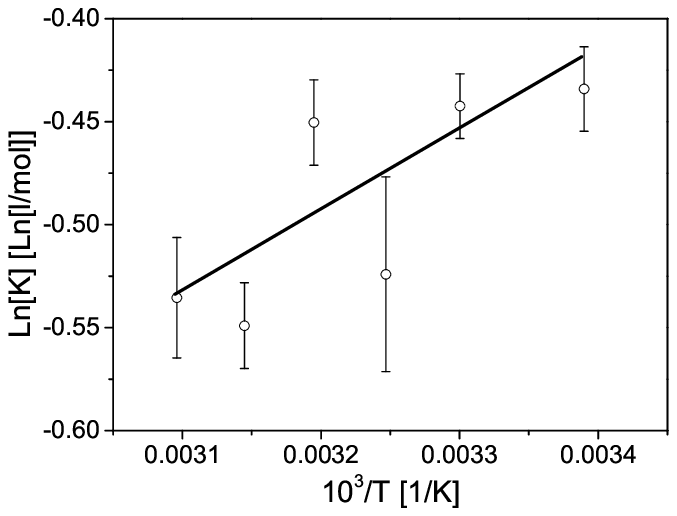}
\caption{Dependence of $\ln K$ on $1/T$ for model 1s.}
\label{model-1s-kfit}
\end{figure}

The fitting results are shown in Figures~\ref{model-1s-22C}
and~\ref{model-1s-kfit}. The value of AICc for the fit in
Figure~\ref{model-1s-22C} is $-503.9$, and the estimates for the model
parameters corresponding to Figure~\ref{model-1s-kfit} are $\ln
C=-1.75\pm 0.5 \,\ln[\text{l/mol}]$ and $\epsilon=-0.78 \pm 0.3
\,\text{kcal/mol}$, with $r^2=0.66$.

The free energy density of hydrogen bonding in model 1 in
terms of the volume fraction of hydrogen bonding molecules $\phi=cv$ and dimensionless association constant $K'=K/v$  has the form
\begin{equation}
f_\text{HB}=m+\phi\ln\frac{\left(\phi-m\right)\left(2\phi-m\right)^2}{4\phi^3},
\end{equation}
where $m$ is a solution of the equation $m/[2\left(\phi-m\right)\left(2\phi-m\right)]=K'$.


\subsection{Model 2}

\begin{figure}
\centering
\includegraphics[width=0.7\textwidth]{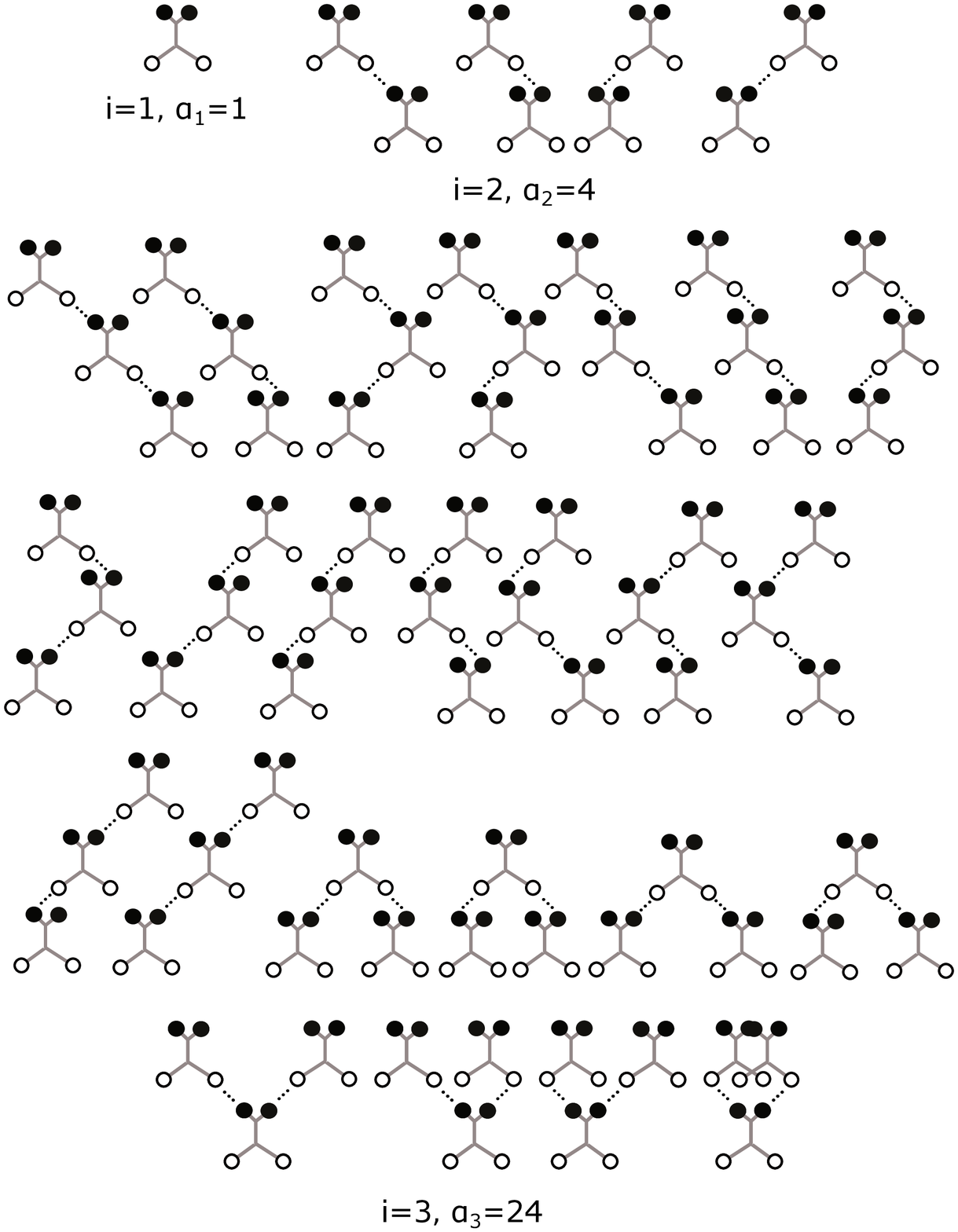}
\caption{Schematic representation of possible aggregates with size up to $i=3$ in model 2.}
\label{model2-scheme}
\end{figure}

In model 2, the following assumptions are made: oxygen can form two
bonds, the $\text{NH}_2$ group can form two bonds, there are no cycles and there is one association constant. The range of allowed aggregates for this case with sizes up to $i=3$ is shown in Figure~\ref{model2-scheme}.

The free energy of hydrogen bonding can be written as 
\begin{equation}
F=M\epsilon-kT\ln\left(p^M\Xi\right),
\label{free-energy2}
\end{equation}
where $M$ is the number of hydrogen bonds and the number of ways to form these bonds is
\begin{equation}
\Xi=\left(\frac{2N!}{\left(2N-M\right)!}\right)^2\frac{1}{M!}.
\end{equation}
Minimization of the free energy yields
\begin{equation}
\frac{m}{\left(2n-m\right)^2}=K.
\label{eq2}
\end{equation}
Next, it is assumed that the concentration of aggregates of size $i$ can be written as
\begin{equation}
c_i=\alpha_iK^{i-1}c_1^i,
\end{equation}
so that the total concentration is 
\begin{equation}
n=\sum_{i=1}^{\infty}ic_i
\label{c-model2}
\end{equation}
and the concentration of bonds is
\begin{equation}
m=\sum_{n=1}^{\infty}\left(n-1\right)c_n.
\label{m-model2}
\end{equation}
Then, we substitute Equations~\ref{c-model2} and~\ref{m-model2} into
Equation~\ref{eq2} written in the form $m=K\left(2n-m\right)^2$ and,
equating coefficients in front of like powers of $c_1$, we can
calculate the first terms of the $\alpha_i$ sequence. The
OEIS tells us that this sequence is
probably that known as A000309, whose $i^{th}$ term is given by
\begin{equation}
\alpha_i=2^i\frac{\left(3i\right)!}{\left(2i+1\right)!\left(i+1\right)!}.
\label{ai-model2}
\end{equation}
In the following section, a proof is given that the sequence A000309
is indeed the set of coefficients $\alpha_i$ that satisfy Equation~\ref{eq2}.
Using this result for the $\alpha_i$, the total concentration can be written down as 
\begin{equation}
n=c_1\cdot _{3}F_{2}\left(\left[\frac{4}{3}, \frac{5}{3}, 2\right],\left[\frac{5}{2}, 3\right],\frac{27Kc_1}{2}\right).
\label{model2-concS}
\end{equation}

\subsubsection{The proof}
\textit{This proof was provided by Mark van Hoeij.}

Let us consider $\alpha_i$, $m$ and $n$ defined according to
Equations~\ref{ai-model2},~\ref{m-model2}, and~\ref{c-model2} respectively. Then, let us define a function
\begin{equation}
y=2n-m=\frac{1}{K}\sum_{i=1}^{\infty}\left(i+1\right)\alpha_i x^i,
\end{equation}
where $x=Kc_1$.
Next, introduce $M=Km$ and $Y=ky$. Then, Equation~\ref{eq2} is
equivalent to
\begin{equation}
M=Y^2.
\end{equation}
If a function $Z$ is defined by
\begin{equation}
Z=1+Y=\sum_{i=0}^{\infty}\left(i+1\right)\alpha_ix^i
\end{equation}
(the difference with $Y$ is that the summation starts from $0$), it can be
verified that
\begin{equation}
\left(27x-2\right)xZ''+\left(54x-3\right)Z'+6Z=0.
\label{eqz}
\end{equation}
Indeed, since
\begin{equation}
3Z'=3\sum_{i=0}^{\infty}\left(i+1\right)i\alpha_ix^{i-1}=3\sum_{i=0}^{\infty}\left(i+1\right)\left(i+2\right)\alpha_{i+1}x^i
\end{equation}
and
\begin{equation}
2xZ''=2\sum_{i=0}^{\infty}\left(i+1\right)i\left(i-1\right)\alpha_ix^{i-1}=2\sum_{i=0}^{\infty}i\left(i+1\right)\left(i+2\right)\alpha_{i+1}x^i,
\end{equation}
Equation~\ref{eqz} is equivalent to 
\begin{equation}
\left(27i^2+27i+6\right)\alpha_i=\left(i+2\right)\left(2i+3\right)\alpha_{i+1},
\end{equation}
which is true for $\alpha_i$ defined by Equation~\ref{ai-model2}.
Differentiating Equation~\ref{eqz} gives
\begin{equation}
\left(27x-2\right)xZ'''+\left(108x-5\right)Z''+60Z'=0,
\end{equation}
or, since $Y'=Z'$,
\begin{equation}
\left(27x-2\right)xY'''+\left(108x-5\right)Y''+60Y'=0,
\label{eqy}
\end{equation}
and it can be found that $u=Y^2$ satisfies the equation
\begin{equation}
\begin{split}
(27x-2)^2x^3u'''''' +2x^2(270x-13)(27x-2)u''''' + 35x(2592x^2-246x+5)u''''+\\
+(204120x^2-13440x+140)u''' + (146160x-5040)u'' + 20160u'  = 0.
\end{split}
\label{equ}
\end{equation}
To show this, we note that, if we take $Y^2$ and differentiate
repeatedly, then all the resulting expressions can be written
in terms of products of $Y$, $Y'$ and $Y''$, because all instances of
$Y'''$ can be eliminated with Equation~\ref{eqy}. Next, we verify that
$M=\sum_{i=1}^{\infty}\left(i-1\right)\alpha_ix^i$ also satisfies
Equation~\ref{equ}. Since both $M$ and $Y^2$ satisfy Equation~\ref{equ}, the functions $M$ and $Y^2$ are equal if the first six
terms in their expansions in powers of $x$ coincide (because the differential equation is sixth order), which we can check by direct computation.

\subsubsection{Model 2m}

\begin{figure}
\centering
\includegraphics{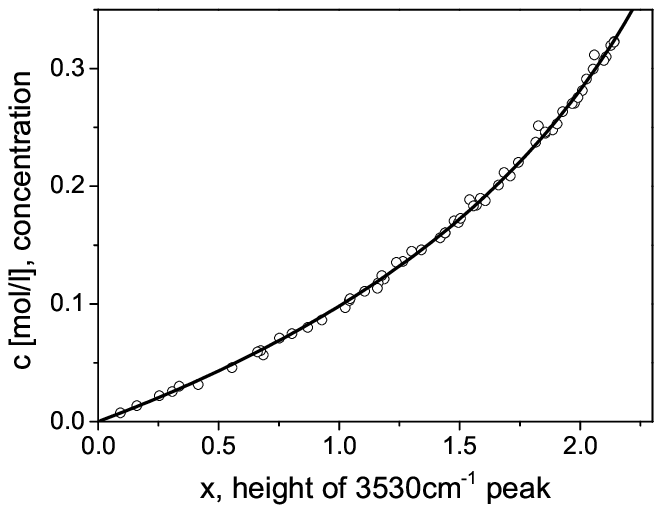}
\caption{Fit of the dependence of the total concentration on the
  height of the $3530\text{cm}^{-1}$ peak at $T=22^\circ\text{C}$ with model 2m.}
\label{model-2m-22C}
\end{figure}

\begin{figure}
\centering
\includegraphics{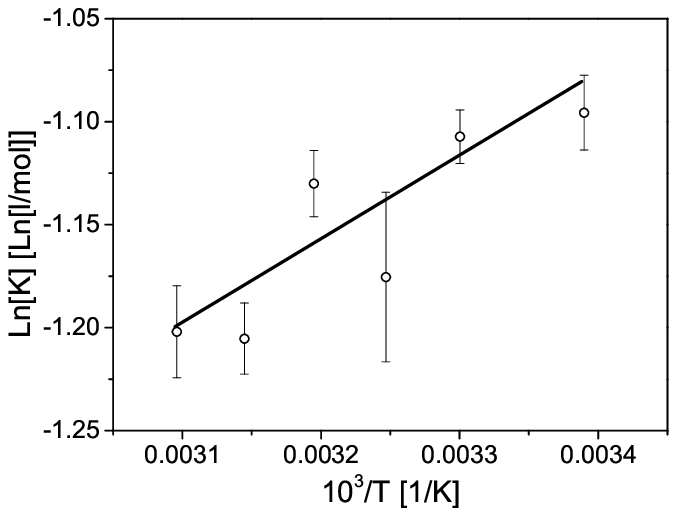}
\caption{Dependence of $\ln K$ on $1/T$ for model 2m.}
\label{model-2m-kfit}
\end{figure}

Let us assume first that the $3530\text{cm}^{-1}$ peak corresponds to
the out-of-phase vibrations of the $\text{NH}_2$ group in unimers. In this case, $c_1=Ax$ where $x$ is the height of the peak and $A$ is some constant. Substituting this in Equation~\ref{model2-concS} gives
\begin{equation}
n=Ax\cdot _{3}F_{2}\left(\left[\frac{4}{3}, \frac{5}{3}, 2\right],\left[\frac{5}{2}, 3\right],\frac{27KAx}{2}\right).
\label{model2m-conc}
\end{equation}
The results of the fit are shown in Figures~\ref{model-2m-22C}
and~\ref{model-2m-kfit}. The value of AICc in
Figure~\ref{model-2m-22C} is $-502.5$, and the estimates of the model
parameters given by the fit in Figure~\ref{model-2m-kfit} are $\ln
C=-2.5\pm 0.3 \,\ln[\text{l/mol}]$ and $\epsilon=-0.8 \pm 0.2
\,\text{kcal/mol}$ with $r^2=0.75$.  

\subsubsection{Model 2g}
Let us assume now that the $3530\text{cm}^{-1}$ peak corresponds to
out-of-phase vibrations of $\text{NH}_2$ free groups. As in the case
of model 1g, we need to include more information about the aggregate
in order to calculate the concentration of free groups. If we
distinguish between molecules with both hydrogens bonded ($N_2$ is the
number of such molecules) and only one bonded hydrogen in the amide group
($N_1$ is the number of such molecules) we can write the number of ways to form bonds as (see model 6 for more details)
\begin{equation}
\begin{split}
\Xi=&\frac{2N!}{\left(2N-M\right)!}\frac{N!2^{N_1}}{\left(N-N_2-N_1\right)!N_1!N_2!}\\
=&\frac{2N!}{\left(2N-M\right)!}\frac{N!2^{2N-M-2N_f}}{N_f!\left(2N-M-2N_f\right)!\left(N_f+M-N\right)!}.
\end{split}
\label{xi-5}
\end{equation}
where $M=N_1+2N_2$ and $N_f=N-N_1-N_2$. Minimizing the free energy and
eliminating $M$ gives
\begin{equation}
c=\frac{n_f}{\left(1-2Kn_f\right)^2}=\frac{Ax}{\left(1-2KAx\right)^2},
\end{equation}
where, according to our peak attribution assumption, $n_f=Ax$.

\begin{figure}
\centering
\includegraphics{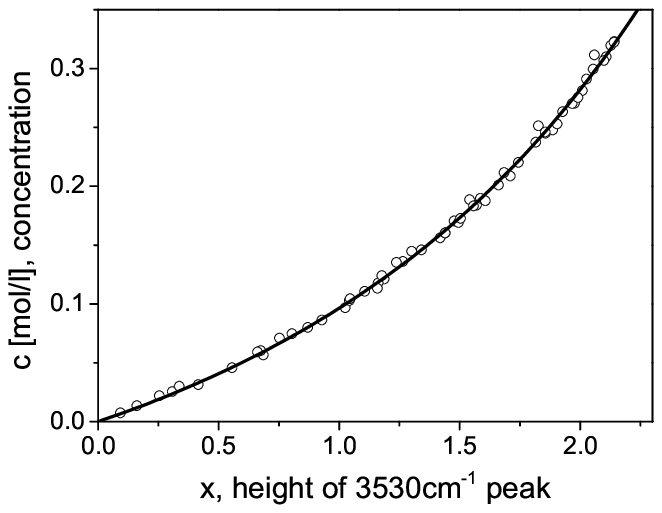}
\caption{Fit of the dependence of the total concentration on the
  height of the $3530\text{cm}^{-1}$ peak at $T=22^\circ\text{C}$ with model 2g.}
\label{model-2g-22C}
\end{figure}

\begin{figure}
\centering
\includegraphics{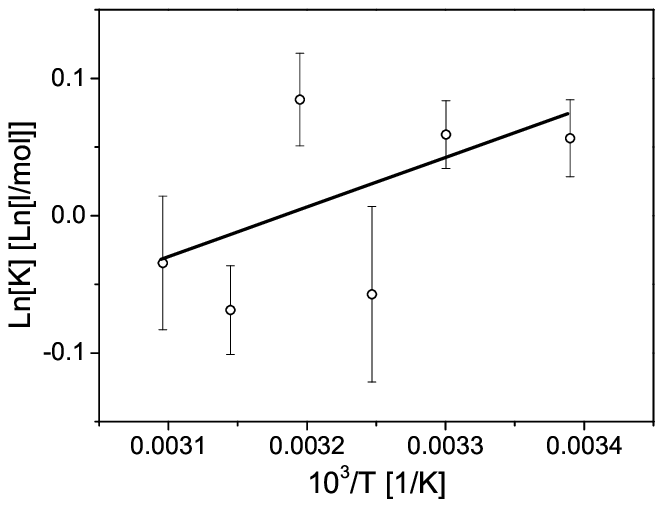}
\caption{Dependence of $\ln K$ on $1/T$ for model 2g.}
\label{model-2g-kfit}
\end{figure}

The results of this fit are shown in Figures~\ref{model-2g-22C} and~\ref{model-2g-kfit}. The value of AICc for the fit in
Figure~\ref{model-2g-22C} is $-499.7$, and the fit in
Figure~\ref{model-2g-kfit} yields estimates for the model parameters
of $\ln C=-1.2\pm 0.7 \,\ln[\text{l/mol}]$ and $\epsilon=-0.7 \pm 0.5
\,\text{kcal/mol}$, with $r^2=0.38$.  

\subsubsection{Model 2s}
\begin{figure}
\centering
\includegraphics{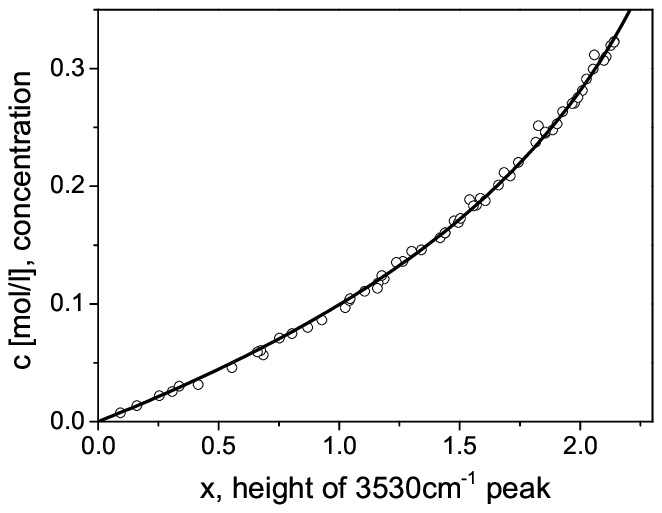}
\caption{Fit of the dependence of the total concentration on the
  height of the $3530\text{cm}^{-1}$ peak at $T=22^\circ\text{C}$ with model 2s.}
\label{model-2s-22C}
\end{figure}

\begin{figure}
\centering
\includegraphics{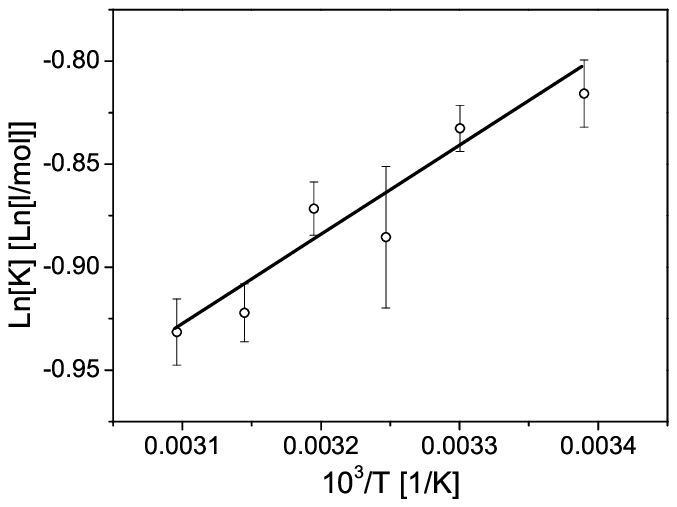}
\caption{Dependence of $\ln K$ on $1/T$ for model 2s.}
\label{model-2s-kfit}
\end{figure}

Let us assume here that the $3530\text{cm}^{-1}$ peak corresponds to
the out-of-phase vibrations of the free $\text{NH}_2$ group in unimers
and dimers. In this case, $Ax=c_1+4Kc_1^2$, where $x$ is the height of
the peak and $A$ is a constant. Substituting this in
Equation~\ref{model2-concS} yields
\begin{equation}
c=\frac{-1+\sqrt{1+16AKx}}{8K}\cdot _{3}F_{2}\left(\left[\frac{4}{3}, \frac{5}{3}, 2\right],\left[\frac{5}{2}, 3\right],\frac{27}{16}\left(-1+\sqrt{1+16AKx}\right)\right).
\label{model2s-conc}
\end{equation}
The results of the fit are shown in Figures~\ref{model-2s-22C} and~\ref{model-2s-kfit}. The value of AICc for the fit shown in Figure~\ref{model-2s-22C} is
$-494.2$, and the fit in Figure~\ref{model-2s-kfit} yields estimates
for the model parameters of $\ln C=0.50\pm 0.2 \,\ln[\text{l/mol}]$ and
$\epsilon=-0.86 \pm 0.1 \,\text{kcal/mol}$, with $r^2=0.92$.

The free energy due to hydrogen bonding in model 2 in terms of the
volume fraction of hydrogen bonding molecules $\phi=cv$ and the dimensionless association constant $K'=K/v$  has the form
\begin{equation}
f_\text{HB}=m+4\phi\ln\frac{\left(2\phi-m\right)}{2\phi},
\end{equation}
where $m$ is a solution of the equation $m/\left(2\phi-m\right)^2=K'$.




\subsection{Model 3}
\begin{figure}
\centering
\includegraphics[width=0.8\textwidth]{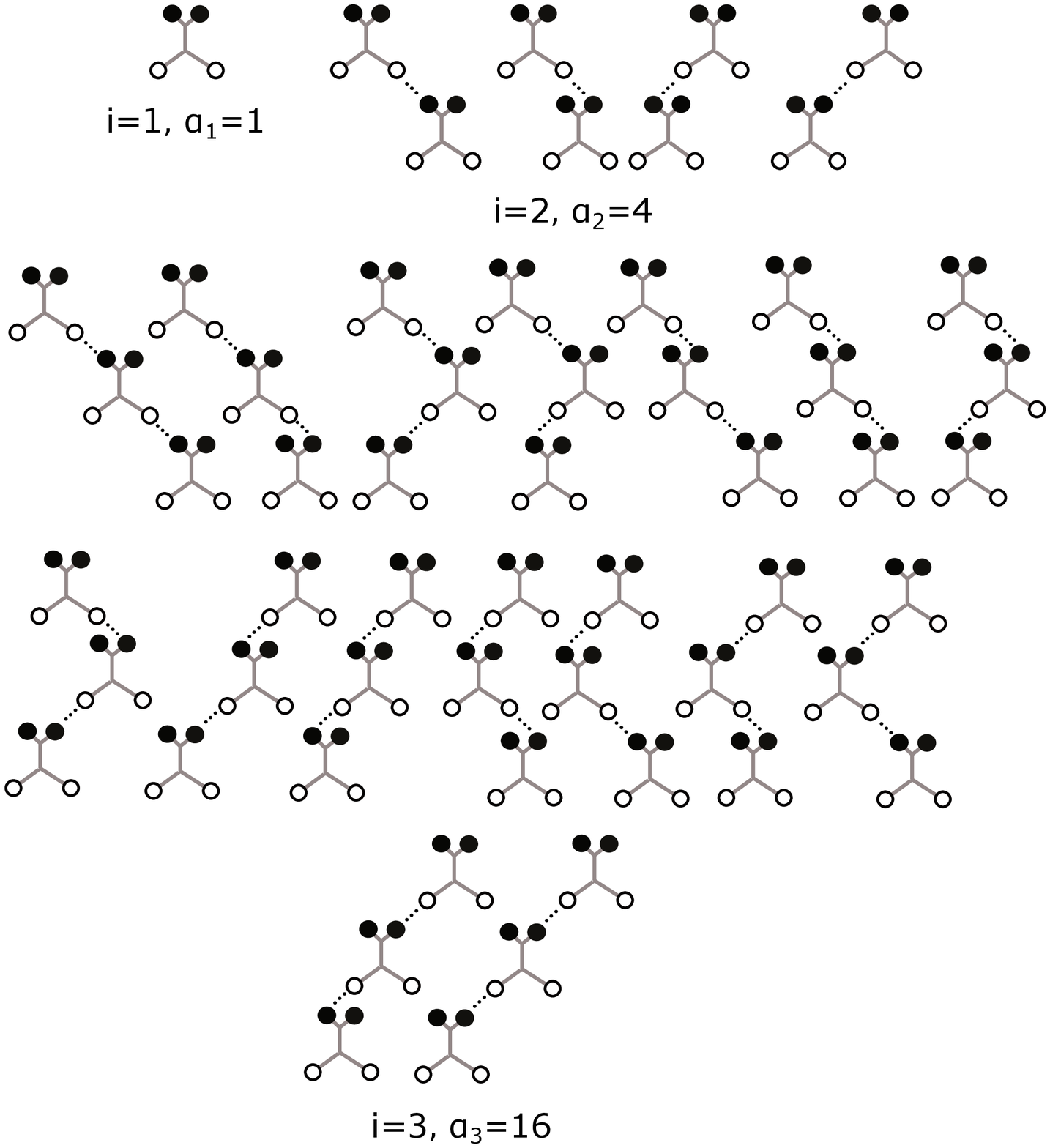}
\caption{Schematic representation of allowed aggregates with size up to $i=3$ in model 3.}
\label{model3-scheme}
\end{figure}

In this model, it is assumed that there is only one bond per oxygen and one bond per $\text{NH}_2$, there are no cyclic dimers and there is only one association constant. The range of allowed aggregates with aggregation numbers up to $i=3$ is shown in Figure~\ref{model3-scheme}; aggregates are chain-like in this case.

The number of ways to form bonds in model 3 can be written as
\begin{equation}
\Xi=\left(\frac{N!}{\left(N-M\right)!}\right)^2\frac{4^M}{M!},
\end{equation}
and minimization of the free energy gives
\begin{equation}
\frac{M}{4\left(N-M\right)^2}=\frac{K}{V}.
\label{eq3}
\end{equation}
In this case, the concentration of aggregates of size $i$ is
\begin{equation}
c_i=4^{i-1}K^{i-1}c_1^i,
\end{equation}
so for the total concentration we have
\begin{equation}
c=\frac{c_1}{\left(1-4Kc_1\right)^2}.
\end{equation}
The concentration of free groups is given by $n_f=n-m$, so the
relation between the total concentration and the concentration of free groups is
\begin{equation}
n=n_f+4Kn_f^2.
\end{equation}

\subsubsection{Model 3m}
Let us first assume that the $3530\text{cm}^{-1}$ peak corresponds to
the out-of-phase vibrations of the $\text{NH}_2$ group in unimers. In
this case, $c_1=Ax$, where $x$ is the height of the peak and $A$ is
some constant. The fitting equation is then
\begin{equation}
c=\frac{Ax}{\left(1-4KAx\right)^2}.
\end{equation}
This expression is the same as the fitting equation for model 2g, but
with an association constant that is two times smaller. This means
that the fitting results are the same, with the values of the model
parameters and statistical measures being $\text{AICc}=-499.7$, $\ln
C=-1.9\pm 0.7 \,\ln[\text{l/mol}]$, $\epsilon=-0.7 \pm 0.5
\,\text{kcal/mol}$, and $r^2=0.38$.

\subsubsection{Model 3g}
In this model, it is assumed that the $3530\text{cm}^{-1}$ peak
corresponds to the out-of-phase vibrations of the free $\text{NH}_2$ groups. Here,
\begin{equation}
c=Ax+4KA^2x^2.
\end{equation}

\begin{figure}
\centering
\includegraphics{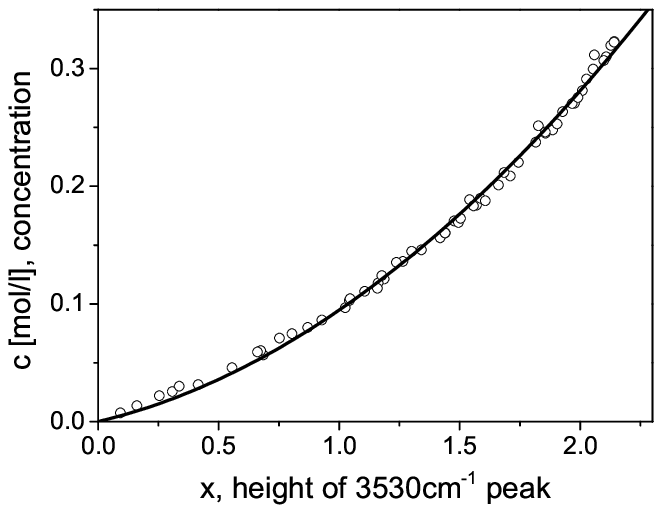}
\caption{Fit of the dependence of the total concentration on the
  height of the $3530\text{cm}^{-1}$ peak at $T=22^\circ\text{C}$ with model 3g.}
\label{model-3g-22C}
\end{figure}

\begin{figure}
\centering
\includegraphics{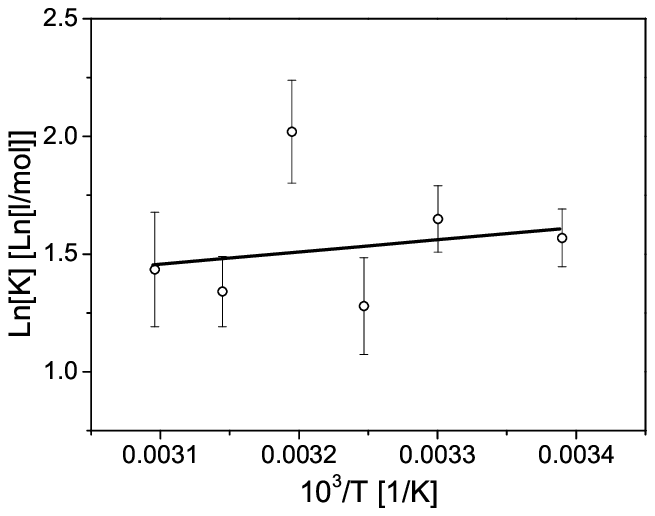}
\caption{Dependence of $\ln K$ on $1/T$ for model 3g.}
\label{model-3g-kfit}
\end{figure}

The results of the fitting procedure are shown in
Figures~\ref{model-3g-22C} and~\ref{model-3g-kfit}. The value of the
AICc parameter for the fit
in Figure~\ref{model-3g-22C} is $-457.6$, and the estimates of the
model parameters given by the fit in~\ref{model-3g-kfit} are $\ln
C=-0.15\pm 3 \,\ln[\text{l/mol}]$ and $\epsilon=-1.0 \pm 1.9
\,\text{kcal/mol}$, with $r^2=0.07$.  

\subsubsection{Model 3s}
\begin{figure}
\centering
\includegraphics{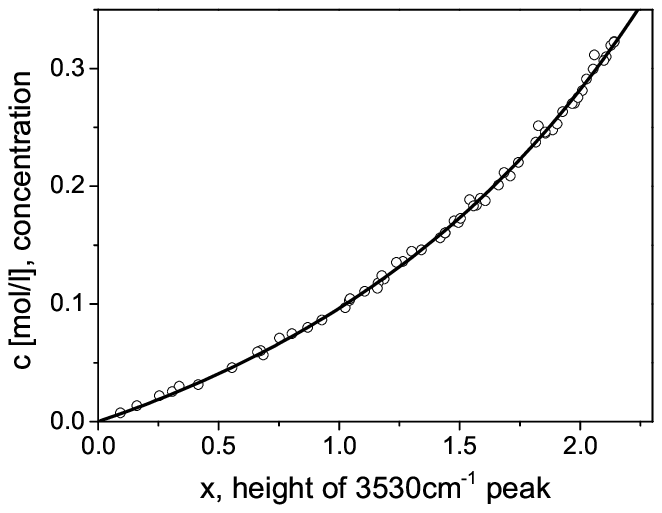}
\caption{Fit of the dependence of the total concentration on the
  height of the $3530\text{cm}^{-1}$ peak at $T=22^\circ\text{C}$ with model 3s.}
\label{model-3s-22C}
\end{figure}

\begin{figure}
\centering
\includegraphics{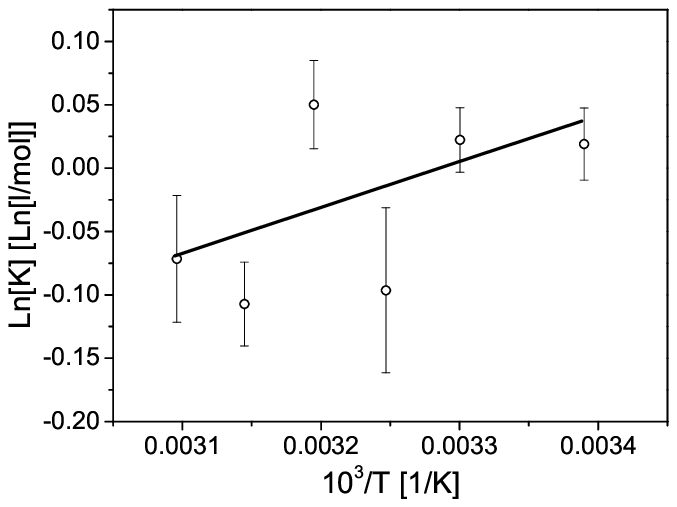}
\caption{Dependence of $\ln K$ on $1/T$ for model 3s.}
\label{model-3s-kfit}
\end{figure}

Let us first assume that the $3530\text{cm}^{-1}$ peak corresponds to the out-of-phase vibrations of free $\text{NH}_2$ groups in unimers and dimers. In this case, $Ax=c_1+4Kc_1^2$, where $x$ is the height of the peak and $A$ is some constant. The fitting equation in this case is
\begin{equation}
c=\frac{-1+\sqrt{1+16KAx}}{2K\left(-3+\sqrt{1+16KAx}\right)^2}.
\end{equation}
The results of the fitting procedure are shown in
Figures~\ref{model-3s-22C} and~\ref{model-3s-kfit}, and the values of
the model parameters and statistical measures are
$\text{AICc}=-499.3$, $\ln C=1.58\pm 0.8 \,\ln[\text{l/mol}]$,
$\epsilon=-0.7 \pm 0.5 \,\text{kcal/mol}$, and $r^2=0.37$.

The free energy density of hydrogen bonding in model 3, written in
terms of the volume fraction of hydrogen bonding molecules $\phi=cv$
and the dimensionless association constant $K'=K/v$, has the form
\begin{equation}
f_\text{HB}=m+2\phi\ln\frac{\left(\phi-m\right)}{\phi},
\end{equation}
where $m$ is a solution of the equation $m/4\left(\phi-m\right)^2=K'$.


\subsection{Model 4}
\begin{figure}
\centering
\includegraphics[width=0.7\textwidth]{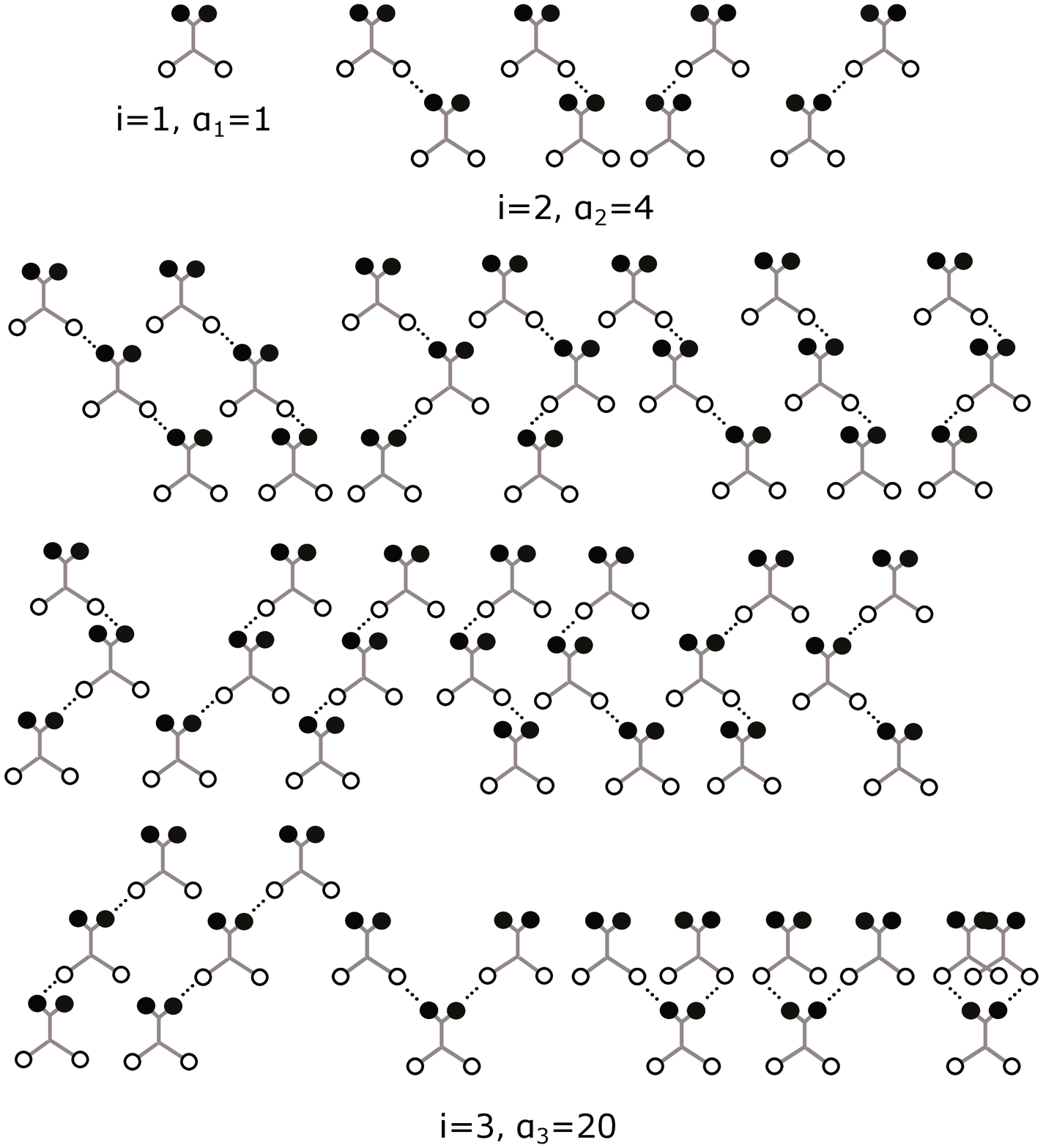}
\caption{Schematic representation of possible aggregates with size up to $i=3$ in model 4.}
\label{models4new-scheme}
\end{figure}

Model 4 is analogous to model 1 but with one bond allowed per $\text{NH}_2$ group and two bonds allowed per O in each acrylamide molecule (see Figure~\ref{models4new-scheme}). 
This means that the expression for the relation between $c$ and $c_1$ is
the same in model 4 as in model 1. However, the number of free
$\text{NH}_2$ groups in model 4 is different from model 1 and is equal
to the number of aggregates as there is only one free $\text{NH}_2$
per aggregate. We can then write that
\begin{equation}
n_f=\sum_{i=1}^{\infty}c_i=\sum_{i=1}^{\infty}\frac{2^{i-1}\left(2i\right)!}{i!\left(i+1\right)!}K^{i-1}c_1^{i}=\frac{1-4c_1K-\sqrt{1-8c_1K}}{8c_1K^2}.
\end{equation}
Substituting this expression into Equation~\ref{model1-conc} gives
\begin{equation}
c=\frac{n_f+2Kn_f^2}{1-2Kn_f}.
\end{equation}

\subsubsection{Model 4m}
First, we assume that the $3530\text{cm}^{-1}$ peak corresponds to
out-of-phase vibrations of free acrylamide molecules. The fitting equation is the same as in model 1m, so we do not repeat the fitting procedure here.

\subsubsection{Model 4g}
With the free groups assumption, we can the write fitting equation as 
\begin{equation}
c=\frac{Ax+2KA^2x^2}{1-2KAx}.
\end{equation}

\begin{figure}
\centering
\includegraphics{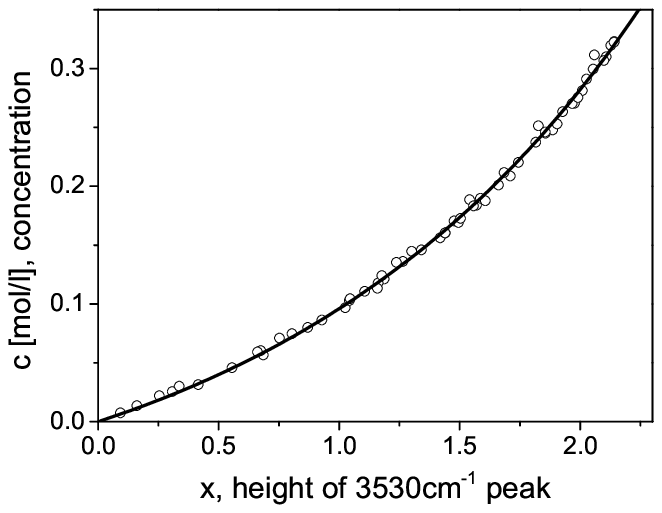}
\caption{Fit of the dependence of the total concentration on the
  height of the $3530\text{cm}^{-1}$ peak at $T=22^\circ\text{C}$ with model 4g.}
\label{model-4g-22C}
\end{figure}

\begin{figure}
\centering
\includegraphics{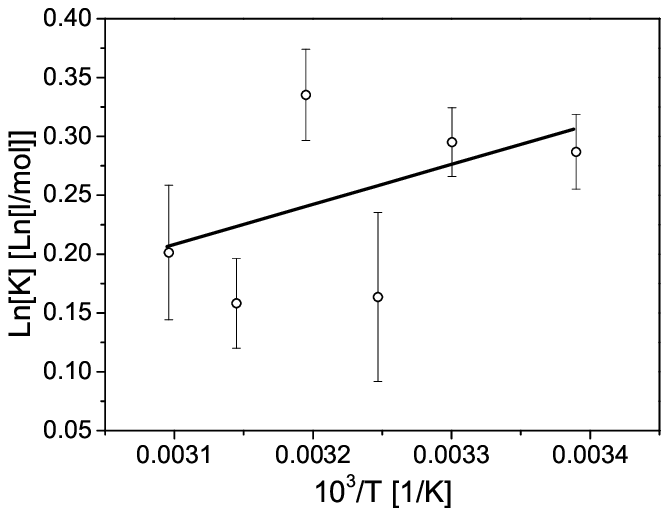}
\caption{Dependence of $\ln K$ on $1/T$ for model 4g.}
\label{model-4g-kfit}
\end{figure}

The results of the fit are shown in Figures~\ref{model-4g-22C}
and~\ref{model-4g-kfit}. The quality of the fit in
Figure~\ref{model-4g-22C} is characterized by the parameter
$\text{AICc}=-495.8$, and the estimates for the model parameters
resulting from the fit in Figure~\ref{model-4g-kfit} are $\ln
C=-0.8\pm 0.9 \,\ln[\text{l/mol}]$ and $\epsilon=-0.7 \pm 0.5
\,\text{kcal/mol}$, with $r^2=0.28$.  

\subsubsection{Model 4s}
In this case, we assume that the $3530\text{cm}^{-1}$ peak corresponds to
out-of-phase vibrations of free $\text{NH}_2$ groups in unimers and
dimers. The fitting equation and all fitting results are the same as for model 1s.



\subsection{Model 5}
\begin{figure}
\centering
\includegraphics[width=0.7\textwidth]{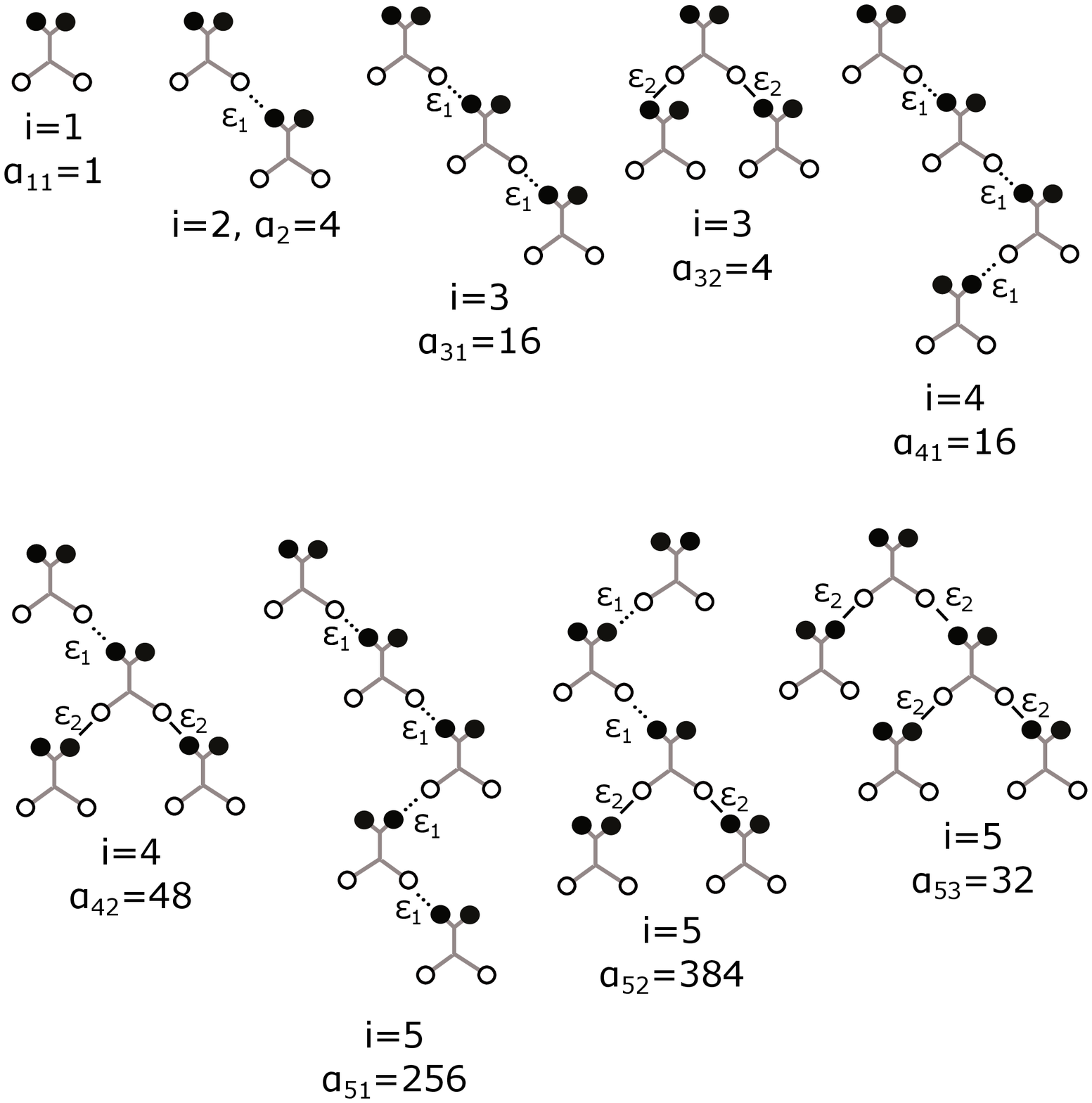}
\caption{Schematic representation of aggregates with size up to $i=5$
  in model 5. In contrast to the figures for one-parameter models, only configurations
  of aggregates with different energies are shown. The quantity
  $a_{ij}$ gives the number of different configurations for an aggregate
  of $i$ molecules that contains $2\left(j-1\right)$ $\epsilon_2$ bonds.}
\label{model5-schemeS}
\end{figure}

Model 5 is the first model with "cooperativity" we consider. In this
model, we allow one bond per oxygen, two bonds per $\text{NH}_2$
group, no cycles and two association constants (corresponding to bond
energies $\epsilon_1$ and $\epsilon_2$) that depend on the
bonding state of the $\text{NH}_2$ group in the donor molecule (see Figure~\ref{model5-schemeS}).

Let us denote the number of molecules with donors involved in
$\epsilon_1$ bonds as $N_1$ (or, in other words, the number of donor
molecules with one bonded hydrogen) and the number of molecules involved in $\epsilon_2$ bonds as $N_2$ (in other words, the number of donor molecules with two bonded hydrogens). Correspondingly, the number of $\epsilon_1$ bonds is $M_1=N_1$ and number of $\epsilon_2$ bonds is $M_2=2N_2$.

The number of ways to form $M_1$ bonds with energy $\epsilon_1$ and
$2N_2$ bonds with energy $\epsilon_2$ can be written as 
\begin{equation}
\Xi=\frac{N!2^{N_1+2N_2}}{\left(N-N_1-2N_2\right)!}\frac{N!2^{N_1}}{\left(N-N_1-N_2\right)!N_1!N_2!},
\end{equation}
where the first factor is the number of ways to choose an acceptor for $M=N_1+2N_2$ bonds, and the second factor is the number of ways to choose $N_1$ and $N_2$ donor groups out of $N$ molecules, taking into account the fact that in molecules with only one bonded hydrogen, this hydrogen can be chosen in two ways.

The free energy of hydrogen bonding in model 5 is
\begin{equation}
F_\text{HB}=\epsilon_1M_1+\epsilon_2M_2-kT\ln\left(p_1^{M_1}p_2^{M_2}\Xi\right),
\end{equation}
and minimizing this with respect to $M_1$ and $M_2$ gives
\begin{equation}
\frac{N_1}{4\left(N-N_1-2N_2\right)\left(N-N_1-N_2\right)}=\frac{K_1}{V}
\label{model5-eq1}
\end{equation}
and
\begin{equation}
\frac{N_2}{4\left(N-N_1-N_2\right)\left(N-N_1-2N_2\right)^2}=\frac{K_2^2}{V^2}.
\label{model5-eq2}
\end{equation}
It is interesting to note that the following equality exists:
\begin{equation}
\frac{\left(2N\right)!}{\left(2N-M\right)!M!}=\sum_{N_2=0}^{M/2}\frac{N!2^{M-2N_2}}{\left(N-M+N_2\right)!\left(M-2N_2\right)!N_2!},
\label{main-equality}
\end{equation}
which shows that model 5 reduces to model 1 when
$\epsilon_1=\epsilon_2$ -- a property that we make use of in our
calculations on model 1g above.

Now, we look for a relation between the total concentration and the concentration of unimers, and assume that concentration of aggregates of size $i$ with $2\left(j-1\right)$ $\epsilon_2$-bonds has the form
\begin{equation}
c_{ij}=\alpha_{ij}K_2^{2\left(j-1\right)}K_1^{i-1-2\left(j-1\right)}c_1^i, 1\leq j\leq \left(i+1\right)/2.
\end{equation}
Substituting this expression in Equations~\ref{model5-eq1}
and~\ref{model5-eq2} gives
\begin{equation}
\alpha_{ij}=\frac{2^{2i-2j}}{(j-1)!j!}\frac{\left(i-1\right)!}{\left(i-2j+1\right)!},
\label{model5-alpha}
\end{equation}
and the dependence of the total solution concentration on the
concentration of unimers can then be calculated to be 
\begin{equation}
c=\sum_{i=1}^{\infty}\sum_{j=1}^{\left(i+1\right)/2}i\alpha_{ij}K_2^{2\left(j-1\right)}K_1^{i-1-2(j-1)}c_1^i=\frac{1-\sqrt{1-\frac{16K_2^2c_1^2}{\left(1-4K_1c_1\right)^2}}}{8K_2^2c_1\sqrt{1-\frac{16K_2^2c_1^2}{\left(1-4K_1c_1\right)^2}}}.
\end{equation}
Alternatively, the coefficients $\alpha_{ij}$ can be found from the
generating function for the family of trees shown in
Figure~\ref{model5-scheme}. If we denote $c_1K_1$ as $z_1$ and
$c_1K_2$ as $z_2$, we can write down the equation for the generating function as
\begin{equation}
G=1+4Gz_1+4z_2^2G^2,
\end{equation}
which can be solved to find
\begin{equation}
G=\frac{1-4z_1\pm\sqrt{\left(1-4z_1\right)^2-16z_2^2}}{8z_2^2}
\end{equation}
where the required expression is that with the negative root. It is
straightforward to verify that expansion of this expression in powers
of $z_1$ and $z_2$ will yield the values of $\alpha_{ij}$ given by
Equation \ref{model5-alpha}. We also can see that if we put $z_1=z_2$ we
will recover the generating function for model 1, as would be expected
from the fact that the two models are equivalent when $\epsilon_1=\epsilon_2$.

The concentration of free groups is given by $n_f=c-n_1-n_2$ and eliminating $n_1$ and $n_2$ from Equations~\ref{model5-eq1} and~\ref{model5-eq2} gives
\begin{equation}
c=\frac{1-4K_1n_f+16K_2^2n_f^2-\left(1-4K_1n_f\right)\sqrt{1+16K_2^2n_f^2}}{8K_2^2n_f}.
\end{equation}

\subsubsection{Model 5m}
Let us assume first that the $3530\text{cm}^{-1}$ peak corresponds to
out-of-phase vibrations of the $\text{NH}_2$ group in free
molecules. In this case, the fitting equation is 
\begin{equation}
c=\frac{1-\sqrt{1-\frac{16A^2K_2^2x^2}{\left(1-4AK_1x\right)^2}}}{8AK_2^2x\sqrt{1-\frac{16A^2K_2^2x^2}{\left(1-4AK_1x\right)^2}}}.
\end{equation}

\begin{figure}
\centering
\includegraphics{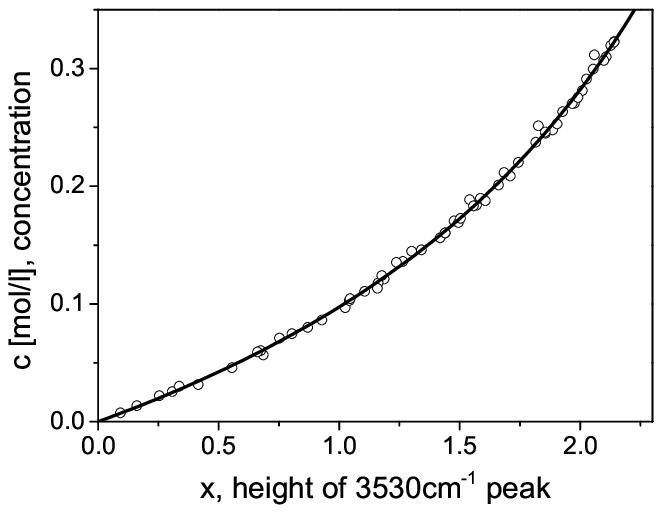}
\caption{Fit of the dependence of the total concentration on the
  height of the $3530\text{cm}^{-1}$ peak at $T=22^\circ\text{C}$ with model 5m.}
\label{model-5m-22C}
\end{figure}

\begin{figure}
\centering
\includegraphics{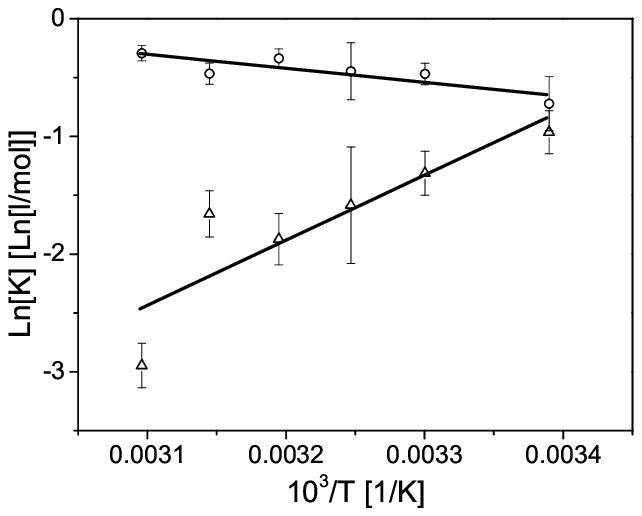}
\caption{Dependence of $\ln K_1$ (triangles) and $\ln K_2$ (circles) on $1/T$ for model 5m.}
\label{model-5m-kfit}
\end{figure}

The results of the fit are shown in Figures~\ref{model-5m-22C}
and~\ref{model-5m-kfit}. The value of the AICc parameter for the
nonlinear fit in Figure~\ref{model-5m-22C} is $-501.7$, and the
estimates of the model parameters given by the fits in
Figure~\ref{model-5m-kfit} are $\ln C_1=-19.6\pm 5 \,\ln[\text{l/mol}]$,
$\epsilon_1=-11 \pm 3 \,\text{kcal/mol}$, $\ln C_2=3.4\pm 1
\,\ln[\text{l/mol}]$, and $\epsilon_2=2.4 \pm 0.7 \,\text{kcal/mol}$. In
all our two-parameter models, we have two coefficients of
determination, which in this case are given by
$r_1^2=0.76$ and $r_2^2=0.73$. 

\subsubsection{Model 5g}
Now we assume that the $3530\text{cm}^{-1}$ peak corresponds to
out-of-phase vibrations of free groups $\text{NH}_2$. In this case the
fitting equation is 
\begin{equation}
c=\frac{1-4AK_1x+16A^2K_2^2x^2-\left(1-4AK_1x\right)\sqrt{1+16A^2K_2^2x^2}}{8AK_2^2x}.
\end{equation}

In this case, we were unable to obtain any results, as the fitting procedure did not converge. 

\subsubsection{Model 5s}
We also check the possibility that the $3530\text{cm}^{-1}$ peak does
not correspond to a single species (such as all free molecules or all free groups), but instead corresponds to absorption by free groups in some subset of aggregates. Here we check the subset composed of unimers and dimers, so $Ax=c_1+4K_1c_1^2$.

\begin{figure}
\centering
\includegraphics{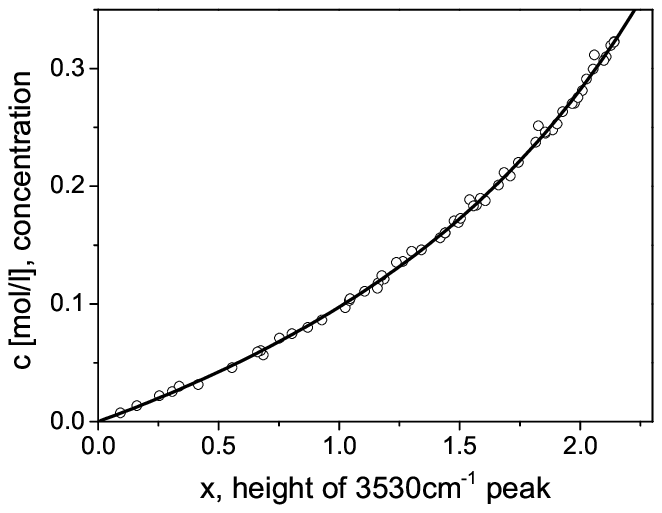}
\caption{Fit of the dependence of the total concentration on the
  height of the $3530\text{cm}^{-1}$ peak at $T=22^\circ\text{C}$ with model 5s.}
\label{model-5s-22C}
\end{figure}

\begin{figure}
\centering
\includegraphics{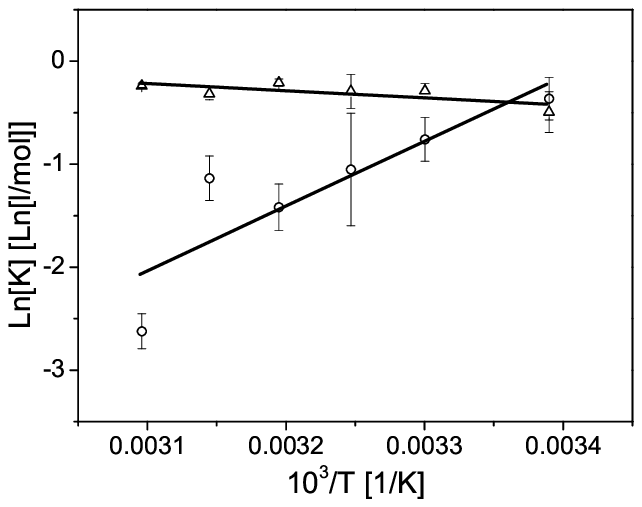}
\caption{Dependence of $\ln K_1$ (triangles) and $\ln K_2$ (circles) on $1/T$ for model 5s.}
\label{model-5s-kfit}
\end{figure}

The results of the fit are shown in Figures~\ref{model-5s-22C}
and~\ref{model-5s-kfit}. The value of AICc for the fit in
Figure~\ref{model-5s-22C} is $-501.7$, and the estimates of the model
parameters given by the fits in~\ref{model-5s-kfit} are $\ln
C_1=-21.5\pm 6 \,\ln[\text{l/mol}]$, $\epsilon_1=-12.5 \pm 3.5
\,\text{kcal/mol}$, $\ln C_2=1.95\pm 1 \,\ln[\text{l/mol}]$, and
$\epsilon_2=1.3 \pm 0.6 \,\text{kcal/mol}$, with $r_1^2=0.76$ and $r_2^2=0.56$. 

\subsection{Model 6}
\begin{figure}
\centering
\includegraphics[width=0.7\textwidth]{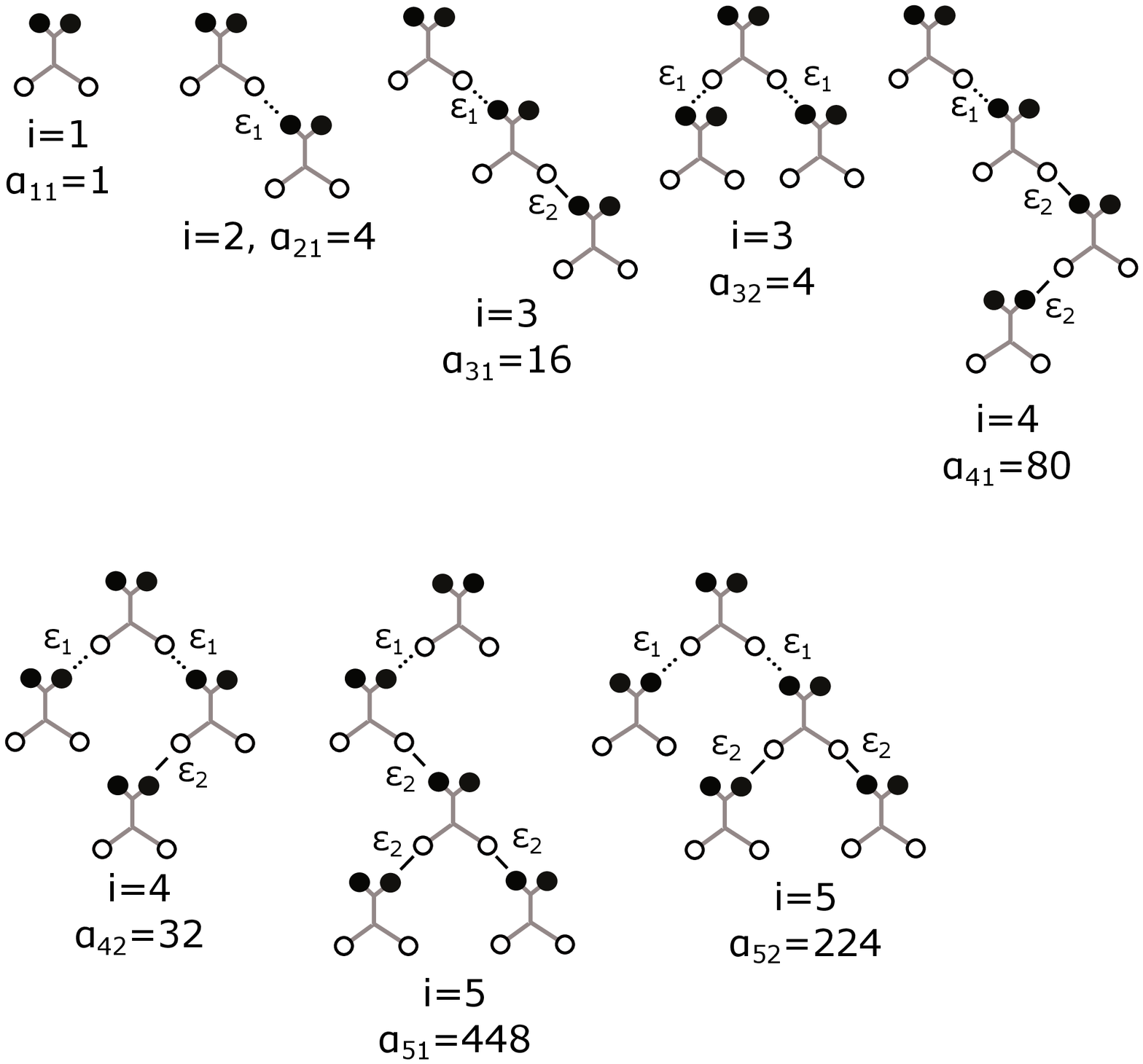}
\caption{Schematic representation of aggregates with size up to $i=5$
  in model 6. Only aggregates with different total bond energies are shown.}
\label{model6-scheme}
\end{figure}

In model 6, we allow one bond per oxygen, two bonds per $\text{NH}_2$
group, no cycles and two association constants. However, in contrast
to model 5, the association constant is now determined by the bonding
state of the acceptor in the donor molecule (see Figure~\ref{model6-scheme}).
Let the bond energy be denoted by $\epsilon_1$ in the case when the
oxygen in the donor amide group is free and by $\epsilon_2$
otherwise. Then, the number of bonds with energy $\epsilon_1$ is $M_1$, the number of bonds with energy $\epsilon_2$ is $M_2$, the number of molecules with donors involved in $\epsilon_1$ bonds is $N_1$, and the number of molecules involved in $\epsilon_2$ bonds is $N_2$.
The number of ways to form bonds can then be written as 
\begin{equation}
\Xi=\frac{N!2^{M_1+M_2}}{\left(N-M_1-M_2\right)!}\frac{\left(2N-2M_1-2M_2\right)!}{\left(2N-3M_1-2M_2\right)!}\frac{\left(2M_1+2M_2\right)!}{\left(2M_1+M_2\right)!}\frac{1}{M_1!M_2!},
\end{equation}
where the first factor in $\Xi$ is the number of ways to choose an
acceptor for $M_1+M_2$ bonds. This uses the assumption that there is
only one bond per oxygen. The second factor is the number of ways to
choose a donor for $\epsilon_1$ bonds, with $2N-2M_1-2M_2$ giving the
number of hydrogens in molecules with free acceptor groups. The third
factor is the number of ways to choose a donor for $\epsilon_2$
bonds. The hydrogens for these bonds should be chosen from molecules
with bonded acceptors. The number of such molecules is $M_1+M_2$ and
they contain $2M_1+2M_2$ hydrogens. As usual, the last
term accounts for the indistinguishability of the bonds. 

Minimization of the free energy gives
\begin{equation}
\frac{M_1\left(2M_1+M_2\right)^2\left(N-M_1-M_2\right)}{2\left(M_1+M_2\right)^2\left(2N-3M_1-2M_2\right)^3}=\frac{K_1}{V}
\label{model7-eq1}
\end{equation}
\begin{equation}
\frac{M_2\left(2M_1+M_2\right)\left(N-M_1-M_2\right)}{2\left(M_1+M_2\right)^2\left(2N-3M_1-2M_2\right)^2}=\frac{K_2}{V}.
\label{model7-eq2}
\end{equation}
We can notice that there are two types of aggregates: those with one
$\epsilon_1$ bond and those with two $\epsilon_1$ bonds. Then, the
concentrations of aggregates with size $i$ and either one or two
$\epsilon_1$ bonds can be written as
\begin{equation}
c_{i1}=\alpha_{i1}K_1K_2^{i-2}c_1^i, i\geq2
\end{equation}
\begin{equation}
c_{i2}=\alpha_{i2}K_1^2K_2^{i-3}c_1^i, i\geq3,
\end{equation}
and the total concentration of acrylamide and concentrations of each
type of bond as
\begin{equation}
c=c_1+\sum_{i=2}^{\infty}ic_{i1}+\sum_{i=3}^{\infty}ic_{i2}
\label{eq73}
\end{equation}
\begin{equation}
m_1=\sum_{i=2}^{\infty}c_{i1}+\sum_{i=3}^{\infty}2c_{i2}
\label{eqm1-m6}
\end{equation}
\begin{equation}
m_2=\sum_{i=2}^{\infty}\left(i-2\right)c_{i1}+\sum_{i=3}^{\infty}\left(i-3\right)c_{i2}.
\label{eqm2-m6}
\end{equation}
Substituting these expressions into Equations~\ref{model7-eq1}
and~\ref{model7-eq2} leads to the following expressions for $\alpha_{i1,2}$:
\begin{equation}
\alpha_{i1}=2^i\frac{\left(2i-2\right)!}{\left(i-1\right)!i!},i\ge 2
\end{equation}
\begin{equation}
\alpha_{i2}=2^{i+1}\frac{\left(2i-3\right)!}{\left(i-3\right)!\left(i+1\right)!},i \ge 3.
\end{equation}
For the dependence of the total concentration on the concentration of unimers we have
\begin{equation}
\begin{split}
c=&\frac{16c_1^2K_1K_2^3-16c_1^2K_2^2K_1^2+K_1^2-4c_1K_1^2K_2}{8c_1K_2^4\sqrt{1-8c_1K_2}}\\
+&\frac{8c_1^2K_2^4-16c_1^2K_1K_2^3-K_1^2+8c_1^2K_1^2K_2^2}{8c_1K_2^4}.
\end{split}
\label{model6-c}
\end{equation}
Another way to determine the coefficients $\alpha_{i1}$ and $\alpha_{i2}$
is to find an expression for the generating function, as was done in
for models 1 and 5. In the current case, we have
\begin{equation}
G\left(z_1,z_2\right)=1+4z_1G_1\left(z_2\right)+4z_1^2G_1^2\left(z_2\right),
\end{equation}
where
\begin{equation}
G_1\left(z_2\right)=\frac{1-4z_2-\sqrt{1-8z_2}}{8z_2^2},
\end{equation}
the generating function for model 1.

\begin{figure}
\centering
\includegraphics[width=0.7\textwidth]{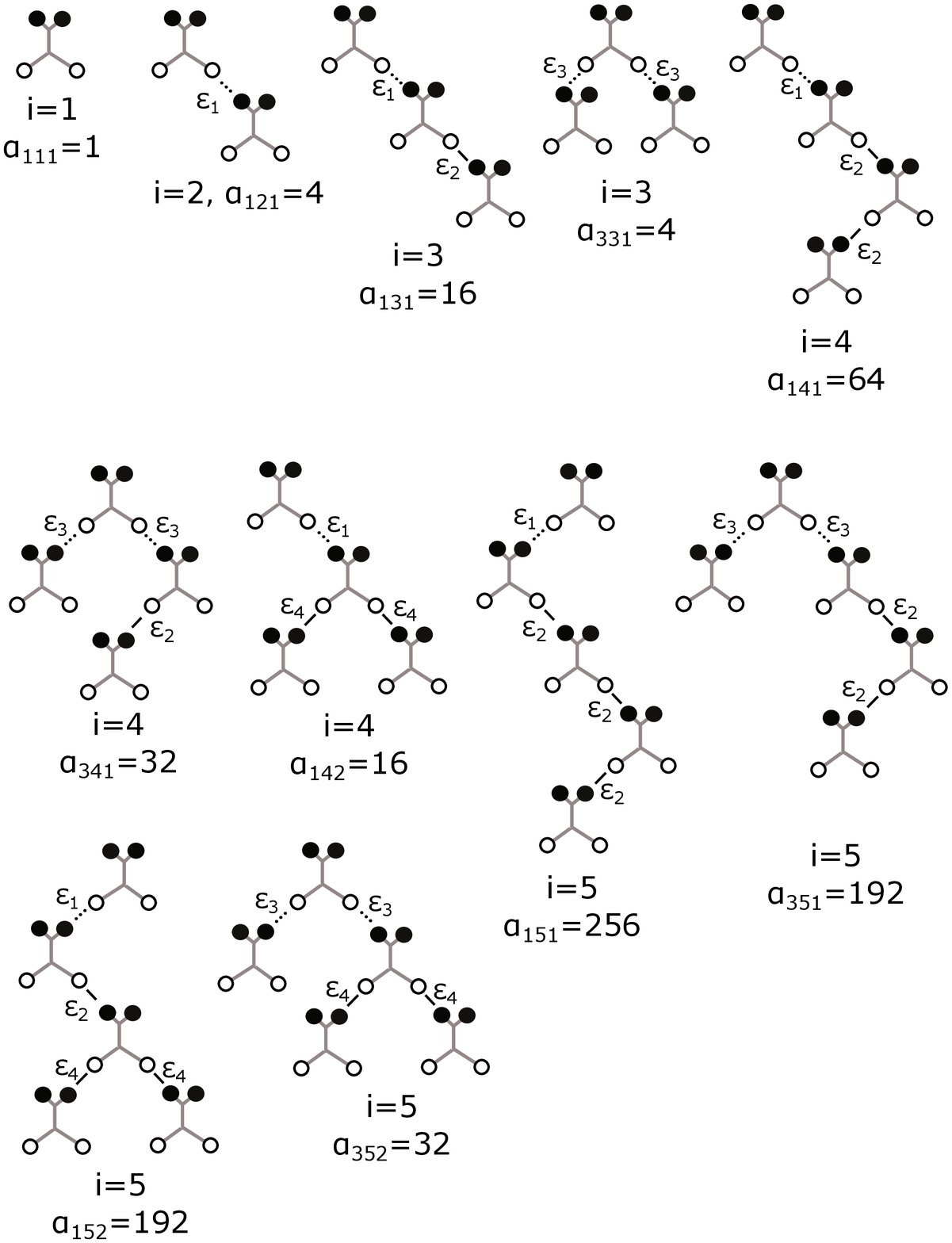}
\caption{Schematic representation of aggregates with size up to $i=5$
  in the modification of model 6 with four association constants. Only
  aggregates with different total bond energies are shown.}
\label{model6-scheme-1}
\end{figure}

Now, let us turn to the calculation of the number of free groups. In
order to do this, it is necessary to distinguish molecules with one
hydrogen bond per $\text{NH}_2$ group and two hydrogen bonds per
$\text{NH}_2$ group. We denote the number of molecules with one
bonded hydrogen and free oxygen by $N_1$
(Figure~\ref{model6-scheme-1}) and the number of molecules with two
bonded hydrogens and free oxygen by $N_3$, so that
$M_1=N_1+2N_3$. Similarly, we denote the number of molecules with one
bonded hydrogen and bonded oxygen by $N_2$, and the number of
molecules with two bonded hydrogens and bonded oxygen, by $N_4$, so
that $M_2=N_2+2N_4$. Then, the number of ways to form bonds is
\begin{equation}
\Xi=\frac{N!2^{M_1+M_2}}{\left(N-M_1-M_2\right)!}\frac{\left(N-M_1-M_2\right)!2^{N_1}}{\left(N-M_1-M_2-N_1-N_3\right)!N_1!N_3!}\frac{\left(M_1+M_2\right)!2^{N_2}}{\left(M_1+M_2-N_2-N_4\right)!N_2!N_4!}.
\end{equation}
After simplification, this becomes
\begin{equation}
\Xi=\frac{N!\left(N_1+N_2+2N_3+2N_4\right)!4^{N_1+N_2+N_3+N_4}}{\left(N-2N_1-3N_3-N_2-2N_4\right)!\left(N_1+2N_3+N_4\right)!N_1!N_2!N_3!N_4!}.
\end{equation}
Also, for simplicity, we assume first that there are four association constants and correspondingly four different bond energies, so we can write for the free energy of hydrogen bonding
\begin{equation}
F_\text{HB}=\epsilon_1N_1+\epsilon_2N_2+2\epsilon_3N_3+2\epsilon_4N_4-kT\ln\left(p_1^{N_1}p_2^{N_2}p_3^{2N_3}p_4^{2N_4}\Xi\right).
\end{equation}

Minimization of this expression gives
\begin{equation}
\frac{N_1\left(N_1+2N_3+N_4\right)}{4\left(N_1+N_2+2N_3+2N_4\right)\left(N-2N_1-3N_3-N_2-2N_4\right)^2}=\frac{K_1}{V}
\label{model7b-eq1}
\end{equation}
\begin{equation}
\frac{N_2}{4\left(N_1+N_2+2N_3+2N_4\right)\left(N-2N_1-3N_3-N_2-2N_4\right)}=\frac{K_2}{V}
\label{model7b-eq2}
\end{equation}
\begin{equation}
\frac{N_3\left(N_1+2N_3+N_4\right)^2}{4\left(N_1+N_2+2N_3+2N_4\right)^2\left(N-2N_1-3N_3-N_2-2N_4\right)^3}=\frac{K_3^2}{V^2}
\label{model7b-eq3}
\end{equation}
\begin{equation}
\frac{N_4\left(N_1+2N_3+N_4\right)}{4\left(N_1+N_2+2N_3+2N_4\right)^2\left(N-2N_1-3N_3-N_2-2N_4\right)^2}=\frac{K_4^2}{V^2},
\label{model7b-eq4}
\end{equation}
and the number of free groups is given by $N_f=N-N_1-N_2-N_3-N_4$.

We notice that, since only one bond per oxygen is allowed, all
aggregates have either one $\epsilon_1$ bond or two $\epsilon_3$
bonds, which lie at the "root" of each aggregate. Furthermore, we note
that the distribution of $\epsilon_2$ and $\epsilon_4$ bonds is very
similar to the original model 6. Then, we can label different aggregates by the set
of three numbers $\left\lbrace 1ij\right\rbrace$ or $\left\lbrace
  3ij\right\rbrace$ where the first letter denotes the type of "root"
(type $1$ bonds or type $3$ bonds), $i$ is the number of molecules in
the aggregate and $j$ the number of type $4$ bonds in aggregate.
The concentrations of aggregates can then be written as
\begin{equation}
c_{1ij}=\alpha_{1ij}K_1K_2^{i-2j}K_4^{2j-2}c_1^i
\end{equation}
\begin{equation}
c_{3ij}=\alpha_{3ij}K_3^2K_2^{i-1-2j}K_4^{2j-2}c_1^i
\end{equation}
and the concentrations of molecules in different bonding states can be
calculated to be
\begin{equation}
c=c_1+\sum_{i=2}^{\infty}\sum_{j=1}^{i/2}ic_{1ij}+\sum_{i=3}^{\infty}\sum_{j=1}^{\left(i-1\right)/2}ic_{3ij}
\end{equation}
\begin{equation}
c_1=\sum_{i=2}^{\infty}\sum_{j=1}^{i/2}c_{1ij}=\sum_{j=1}^{\infty}\sum_{i=2j}^{\infty}c_{1ij}
\end{equation}
\begin{equation}
c_3=\sum_{i=3}^{\infty}\sum_{j=1}^{\left(i-1\right)/2}c_{3ij}=\sum_{j=1}^{\infty}\sum_{i=2j+1}^{\infty}c_{3ij}
\end{equation}
\begin{equation}
c_2=\sum_{j=1}^{\infty}\sum_{i=2j}^{\infty}\left(i-2j\right)c_{1ij}+\sum_{j=1}^{\infty}\sum_{i=2j+1}^{\infty}\left(i-1-2j\right)c_{3nj}
\end{equation}
\begin{equation}
c_4=\sum_{j=1}^{\infty}\sum_{i=2j}^{\infty}\left(j-1\right)c_{1ij}+\sum_{j=1}^{\infty}\sum_{i =2j+1}^{\infty}\left(j-1\right)C_{3ij}.
\end{equation}
Additionally, we can write that 
\begin{equation}
n_f=c_1+\sum_{j=1}^{\infty}\sum_{i=2j}^{\infty}jc_{1ij}+\sum_{j=1}^{\infty}\sum_{i=2j+1}^{\infty}\left(j+1\right)c_{3ij}.
\end{equation}
Substituting these expressions into
Equations~\ref{model7b-eq1},~\ref{model7b-eq2},~\ref{model7b-eq3},~\ref{model7b-eq4} gives
\begin{equation}
\alpha_{1ij}=2^{2i-2j}\frac{1}{j!\left(j-1\right)!}\frac{\left(i-2\right)!}{\left(i-2j\right)!}, i\geq 2j
\end{equation}
\begin{equation}
\alpha_{3ij}=2^{2i-2j-1}\frac{1}{\left(j+1\right)!\left(j-1\right)!}\frac{\left(i-2\right)!}{\left(i-2j-1\right)!}, i\geq 2j+1.
\end{equation}

We can now evaluate the sums and calculate the dependence of $n_f$ on $c_1$. 

Since we assumed that equilibrium is described by two association
constants in model 6, we put $K_3=K_1$ and $K_4=K_2$ and find
\begin{equation}
n_f=\frac{c_1\left(4c_1K_1K_2^2+K_2^2\sqrt{1-8c_1K_2}+K_1^2\left(1-4c_1K_2-\sqrt{1-8c_1K_2}\right)\right)}{K_2^2\sqrt{1-8c_1K_2}}.
\end{equation}
As we know the dependence of both $c$ and $n_f$ on $c_1$, we have
parametrically defined a function $c\left(n_f\right)$, which we can use to fit experimental data in case of the free groups assumption.  

\subsubsection{Model 6m}
Here, it is assumed that the $3530\text{cm}^{-1}$ peak corresponds to
the out-of-phase vibrations of the $\text{NH}_2$ group in free
molecules. In this case, we put $c_1=Ax$ and substitute it into Equation~\ref{model6-c}. 

\begin{figure}
\centering
\includegraphics{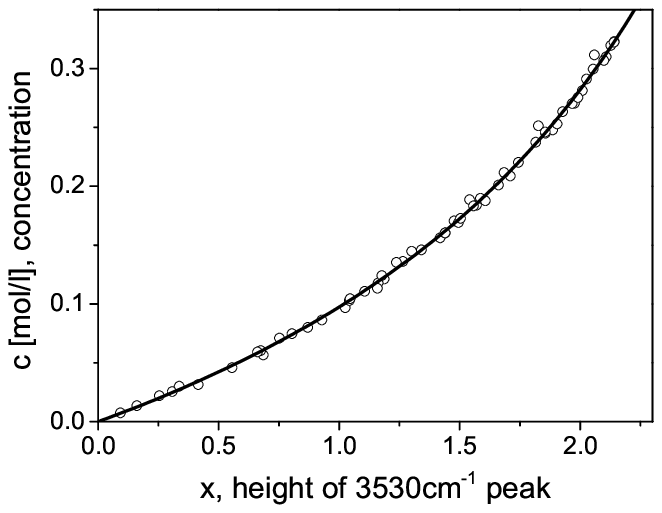}
\caption{Fit of the dependence of the total concentration on the
  height of the $3530\text{cm}^{-1}$ peak at $T=22^\circ\text{C}$ with model 6m.}
\label{model-6m-22C}
\end{figure}

\begin{figure}
\centering
\includegraphics{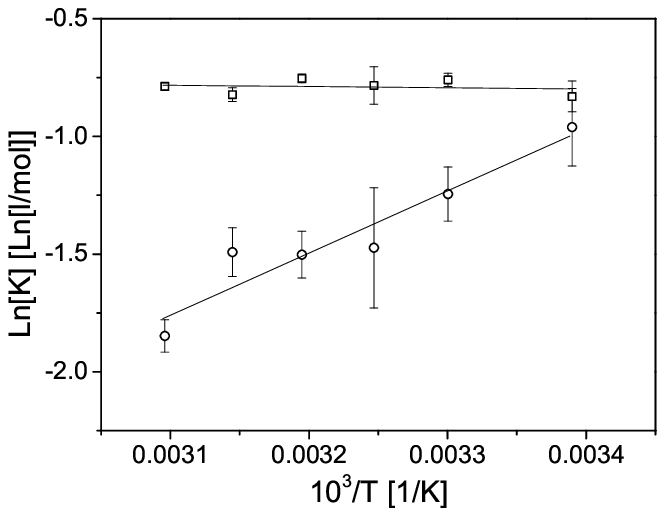}
\caption{Dependence of $\ln K_1$ (circles) and $\ln K_2$ (squares) on $1/T$ for model 6m.}
\label{model-6m-k}
\end{figure}

The fitting results are shown in Figures~\ref{model-6m-22C}
and~\ref{model-6m-k}. The value of the information criterion for
the fit shown in Figure~\ref{model-6m-22C} is $\text{AICc}=-501.7$,
and the estimates of the model parameters given by the fits in
Figure~\ref{model-6m-k} are $\ln C_1=-9.9\pm 1 \,\ln[\text{l/mol}]$,
$\epsilon_1=-5.2 \pm 0.8 \,\text{kcal/mol}$, $\ln
C_2=-0.62\pm 0.5 \,\ln[\text{l/mol}]$, and $\epsilon_2=0.1 \pm 0.2
\,\text{kcal/mol}$, with $r_1^2=0.91$ and $r_2^2=0.03$. 

\subsubsection{Model 6g}
\begin{figure}
\centering
\includegraphics{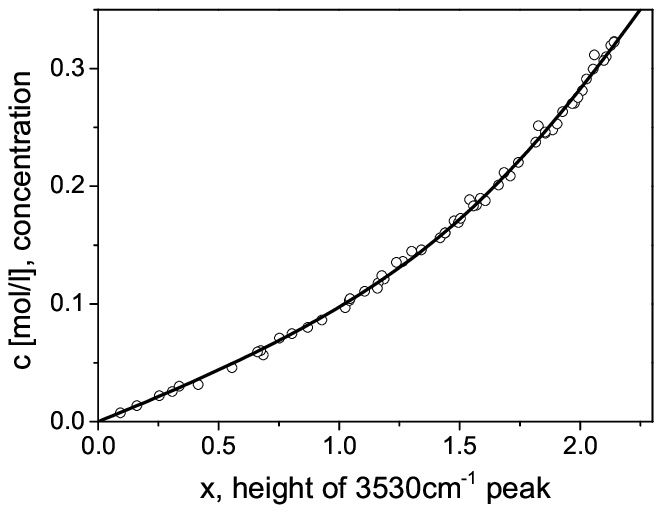}
\caption{Fit of the dependence of the total concentration on the
  height of the $3530\text{cm}^{-1}$ peak at $T=22^\circ\text{C}$ with model 6g.}
\label{model-6g-22C}
\end{figure}

\begin{figure}
\centering
\includegraphics{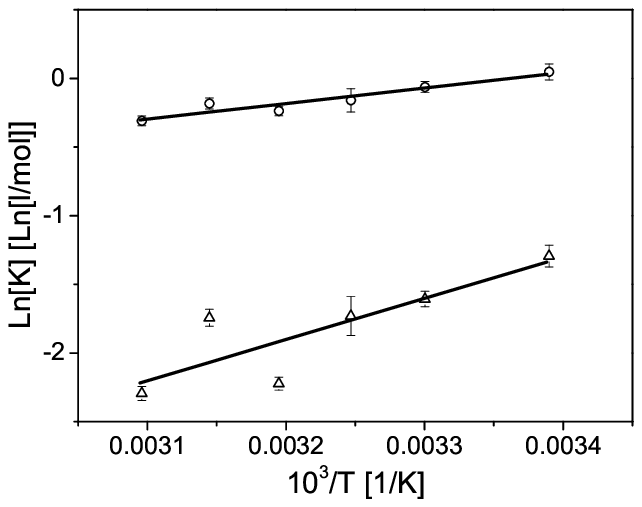}
\caption{Dependence of $\ln K_1$ (triangles) and $\ln K_2$ (circles) on $1/T$ for model 6g.}
\label{model-6g-k}
\end{figure}

In this section, it is assumed that the $3530\text{cm}^{-1}$ peak
corresponds to absorption by free groups, so that $n_f=Ax$. The
results of the fitting procedure are shown in
Figures~\ref{model-6g-22C} and~\ref{model-6g-k}. For the fit in
Figure~\ref{model-6g-22C}, we have that $\text{AICc}=-493.8$. The
estimates of the model parameters for the fits in
Figure~\ref{model-6g-k} are $\ln C_1=-11.5\pm 5 \,\ln[\text{l/mol}]$,
$\epsilon_1=-5.9 \pm 3 \,\text{kcal/mol}$, $\ln C_2=3.5\pm 0.6
\,\ln[\text{l/mol}]$, and $\epsilon_2=-2.2 \pm 0.3 \,\text{kcal/mol}$, with $r_1^2=0.71$
and $r_2^2=0.90$. 

\subsubsection{Model 6s}
\begin{figure}
\centering
\includegraphics{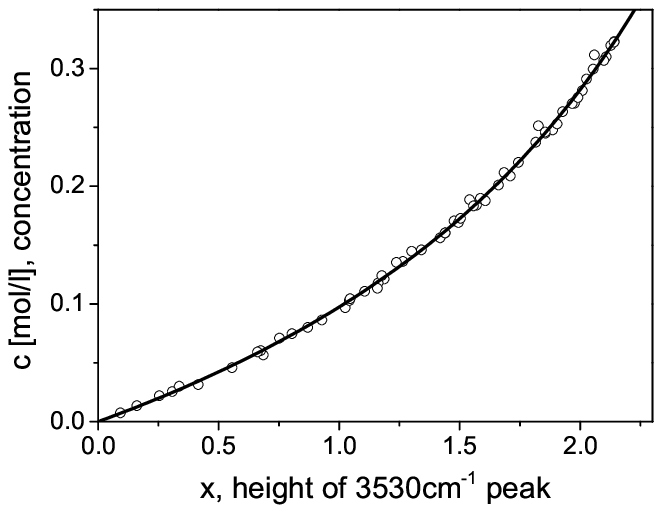}
\caption{Fit of the dependence of the total concentration on the
  height of the $3530\text{cm}^{-1}$ peak at $T=22^\circ\text{C}$ with model 6s.}
\label{model-6s-22C}
\end{figure}

\begin{figure}
\centering
\includegraphics{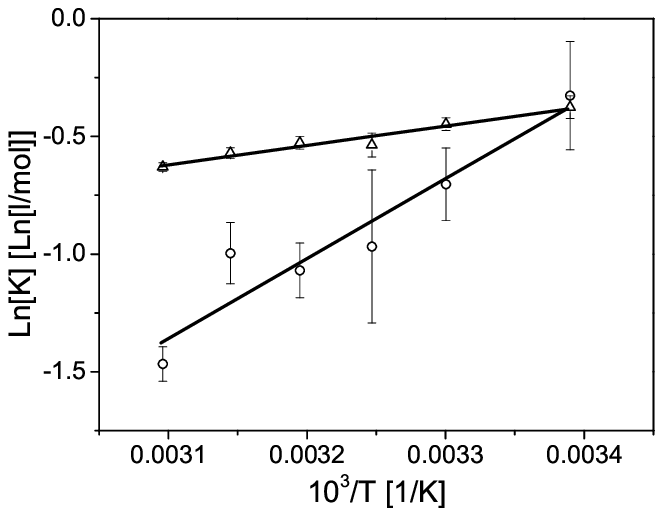}
\caption{Dependence of $\ln K_1$ (circles) and $\ln K_2$ (squares) on $1/T$ for model 6s.}
\label{model-6s-k}
\end{figure}

For the assumption  $Ax=c_1+4K_1c_1^2+4K_1^2c_1^3$ ($\text{NH}_2$
groups in unimers, dimers and  trimers without $\epsilon_2$-bonds), we
have the fitting results shown in Figures~\ref{model-6s-22C} and~\ref{model-6s-k}. 
The quality of fit in Figure~\ref{model-6s-22C} can be characterized by
$\text{AICc}=-501.7$. The estimates of parameters for the fits in
Figure~\ref{model-6s-k} are $\ln C_1=-11.89\pm 2 \,\ln[\text{l/mol}]$,
$\epsilon_1=-6.7 \pm 1.1 \,\text{kcal/mol}$, $\ln C_2=-3.18\pm 0.3
\,\ln[\text{l/mol}]$, and $\epsilon_2=-1.6 \pm 0.2 \,\text{kcal/mol}$,
with $r_1^2=0.91$ and $r_2^2=0.95$. 

The free-energy density of hydrogen bonding in model 6 in terms of the
volume fraction of hydrogen bonding molecules $\phi=cv$ and dimensionless association constants $K_1'=K_1/v$, $K_2'=K_2/v$ has the form
\begin{equation}
f_\text{HB}=m_1+m_2+\phi\ln\frac{\left(2\phi-3m_1-2m_2\right)^2}{4\phi\left(\phi-m_1-m_2\right)},
\end{equation}
where $m_1$ and $m_2$ can be calculated by taking the
sums in Equations~\ref{eqm1-m6} and~\ref{eqm2-m6} to be functions of the
association constants and $c_1$. Then, we can consider $c_1$ as a
parameter and now have a parametrically defined function $f_\text{HB}\left(\phi\right)$.

\subsection{Model 7}
\begin{figure}
\centering
\includegraphics[width=0.7\textwidth]{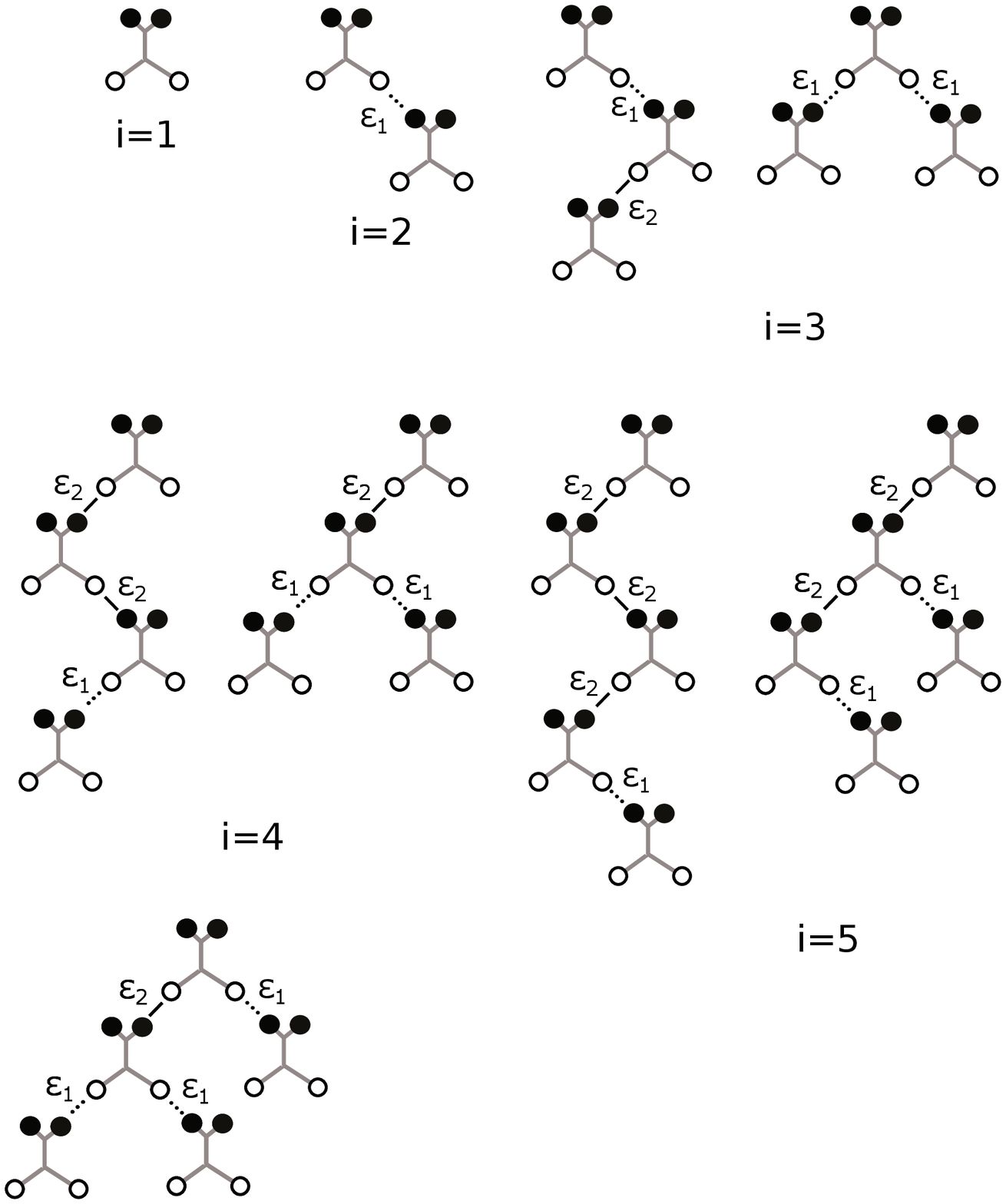}
\caption{Schematic representation of aggregates with size up to $i=5$ in model 7. Only configurations corresponding to different energies are shown.}
\label{model7-scheme}
\end{figure}

In model 7, we again allow one bond per oxygen, two bonds per
$\text{NH}_2$ group, no cyclic dimers and two association
constants. However, in contrast to models 5 and 6, the association
constant is now determined by the bonding state of the $\text{NH}_2$
group in the acceptor molecule (see Figure~\ref{model7-scheme}).

\begin{figure}
\centering
\includegraphics[width=0.7\textwidth]{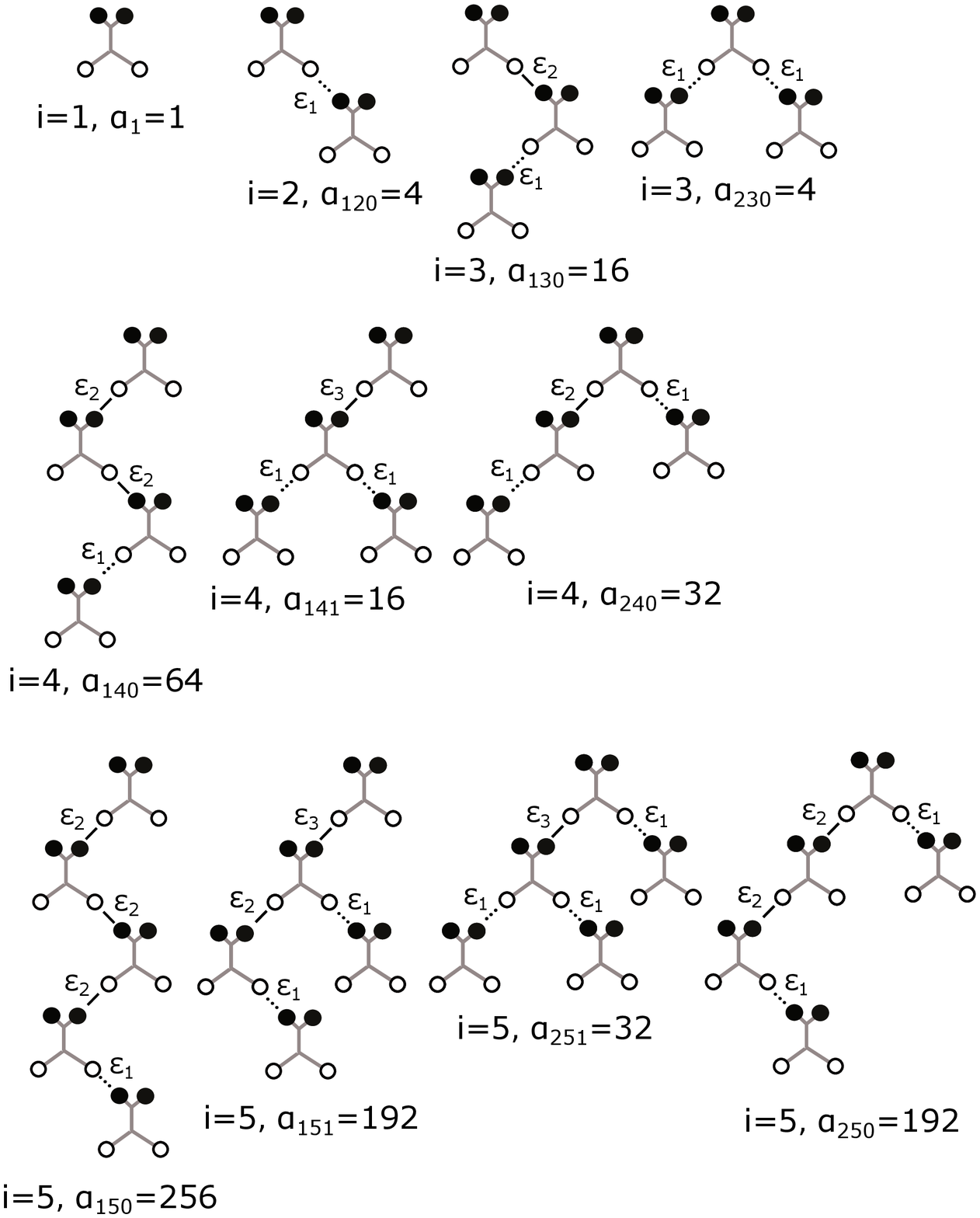}
\caption{Schematic representation of aggregates with size up to $i=5$ in model 7. Only configurations corresponding to different energies are shown.}
\label{model7-3const}
\end{figure}

We denote the energy of a bond formed by an acceptor with a free
donor group by $\epsilon_1$ and the energy of a
bond formed by an acceptor molecule with one or two bonded
hydrogens by $\epsilon_2$. As in some previous models, we will first consider a more
detailed case (with three association constants), in which the situations
corresponding to one bonded hydrogen in an acceptor molecule
($\epsilon_2$) and to two bonded hydrogens in an acceptor molecule
($\epsilon_3$) differ from each other (see
Figure~\ref{model7-3const}). We write $M_1$ for the number of bonds
with energy $\epsilon_1$, $M_2$ for the number of bonds with energy
$\epsilon_2$, and $M_3$ for the number of bonds with energy $\epsilon_3$
(see Figure \ref{model7-3const}). Similarly, we write $N_1$ for the
number of free molecules, $N_2$ for the number of molecules with a free
oxygen and one bonded hydrogen, $N_3$ for the number of molecules with
a free oxygen and two bonded hydrogens, $N_4$ for the number of molecules
with a bonded oxygen and both hydrogens free, $N_5$ for the number of
molecules with a bonded oxygen and one bonded hydrogen, and $N_6$ for the
number of molecules with a bonded oxygen and two bonded hydrogens. Among these values, the following relations exist: $M_1=N_4$, $M_2=N_5$, $M_3=N_6$, $N=\sum_{i=1}^{6}N_i$, $M_1+M_2+M_3=N_2+2N_3+N_5+2N_6$.

In this case the number of ways to choose acceptors is
\begin{equation}
\frac{N!2^{N_4+N_5+N_6}}{\left(N-N_4-N_5-N_6\right)!}.
\end{equation}
The number of ways to choose donors is (this expression is effectively
the same thing as Equation~\ref{main-equality})
\begin{equation}
\frac{\left(N-N_4-N_5-N_6\right)!2^{N_2}}{N_1!N_2!N_3!}\frac{\left(N_4+N_5+N_6\right)!2^{N_5}}{N_4!N_5!N_6!},
\end{equation}
and finally we have
\begin{equation}
\Xi=\frac{N!2^{2N-2N_1-2N_4}\left(N_4+N_5+N_6\right)!}{N_1!\left(2N-2N_1-3N_4-2N_5-N_6\right)!\left(N_1+2N_4+N_5-N\right)!N_4!N_5!N_6!}.
\end{equation}
Minimization of the free energy yields the following set of equations:
\begin{equation}
\frac{4N_1\left(N_1+2N_4+N_5-N\right)}{\left(2N-2N_1-3N_4-2N_5-N_6\right)^2}=1
\end{equation}
\begin{equation}
\frac{4N_4\left(N_1+2N_4+N_5-N\right)^2}{\left(2N-2N_1-3N_4-2N_5-N_6\right)^3\left(N_4+N_5+N_6\right)}=\frac{K_1}{V}
\end{equation}
\begin{equation}
\frac{N_5\left(N_1+2N_4+N_5-N\right)}{\left(2N-2N_1-3N_4-2N_5-N_6\right)^2\left(N_4+N_5+N_6\right)}=\frac{K_2}{V}
\end{equation}
\begin{equation}
\frac{N_6}{\left(2N-2N_1-3N_4-2N_5-N_6\right)\left(N_4+N_5+N_6\right)}=\frac{K_3}{V}.
\end{equation}


For the concentrations of aggregates, we may write
\begin{equation}
C_{ijk}=\alpha_{ijk}K_1^{i-1-j-k}K_2^{j}K_3^{k}c_1^i,
\label{model16_c}
\end{equation}
where $0 \leq j \leq i-2$ and there are two series of $k$ values that satisfy $2k+j+2=i$ for $j$ and $i$ both odd or even and $2k+j+3=i$ otherwise.
Based on these relations, one index may be removed and two series of
concentrations introduced instead:
\begin{equation}
C_{1ik}=\alpha_{1ik}K_1^{k+1}K_2^{i-2k-2}K_3^{k}c_1^i, 0\le k \le \infty, 2k+2 \le i \le \infty
\end{equation}
\begin{equation}
C_{2ik}=\alpha_{2ik}K_1^{k+2}K_2^{i-2k-3}K_3^{k}c_1^i, 0\le k \le \infty, 2k+3 \le i \le \infty.
\end{equation}

Therefore, we can write
\begin{equation}
c=c_1+\sum_{k=0}^{\infty}\sum_{i=2k+2}^{\infty}i\alpha_{1ik}K_1^{k+1}K_2^{i-2k-2}K_3^{k}c_1^i+\sum_{k=0}^{\infty}\sum_{i=2k+3}^{\infty}i\alpha_{2ik}K_1^{k+2}K_2^{i-2k-3}K_3^{k}c_1^i
\end{equation}

\begin{equation}
n_1=c_1
\end{equation}

\begin{equation}
n_4=\sum_{k=0}^{\infty}\sum_{i=2k+2}^{\infty}\left(k+1\right)\alpha_{1ik}K_1^{k+1}K_2^{i-2k-2}K_3^{k}c_1^i+\sum_{k=0}^{\infty}\sum_{i=2k+3}^{\infty}\left(k+2\right)\alpha_{2ik}K_1^{k+2}K_2^{i-2k-3}K_3^{k}c_1^i
\end{equation}

\begin{equation}
\begin{split}
n_5=&\sum_{k=0}^{\infty}\sum_{i=2k+2}^{\infty}\left(i-2k-2\right)\alpha_{1ik}K_1^{k+1}K_2^{i-2k-2}K_3^{k}c_1^i\\
+&\sum_{k=0}^{\infty}\sum_{i=2k+3}^{\infty}\left(i-2k-3\right)\alpha_{2ik}K_1^{k+2}K_2^{i-2k-3}K_3^{k}c_1^i
\end{split}
\end{equation}

\begin{equation}
n_6=\sum_{k=0}^{\infty}\sum_{i=2k+2}^{\infty}k\alpha_{1ik}K_1^{k+1}K_2^{i-2k-2}K_3^{k}c_1^i+\sum_{k=0}^{\infty}\sum_{i=2k+3}^{\infty}k\alpha_{2ik}K_1^{k+2}K_2^{i-2k-3}K_3^{k}c_1^i
\end{equation}

\begin{equation}
\alpha_{1ik}=4^{i-1-k}\frac{\left(i-2\right)!}{\left(i-2-2k\right)!k!\left(k+1\right)!}
\end{equation}
\begin{equation}
\alpha_{2ik}=4^{i-2-k}\frac{\left(i-2\right)!\left(2k+2\right)}{\left(i-3-2k\right)!\left(k+1\right)!\left(k+2\right)!}
\end{equation}

By summing the series, assuming that $K_2=K_3$, it can be shown that

\begin{equation}
\begin{split}
c=\frac{1-4c_1K_2}{8c_1K_2^2\sqrt{1-8c_1K_2-16c_1^2\left(K_1-K_2\right)K_2}}+\frac{8c_1^2K_2\left(K_2-K_1\right)-1}{8c_1K_2^2}.
\label{model7-c}
\end{split}
\end{equation}

Finally, the number of free groups, which is the sum of $c_1$ and
$n_4$, can be calculated as
\begin{equation}
n_f=c_1-\frac{c_1K_1}{K_2}+\frac{c_1K_1}{K_2\sqrt{1-8c_1K_2-16c_1^2\left(K_1-K_2\right)K_2}}.
\end{equation}

\subsubsection{Model 7m}
Here, it is assumed that the $3530\text{cm}^{-1}$ peak corresponds to
the out-of-phase vibrations of the $\text{NH}_2$ group in free
molecules. In this case, we put $c_1=Ax$ and substitute it into Equation~\ref{model7-c}. 

\begin{figure}
\centering
\includegraphics{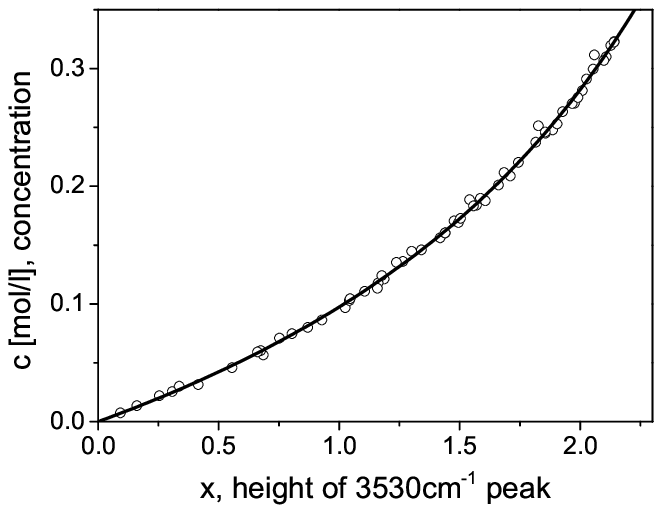}
\caption{Fit of the dependence of the total concentration on the
  height of the $3530\text{cm}^{-1}$ peak at $T=22^\circ\text{C}$ with model 7m.}
\label{model-7m-22C}
\end{figure}

\begin{figure}
\centering
\includegraphics{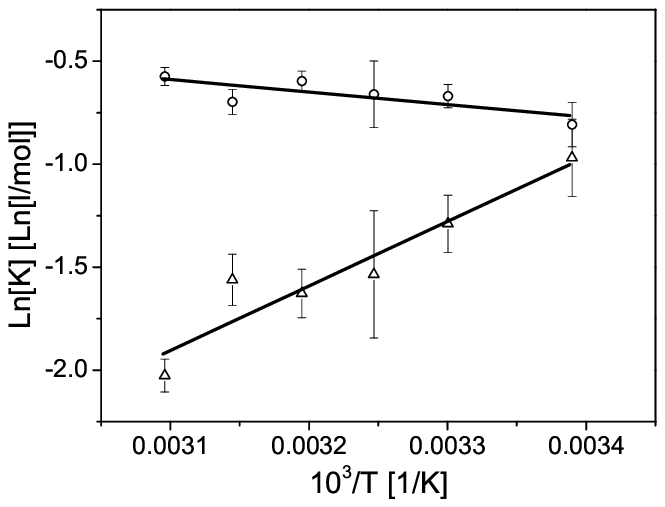}
\caption{Dependence of $\ln K_1$ (triangles) and $\ln K_2$ (circles) on $1/T$ for model 7m.}
\label{model-7m-k}
\end{figure}

The fitting results are shown in Figures~\ref{model-7m-22C}
and~\ref{model-7m-k}. The quality of fit in Figure~\ref{model-7m-22C}
can be characterized by $\text{AICc}=-501.7$, and the estimates of the
model parameters for the fits in Figure~\ref{model-7m-k} are $\ln
C_1=-11.6\pm 2 \,\ln[\text{l/mol}]$, $\epsilon_1=-6.2 \pm 1
\,\text{kcal/mol}$, $\ln C_2=1.3\pm 0.8 \,\ln[\text{l/mol}]$, and
$\epsilon_2=1.2 \pm 0.5 \,\text{kcal/mol}$, with $r_1^2=0.9$ and $r_2^2=0.62$. 

\subsubsection{Model 7g}
Next, it is assumed that the $3530\text{cm}^{-1}$ peak corresponds to absorption by free groups, so that $n_f=Ax$.

\begin{figure}
\centering
\includegraphics{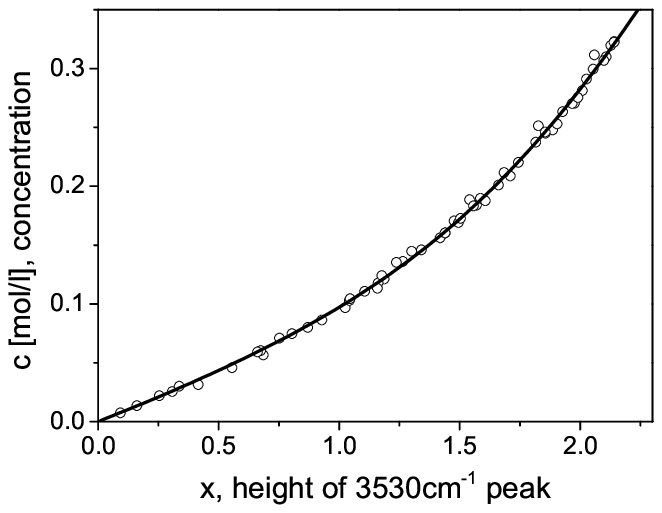}
\caption{Fit of the dependence of the total concentration on the
  height of the $3530\text{cm}^{-1}$ peak at $T=22^\circ\text{C}$ with model 7g.}
\label{model-7g-22C}
\end{figure}

\begin{figure}
\centering
\includegraphics{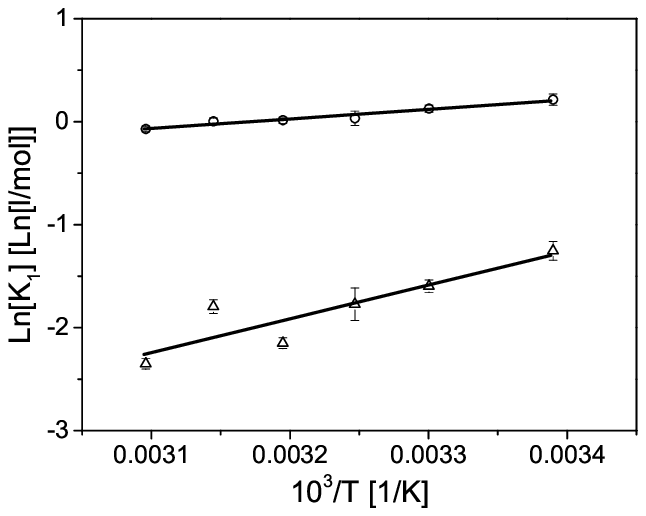}
\caption{Dependence of $\ln K_1$ (triangles) and $\ln K_2$ (circles) on $1/T$ for model 7g.}
\label{model-7g-k}
\end{figure}

The results of the fitting procedure are shown in
Figures~\ref{model-7g-22C} and~\ref{model-7g-k}. The value of AICc for
the fit in Figure~\ref{model-7g-22C} is $-497.4$. The estimates of the
model parameters for the fits in Figure~\ref{model-7g-k} are $\ln
C_1=-12.4\pm 3 \,\ln[\text{l/mol}]$, $\epsilon_1=-6.5 \pm 1.6
\,\text{kcal/mol}$, $\ln C_2=-2.9\pm 0.3 \,\ln[\text{l/mol}]$, and
$\epsilon_2=-1.8 \pm 0.2 \,\text{kcal/mol}$, with $r_1^2=0.80$ and $r_2^2=0.95$. 

\subsubsection{Model 7s}
\begin{figure}
\centering
\includegraphics{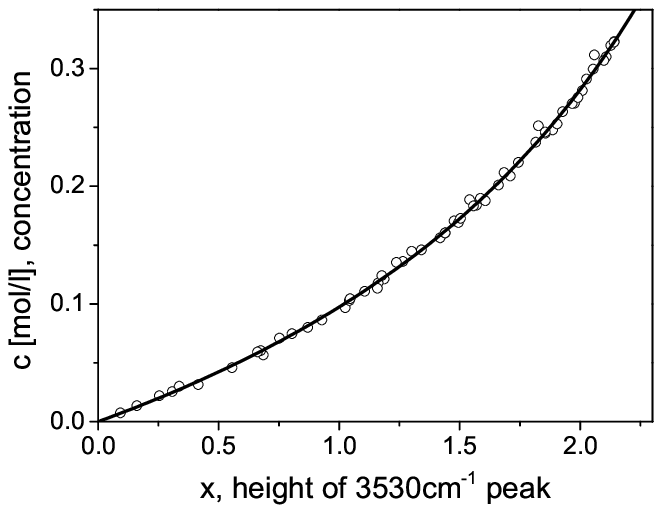}
\caption{Fit of the dependence of the total concentration on the
  height of the $3530\text{cm}^{-1}$ peak at $T=22^\circ\text{C}$ with model 7s.}
\label{model-7s-22C}
\end{figure}

\begin{figure}
\centering
\includegraphics{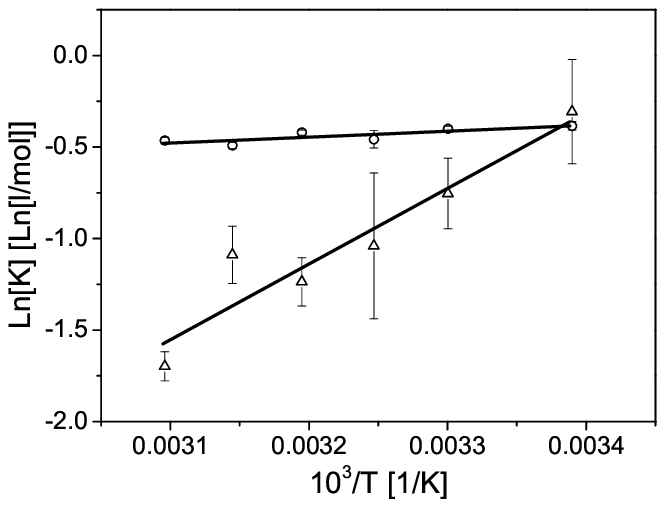}
\caption{Dependence of $\ln K_1$ (triangles) and $\ln K_2$ (circles) on $1/T$ for model 7s.}
\label{model-7s-k}
\end{figure}

In the case of the assumption  $Ax=c_1+4K_1c_1^2+4K_1^2c_1^3$
($\text{NH}_2$ groups in unimers, dimers and  trimers without
$\epsilon_2$-bonds) the fitting results shown in
Figures~\ref{model-7s-22C} and~\ref{model-7s-k} are found. 
The value of the information criterion for the fit in
Figure~\ref{model-7s-22C} is $\text{AICc}=-501.7$, and the estimates
of the model parameters given by the fits in Figure~\ref{model-7s-k}
are $\ln C_1=-14.4\pm 2 \,\ln[\text{l/mol}]$, $\epsilon_1=-8.2 \pm 1.4
\,\text{kcal/mol}$, $\ln C_2=-1.5\pm 0.3 \,\ln[\text{l/mol}]$, and
$\epsilon_2=-0.6\pm 0.2 \,\text{kcal/mol}$, with $r_1^2=0.9$ and $r_2^2=0.69$. 

\subsection{Model 8}
\begin{figure}
\centering
\includegraphics[width=0.7\textwidth]{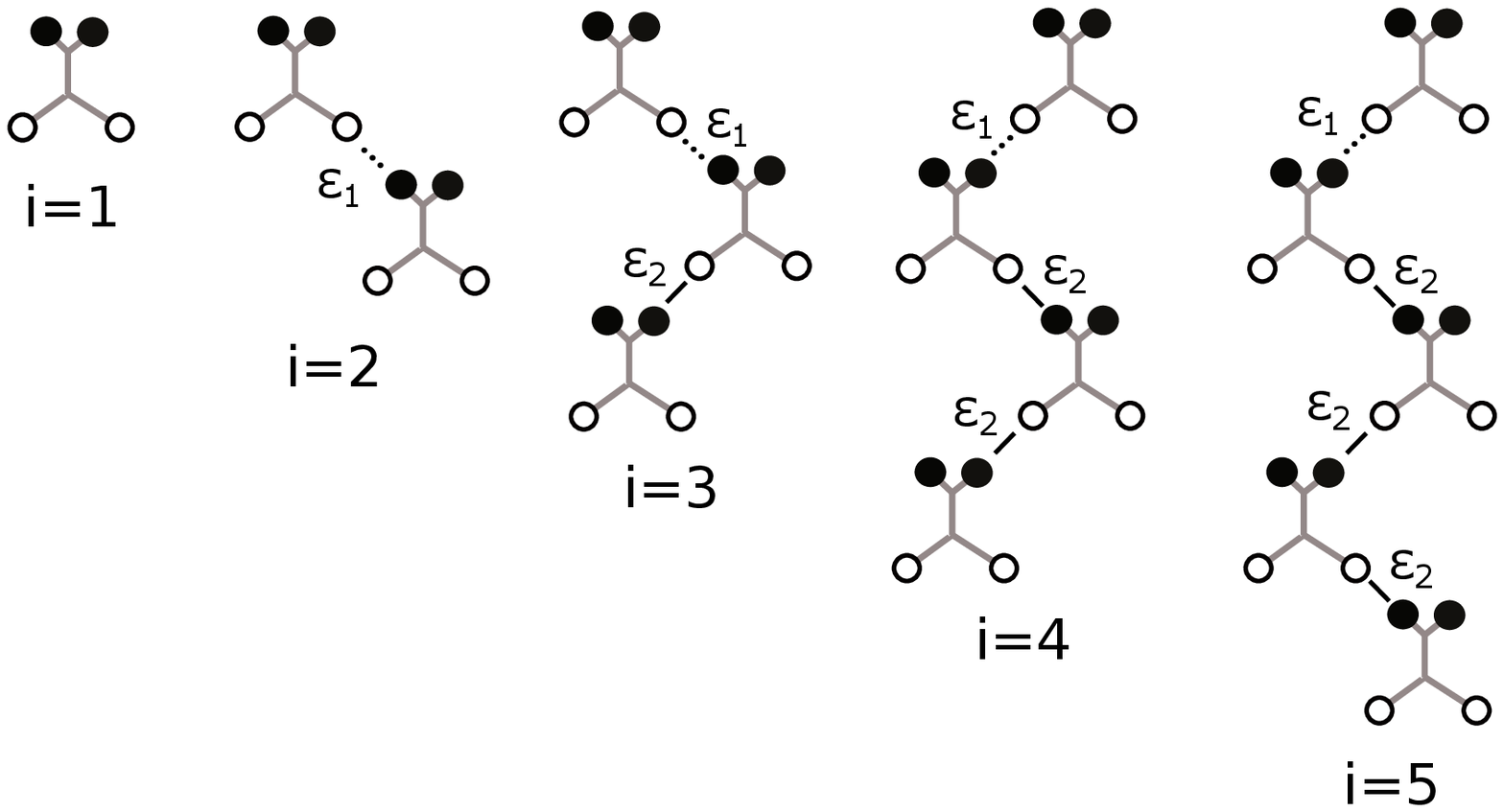}
\caption{Schematic representation of aggregates with size up to $i=5$ in model 8. Only configurations corresponding to different energies are shown.}
\label{model8-scheme}
\end{figure}

In model 8, we allow only one bond per oxygen and one bond per
$\text{NH}_2$ group, so all aggregates are supposed to be linear. We
also assume that the association equilibrium is described by two
association constants and there are no cyclic dimers. We suppose that
the bond has energy $\epsilon_1$ if the acceptor of donor molecule is free and $\epsilon_2$ otherwise (see Figure~\ref{model8-scheme}). 
It is worth mentioning that this model does not change if we assume
that the energy of the bond is defined by the bonding state of the donor
group in the acceptor molecule. 
The number of free molecules will be denoted by $N_0$, the number of molecules with one bonded hydrogen and free oxygen by $N_1$, the number of molecules with one bonded hydrogen and one bond per oxygen by $N_2$, and the number of molecules with one bond per oxygen and free hydrogens by $N_3$. Then, the number of ways to form bonds is
\begin{equation}
\Xi=\frac{N!2^{N_3+N_2}}{\left(N-N_3-N_2\right)!}\frac{\left(N_0+N_1\right)!2^{N_1}}{N_1!N_0!}\frac{\left(N_3+N_2\right)!2^{N_2}}{N_2!N_3!},
\end{equation}
where the first factor is the number of ways to choose acceptors, the
next factor is the number of ways to choose donors for bonds with
energy $\epsilon_1$, and the following factor is the number of ways to
choose  bonds with energy $\epsilon_2$.

Taking into account that $N_3=N_1$ and $N_0=N-2N_1-N_2$, we have
\begin{equation}
\Xi=\frac{N!\left(N_1+N_2\right)!2^{2N_1+2N_2}}{\left(N-2N_1-N_2\right)!\left(N_1!\right)^2N_2!}.
\end{equation}
Minimization of the free energy gives
\begin{equation}
\frac{N_1^2}{4\left(N_1+N_2\right)\left(N-2N_1-N_2\right)^2}=\frac{K_1}{V}
\end{equation}
\begin{equation}
\frac{N_2}{4\left(N_1+N_2\right)\left(N-2N_1-N_2\right)}=\frac{K_2}{V}.
\end{equation}
In this model it is straightforward to guess that 
\begin{equation}
c_i= c_1^i 4^{i-1}K_1K_2^{i-2}.
\end{equation}
Then, the total concentration can be calculated as
\begin{equation}
c=\sum_{i=1}^{\infty}ic_i=c_1+\frac{8c_1^2K_1\left(1-2c_1K_2\right)}{\left(1-4c_1K_2\right)^2}.
\label{model8-c}
\end{equation}
For the concentration of the free groups, $n_f=c-n_1-n_2$, we have
\begin{equation}
n_f=c_1+\frac{4K_1c_1^2}{1-4K_2c_1}.
\end{equation}

\subsubsection{Model 8m}
Let us assume first that the $3530\text{cm}^{-1}$ peak corresponds to
the out-of-phase vibrations of the $\text{NH}_2$ group in free
molecules. In this case, we put $c_1=Ax$ and substitute it into Equation~\ref{model8-c}. 

\begin{figure}
\centering
\includegraphics{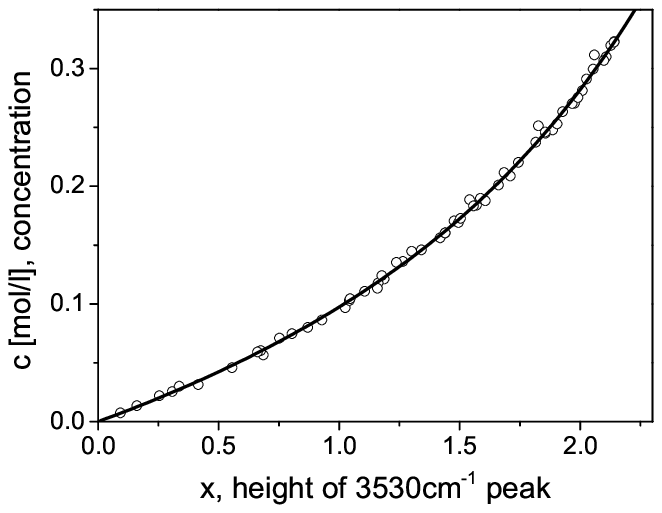}
\caption{Fit of the dependence of the total concentration on the
  height of the $3530\text{cm}^{-1}$ peak at $T=22^\circ\text{C}$ with model 8m.}
\label{model-8m-22C}
\end{figure}

\begin{figure}
\centering
\includegraphics{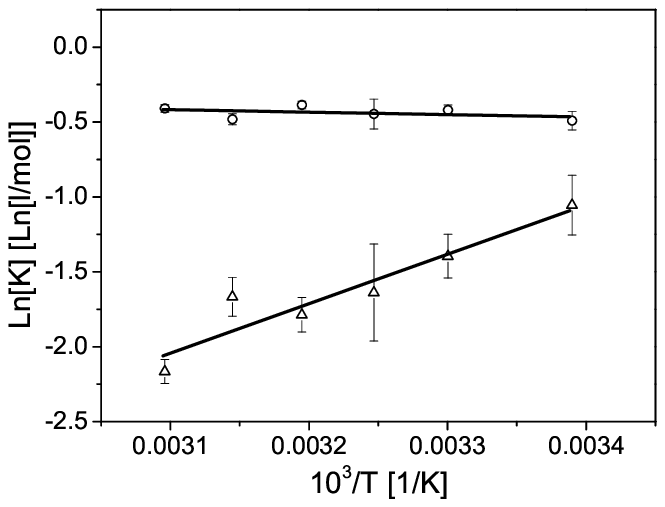}
\caption{Dependence of $\ln K_1$ (triangles) and $\ln K_2$ (circles) on $1/T$ for model 8m.}
\label{model-8m-k}
\end{figure}

The fitting results are shown in Figures~\ref{model-8m-22C}
and~\ref{model-8m-k}. The quality of fit in Figure~\ref{model-8m-22C}
can be characterized by $\text{AICc}=-501.5$, and the estimates of the
model parameters yielded by the fits in Figure~\ref{model-8m-k} are
$\ln C_1=-12.3\pm 2 \,\ln[\text{l/mol}]$, $\epsilon_1=-6.5 \pm 1
\,\text{kcal/mol}$, $\ln C_2=0.1\pm 0.6 \,\ln[\text{l/mol}]$, and
$\epsilon_2=0.3 \pm 0.5 \,\text{kcal/mol}$, with $r_1^2=0.89$ and $r_2^2=0.18$. 

\subsubsection{Model 8g}
Next, it is assumed that the $3530\text{cm}^{-1}$ peak corresponds to
absorption by free groups, so that $n_f=Ax$.

\begin{figure}
\centering
\includegraphics{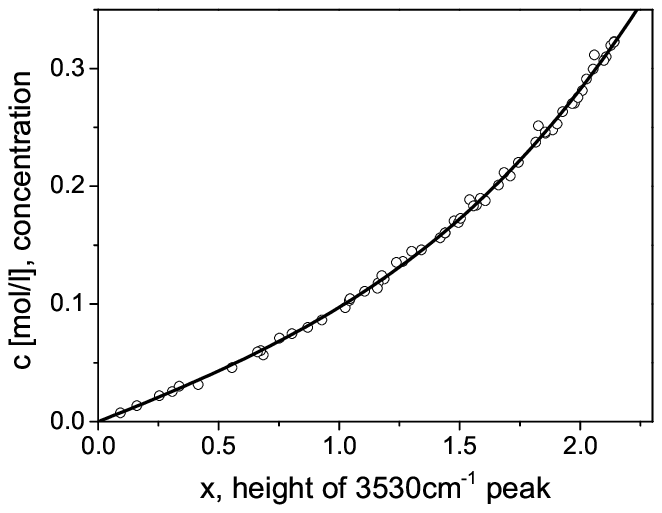}
\caption{Fit of the dependence of the total concentration on the
  height of the $3530\text{cm}^{-1}$ peak at $T=22^\circ\text{C}$ with model 8g.}
\label{model-8g-22C}
\end{figure}

\begin{figure}
\centering
\includegraphics{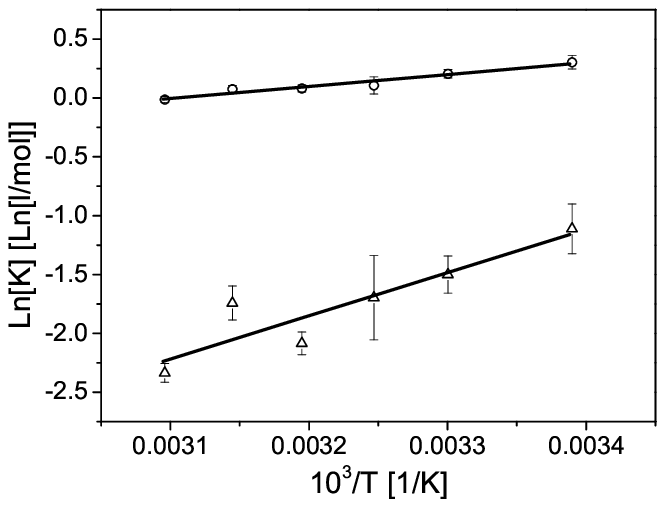}
\caption{Dependence of $\ln K_1$ (triangles) and $\ln K_2$ (circles) on $1/T$ for model 8g.}
\label{model-8g-k}
\end{figure}

The fitting results are shown in Figures~\ref{model-8g-22C}
and~\ref{model-8g-k}. The value of the AICc parameter characterizing
the quality of fit in Figure~\ref{model-8g-22C} is $-499.5$, and the
estimates of the model parameters resulting from the fits in
Figure~\ref{model-8g-k} are $\ln C_1=-13.6\pm 3 \,\ln[\text{l/mol}]$,
$\epsilon_1=-7.3 \pm 1.6 \,\text{kcal/mol}$, $\ln C_2=-3.2\pm 0.4
\,\ln[\text{l/mol}]$, and $\epsilon_2=-2.0 \pm 0.2 \,\text{kcal/mol}$,
with $r_1^2=0.83$ and $r_2^2=0.95$. 

\subsubsection{Model 8s}
\begin{figure}
\centering
\includegraphics{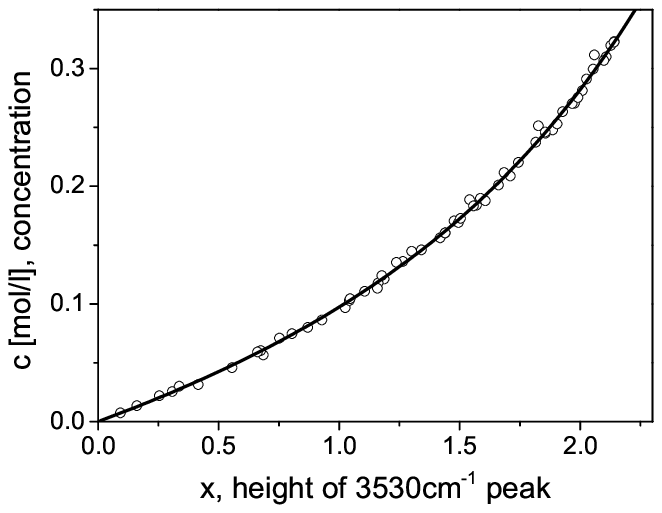}
\caption{Fit of the dependence of the total concentration on the
  height of the $3530\text{cm}^{-1}$ peak at $T=22^\circ\text{C}$ with model 8s.}
\label{model-8s-22C}
\end{figure}

\begin{figure}
\centering
\includegraphics{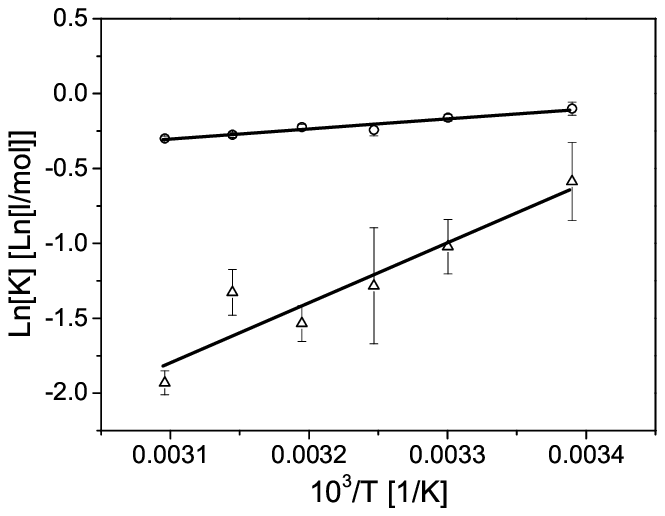}
\caption{Dependence of $\ln K_1$ (triangles) and $\ln K_2$ (circles) on $1/T$ for model 8s.}
\label{model-8s-k}
\end{figure}

In the case of the assumption  $Ax=c_1+4K_1c_1^2$ ($\text{NH}_2$
groups in unimers and dimers without $\epsilon_2$-bonds), the fitting
results shown in Figures~\ref{model-8s-22C} and~\ref{model-8s-k} are found. 
The quality of fit in Figure~\ref{model-8s-22C} can be characterized
by $\text{AICc}=-501.4$. The estimates of the model parameters given
by the fits in Figure \ref{model-8s-k} are $\ln C_1=-14.2\pm 2
\,\ln[\text{l/mol}]$, $\epsilon_1=-7.9 \pm 1.4 \,\text{kcal/mol}$, $\ln
C_2=-2.4\pm 0.3 \,\ln[\text{l/mol}]$, and $\epsilon_2=-1.3\pm 0.2
\,\text{kcal/mol}$, with $r_1^2=0.88$ and $r_2^2=0.93$. 

\subsection{Model 9}
In this model, we allow two bonds per oxygen and only one bond per
$\text{NH}_2$ group. We assume that the energy of the bond is determined
by the bonding state of the oxygen. If the oxygen forms only one bond,
then its energy is $\epsilon_1$, and if it forms two then its energy is
$\epsilon_2$. This means that this model is analogous to model 5 with
the roles of the acceptor and donor group exchanged. Therefore, all
equations are the same, apart from the value of the number of free $\text{NH}_2$ groups. It is clear that in case of model 9 there is only one free group per each aggregate, so we have
\begin{equation}
n_f=\frac{1-4c_1K_1-\sqrt{1-8c_1K_1+16c_1^2K_1^2-16c_1^2K_2^2}}{8c_1K_2^2}.
\end{equation}

We do not repeat the fitting for models 9m and 9s as these cases are equivalent to models 5m and 5s.

\subsubsection{Model 9g}
We assume that the $3530\text{cm}^{-1}$ peak corresponds to absorption
by free groups, so that $n_f=Ax$. The fitting results are not shown,
because at higher temperatures they give a negative value of $K_1$.

\subsection{Model 10}

In this model, we allow two bonds per oxygen and only one bond per
$\text{NH}_2$ group. We assume that the energy of the bond is determined
by the bonding state of the $\text{NH}_2$ group in the acceptor
molecule. If the hydrogen in the acceptor molecule is free, then the bond energy is $\epsilon_1$; otherwise, it is $\epsilon_2$.

Therefore, this model is analogous to model 6 with the roles of the acceptor
and donor groups exchanged, and all equations are the same, the only
difference being the
value of the number of free $\text{NH}_2$ groups. It is clear that,
in the case of model 9, there is only one free group per each aggregate, so we have
\begin{equation}
\begin{split}
n_f=&c_1+\frac{K_1\left(1-4K_2c_1+\sqrt{1-8c_1K_2}\right)}{2K_2^2}\\
+&\frac{K_2^2\left(1-8c_1K_2+8c_1^2K_2^2-\sqrt{1-8c_1K_2}+4c_1K_2\sqrt{1-8c_1K_2}\right)}{8c_1K_2^4}.
\end{split}
\end{equation}

\subsubsection{Model 10g}
Here, we assume that the $3530\text{cm}^{-1}$ peak corresponds to absorption by free groups, so that $n_f=Ax$.

\begin{figure}
\centering
\includegraphics{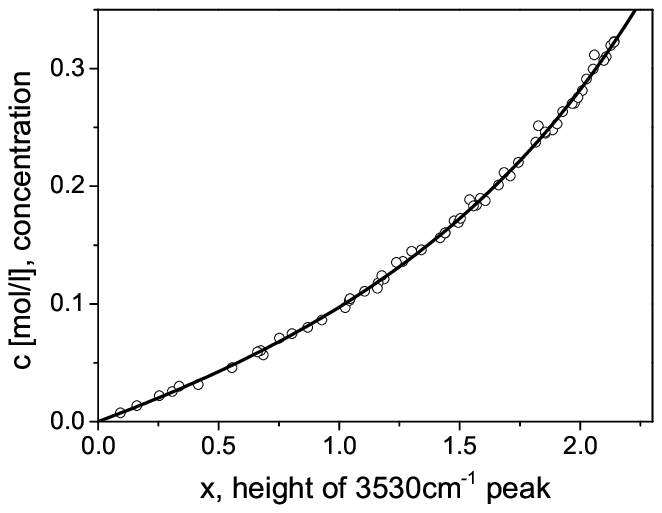}
\caption{Fit of the dependence of the total concentration on the
  height of the $3530\text{cm}^{-1}$ peak at $T=22^\circ\text{C}$ with model 10g.}
\label{model-10g-22C}
\end{figure}

\begin{figure}
\centering
\includegraphics{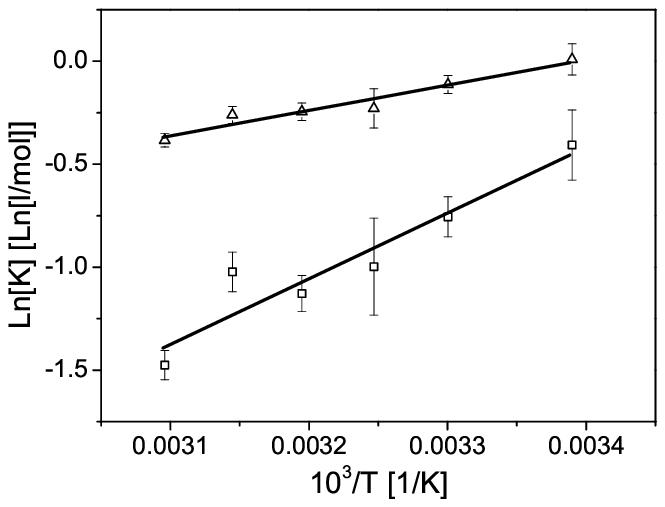}
\caption{Dependence of $\ln K_1$ (triangles) and $\ln K_2$ (circles) on $1/T$ for model 10g.}
\label{model-10g-k}
\end{figure}

The results of the fitting procedure are shown in
Figures~\ref{model-10g-22C} and~\ref{model-10g-k}. The quality of fit
in Figure~\ref{model-10g-22C} can be characterized by
$\text{AICc}=-501.1$. The estimates of the model parameters for the
fits in Figure~\ref{model-10g-k} are $\ln C_1=-11.3\pm 2
\,\ln[\text{l/mol}]$, $\epsilon_1=-6.3 \pm 1.0 \,\text{kcal/mol}$, $\ln
C_2=-4.2\pm 0.5 \,\ln[\text{l/mol}]$, and $\epsilon_2=-2.4 \pm 0.3
\,\text{kcal/mol}$,  with $r_1^2=0.90$ and $r_2^2=0.95$. 

\subsection{Model 11}

In this model we allow two bonds per oxygen and only one bond per
$\text{NH}_2$ group. We assume that the energy of the bond is determined
by the bonding state of the acceptor in the donor molecules. If the
oxygen in the donor molecule is free then the bond energy is $\epsilon_1$; otherwise, it is $\epsilon_2$.

Therefore, this model is analogous to model 7 with the roles of the
acceptor and donor group exchanged. All equations are the same, except
for the value of the number of free $\text{NH}_2$ groups. Again, as in
the two previous models, there is only one free group per aggregate
\begin{equation}
n_f=\frac{1-4K_2c_1-8K_2c_1^2\left(K_2-K_1\right)-\sqrt{1-8K_2c_1-16K_2\left(K_1-K_2\right)c_1^2}}{8K_2^2c_1}.
\end{equation}

\subsubsection{Model 11g}
Here we assume that the $3530\text{cm}^{-1}$ peak corresponds to
absorption by free groups, so that $n_f=Ax$. However, in this case, the fits do not converge.

\section{The best model selection}

To compare the quality of fit of different models we use Akaike's
information criterion. It states that the best model is that with the
smallest value of
\begin{equation}
AIC=2k-2\ln\left(L\left(A,K_1,K_2,\sigma\right)\right),
\end{equation}
where $L\left(A,K_1,K_2,\sigma\right)$ is a likelihood function and
$k$ is the number of model parameters (4 in two-parameter models and 3
in one-parameter models because $\sigma$ is also included in the
context of the likelihood function). The likelihood function for the
model with parameters obtained by minimization of the sum of squared deviations is usually written in the form
\begin{equation}
\begin{split}
L\left(A,K_1,K_2,\sigma\right)=\Pi_{i=1}^{n}\frac{1}{\sqrt{2\pi \sigma^2}} \exp\left\lbrace-\frac{\left(y_i-\text{model}\left(x_i,A,K_1,K_2\right)\right)^2}{2\sigma^2}\right\rbrace.
\end{split}
\end{equation}
In this case, minimization of the sum of squared deviations is
equivalent to maximization of the logarithm of the likelihood
function. Then, for the AIC value we have
\begin{equation}
\begin{split}
\text{AIC}=&2k+n\ln\left(2\pi\right)+n\ln\left(\hat{\sigma}^2\right)+\frac{\left(y_i-\text{model}\left(x_i\right)\right)^2}{2\hat{\sigma}^2}\\
=&2k+n\ln\left(2\pi\right)+n\ln\frac{\text{RSS}}{n}+n,
\label{AIC}
\end{split}
\end{equation}
where the estimate for $\hat{\sigma}$ is
$\hat{\sigma}^2=\text{RSS}/n$ and the sum of the squares of the residuals is \\
$\text{RSS}=\sum_{i=1}^n\left(y_i-\text{model}\left(x_i\right)\right)^2$.

For a finite sample size there exists the following correction:
\begin{equation}
\text{AICc}=\text{AIC}+\frac{2\left(k+1\right)\left(k+2\right)}{n-k-2},
\end{equation}
which is useful when we compare models with different number of parameters.

The relative probabilities of the two models with values of the
information criterion given by $\text{AIC}_1$ and $\text{AIC}_2$ respectively can be estimated as
\begin{equation}
\exp\left(\frac{\text{AIC}_1-\text{AIC}_2}{2}\right)
\end{equation}
So, if $\text{AIC}_1-\text{AIC}_2=-501.7+493.9=-7.8$ then model with
value $\text{AIC}_1$ is $49.4$ times more probable than the model with
value $\text{AIC}_2$. 

\clearpage
\end{document}